\documentclass[10pt]{article}
\usepackage[margin=1in]{geometry}
\usepackage{amsfonts,amsmath,amssymb}
\usepackage[none]{hyphenat}
\usepackage{fancyhdr}
\usepackage{graphicx}
\usepackage{float}
\usepackage[nottoc,notlot,notlof]{tocbibind}

\usepackage{amssymb,amsmath,tikz}
\usepackage{pgfplots} 
\usepackage{amsthm}
\usepackage{mathrsfs}
\usepackage{pifont}
\usepackage{slashed}
\usepackage{mathtools}
\usepackage{cancel}

\showoutput
\DeclareMathOperator*{\p}{p}
\DeclareMathOperator*{\x}{x}
\DeclareMathOperator*{\y}{y}
\DeclareMathOperator*{\q}{k}

\newcommand*{\ud}{\mathrm{\,d}}

\theoremstyle{plain}

\newtheorem*{twr*}{Theorem}
\newtheorem*{lem*}{Lemma}
\newtheorem{twr}{Theorem}
\newtheorem{lem}{Lemma}
\newtheorem{defin}{Definition}
\newtheorem*{defin*}{Definition}

\newtheorem*{rem*}{Remark}

\newtheorem{cor*}{Corollary}

\newtheorem*{notn*}{Notation}
\newtheorem*{wiener-ito*}{Wiener-It\^o-Segal Decomposition}
\newtheorem*{prop*}{Proposition}

\DeclareMathAlphabet{\mathpzc}{OT1}{pzc}{m}{it}

\begin{document}

\begin{titlepage}
\begin{center}
\vspace*{1cm}
\large{\textbf{CAUSAL PERTURBATIVE QED AND WHITE NOISE.}}\\
\large{\textbf{THE INDUCTIVE STEP}}\\
\vspace*{2cm}
\small{JAROS{\L}AW WAWRZYCKI}\\[1mm]
\tiny{Bogoliubov Labolatory of Theoretical Physics}
\\
\tiny{Joint Institute of Nuclear Research, 141980 Dubna, Russia}
\\
\tiny{jaroslaw.wawrzycki@wp.pl}
\\
\vspace*{1cm}
\tiny{\today}\\
\vfill
\begin{abstract}
We present the Bogoliubov's causal perturbative QFT, which includes 
only one refinement: the creation-annihilation operators at a point, \emph{i.e.} 
for a specific momentum, are mathematically 
interpreted as the Hida operators from the white noise analysis. 
We leave the rest of the theory completely unchanged. This allows avoiding 
infrared-- and ultraviolet -- divergences in the transition 
to the adiabatic limit for interacting fields. 
Here we present the inductive step. We are especially
concentrated with the analysis of the second order contribution to the scattering operator, and give explicit computations
using the calculus of integral kernel operators, \emph{i.e} contractions of their vector-valued kernels,
and splittings of the contractions into advanced and retarded parts. Explicit example of QED is provided, together with
the detailed analysis of the key point in avoiding UV divergence.   
\end{abstract}
\vspace*{0.5cm}
\tiny{{\bf Keywords}: scattering operator, causal perturbative method in QFT, \\
white noise, Hida operators, integral kernel operators, Fock expansion}
\end{center}
\vfill
\end{titlepage}

\tableofcontents
\thispagestyle{empty}
\clearpage

\setcounter{page}{1}

\section{Introduction}

In \cite{Bogoliubov_Shirkov} there was initiated a rigorous formulation of the renormalization
procedure in perturbative QFT, based on the so-called Bogoliubov cusality axioms for the scattering 
operator. However, \cite{Bogoliubov_Shirkov} left some freedom as to the rigorous mathematical interpretation
of the generalized operators representing free fields, their Wick products and higher order contributions
$S_n$ to the scattering operator, indicatng only that they should be some kind of generalized operators. 
Epstein and Glaser \cite{Epstein-Glaser} developed these ideas, and make them rigorous,
by adding the assumption that the generalized free fields, their Wick products, and higher order 
contributions $S_n$ are understood precisely as the operator valued distributions in the Wightman sense. 
They were able to construct rigorously and iductively all higher order contributions $S_n$, using
only the Bogliubov axioms. Instead of the formal procedure of multiplying by the step theta function, and then
renormalization of the ill-defined products of distributions by the step function, they construct
the advanced and retarded parts of the causally supported operator distributions, using the splitting of causlly 
supported tempered distributions into the advanced and retarded parts. The only additional axiom they 
used during the splitting is that the retarded and advanced parts should have the same singularity order at zero
(in space-time coordinates) as the splitted causal distribution. In this way they make the perturbative
QFT mathematically transparent and remove all ultraviolet (UV) divergences. Besides the mathematical transparency,
this method substantially simplifies the investigation of renormalizabilty \cite{Epstein-Glaser}. 
Recall that the splitting is not unique, and depends on the singularity degree at zero (in space-time)
of the causal distribution. Theory is renormalizable if the singularity degrees of all higher order contributions
remain bouned, with the bound independent of the order $n$. In this case theory is fixed by a finite set of
arbitrary constants. However, in theories with the interacion Lagrangian $\mathcal{L}(x)$ containing massless fields some
infrared (IR) problems remained unsolved: in general the limit $g \rightarrow 1$ does not exists for the higher order contributions
to the scattering operator as well as for the interacting fields (``adiabatic limit probem'') . This is e.g. the case for QED. 

In \cite{WN} and \cite{IF} we have proposed to improve the Bogoliubov-Epstein-Glaser perturbative QFT by reinterpreting 
the generalized operators (the free fields, their Wick products, higher order contributions $S_n$)
and regard them not as the Wightman's operator distributions, but as the (finite sums) of generalized
integral kernel operators with vector-valued distributional kernels in the white noise sense \cite{obataJFA}.
There are several advantages of this approach. 
First, this approach allows for a deeper examination of the adiabatic problem.
Namely, it allows to
prove that the adiabatic limit for interacting fields does exist in various QED's with the $U(1)$-coupling and massive charged field
 if and only if the charged field is massive and the splitting into advanced and retarded parts of the causal distributions 
is ``natural'' \cite{IF}. After introducing necessary definitions and theorems we give a rigorous formulation 
of this statement at the end of Section \ref{idea}. 
Thus, the condition of existence of the adiabatic limit puts additional and non-trivial conditions on the arbitrary constants
inherent in the splitting and on the mass of the charged field. For example: in QED with massless charged fields, the adiabatic limit does not exist
for no choice of the normalization in the splittig. Therefore, it allows proving some theorems which were impossible 
to prove within the more traditional approach, e.g. theorem that the charged fields are massive, or theorems concerning IR asymptotics
of QED. Second, this approach throws some new light on the nonrenormalizable case, when we have massless fields
in the interaction Lagrangian $\mathcal{L}(x)$. In the traditional approach based on Wightman's operator vaued distributions, the number of arbitrary constants 
associated with the choice of the splitting grows together with the order. In this context, it is therefore important that
the requirement of the existence of the adiabatic limit, puts non-trivial conditions on the arbitrary constants which 
appear in the splitting, compare Section \ref{Ren}. It should be emphasized, that these results are obtained without imposing any \emph{ad hoc}
additional assumptions, except the over whole assumption that the annihilation and creation operators of the free fields
are identified with the Hida operators!

With this motivation in mind, we show in this paper that Bogliubov axioms with Hida operators can serve as a practical computational basis for $S_n$, 
and give the Epstein-Glaser inductive step \cite{Epstein-Glaser} in terms of the calculus of integral kernel operators of Hida, Obata and Sait\^o \cite{obataJFA}.
We show how the Epstein-Glaser inductive step simplifies the computation of the higher order terms in comparison to the standard approach, 
based on renormalization procedure.
In this paper we are mainly concentrated in detailed analysis and practical computation of the second order contribution $S_2(x,y)$
for general $\mathcal{L}$, including spinor QED. General existence proofs for $S_n$ we have already presented in \cite{WN},
but in order to make the present paper maximally independent, we have decided to give all details for $S_2$ 
in the language of integral kernel operators (in the sense of \cite{obataJFA}),
as this approach is not used in QFT, and practically unknown by mathematical physicists working with perturbative QFT. 
Thus, we have repeated existence proofs for $S_2(x,y)$ in the language of the vector-valued kernels, their contractions,
and with adaptation of the Epstein-Glaser splitting to the vector-valued distributional kernels. 
We are giving practical formulas for the computation of the
retarded and advanced parts of contractions of the kernels of the Wick decomposition 
of the products $\mathcal{L}(x)\mathcal{L}(y)$ of the interaction Lagrangian density operator $\mathcal{L}$ 
together with their analysis. This has not been presented in \cite{WN}, \cite{IF}.
At the end of Section \ref{idea}, after giving in more details the whole idea,  we provide a more detailed
and technical plan for the whole paper.

\section{The main idea}\label{idea}

We are using the Hida white noise operators 
(here $\boldsymbol{\p}$ subsumes the spatial momenta components and the corresponding discrete spin
components in order to simplify notation) 
\[
\partial_{\boldsymbol{\p}}^{*}, \,\,\, \partial_{\boldsymbol{\p}}
\]
which respect the canonical commutation or anticommutation relations
\[
\big[\partial_{\boldsymbol{\p}}, \partial_{\boldsymbol{\q}}^{*}\big]_{{}_{\mp}} = \delta(\boldsymbol{\p}-\boldsymbol{\q}),
\]
as the creation-annihilation operators 
\[
a(\boldsymbol{\p})^{+}, \,\,\, a(\boldsymbol{\p}),
\]
of the free fields in the causal perturbative Bogoliubov-Epstein-Glaser QFT, \cite{Bogoliubov_Shirkov}, \cite{Epstein-Glaser}, 
leaving all the rest of the axioms of the theory completely unchanged.
\emph{I.e.}
we are using the standard Gelfand triple
\[
\left. \begin{array}{ccccc} 
E & \subset & \mathcal{H} & \subset & E^* 
\end{array}\right.,
\]
over the single particle Hilbert space $\mathcal{H}$ of the total system of free fields 
determined by the corresponding standard self-adjoint operator $A$ in $\mathcal{H}$ (with some positive power $A^r$ being nuclear), 
and its lifting to the standard Gelfand triple 
\[
\left. \begin{array}{ccccc} 
(E) & \subset & \Gamma(\mathcal{H}) & \subset & (E)^* 
\end{array}\right.,
\]
over the total Fock space $\Gamma(\mathcal{H})$ of the total system of free fields underlying the actual QFT, e.g. QED, 
which is naturally determined by the self-adjoint and standard operator $\Gamma(A)$, and
with the nuclear Hida test space $(E)$ and its strong dual $(E)^*$.  Here $E$ is the single particle test space in the total single particle space
in the total Fock space, and is equal to the direct sum of the particular single particle test spaces $E_1, E_2, \ldots$ of the particular free fields
$\mathbb{A}^{(1)}, \mathbb{A}^{(2)}, \ldots$ undelying the QFT in question,
which together with their strong duals $E_{1}^{*}, \ldots $ and the total single particle Hilbert spaces
compose Gelfand triples. (The respective single particle test spaces, the direct summands of $E$,
of the corresponding free fields $\mathbb{A}^{(1)}, \mathbb{A}^{(2)}, \ldots$,  will be denoted by 
$E_1, E_2, \ldots$ or $E_{{}_{'}}, E_{{}_{''}}, \ldots$. 
\[
\left. \begin{array}{ccccc} 
E & \subset & \mathcal{H} & \subset & E^* \\
\parallel & & \parallel & & \parallel \\
E_1 \oplus \ldots \oplus E_N & & \mathcal{H}_1 \oplus \ldots \oplus \mathcal{H}_N & & E_{1}^{*} \oplus \ldots \oplus E_{N}^{*}
\end{array}\right.,
\]
In general the nuclear spaces $E_1, E_2, \ldots$ are initially defined as linear spaces of smooth sections of smooth vector bundles. 
In each case they are naturally unitary isomorphic to the
nuclear spaces of test $\mathbb{C}^{d_1}$, $\mathbb{C}^{d_2}$, $\ldots$-valued functions on the corresponding orbits 
$\mathscr{O} = \{p: p\cdot p =m, p_0\geq 0  \}$ in momentum space for the fields of mass $m$, which can be represented as test function
spaces  $E_1 = \mathcal{S}(\mathbb{R}^3; \mathbb{C}^{d_1}), \ldots$ in massive case, or as  
$E_1 = \mathcal{S}^{0}(\mathbb{R}^3; \mathbb{C}^{d_1}), \ldots$, in massless case $m=0$, and are restrictions to the correponding orbits
$\mathscr{O}$ of the Fourier transforms of the scalar, vector, spinor, $\ldots$, (depending on the kind of field $\mathbb{A}^{(1)}, \ldots$) 
functions lying, respectively, in the Fourier inverse images of   $\mathcal{S}(\mathbb{R}^4; \mathbb{C}^{d_1}), \ldots$ or
$\mathcal{S}^{0}(\mathbb{R}^4; \mathbb{C}^{d_1}), \ldots$. Here $\mathcal{S}^{0}(\mathbb{R}^n; \mathbb{C}^{d})$ is the closed subspace
of $\mathcal{S}(\mathbb{R}^n; \mathbb{C}^{d})$ of all those functions who's all derivatives vanish at zero.
The natural unitary isomorphism is in each case defined through the smooth idempotent associated to the smooth bundle 
$E_1, E_2, \ldots$, and which is also continuous in the nuclear topologies (recall e.g. the ``projection'', or rather idempotent,
defining the single particle space of the free Dirac field):
\[
\left. \begin{array}{ccccc} \mathcal{S}_{A}(\sqcup \mathbb{R}^3; \mathbb{C}) & \subset & L^2(\sqcup \mathbb{R}^3;\mathbb{C}) & \subset & \mathcal{S}_{A}(\sqcup \mathbb{R}^3; \mathbb{C})^* \\
\downarrow \uparrow & & \downarrow \uparrow & & \downarrow \uparrow \\
E & \subset & \mathcal{H} & \subset & E^*
\end{array}\right..
\]
This means that the space $E$ consisting of direct sums of restrictions of the Fourier transforms of space-time test $\mathbb{C}^d$-valued 
functions (or rather smooth sections -- their images under a smooth idempotent in case of higher spin charged fields)
to the respective orbits $\mathscr{O}$ in momenta defining the representation of $T_4 \ltimes SL(2, \mathbb{C})$ in the Mackey's classification
(acting in the single particle Hilbert space for each corresponding free field) is given the standard realization with the help of a standard operator
$A$ in $L^2(\sqcup \mathbb{R}^3; \mathbb{C}) \cong \mathcal{H}$ (with standard $A$, \emph{i.e.} 
self-adjoint positive, with some negative power of which being nuclear, or trace-class,
and with the minimal spectral value greater than $1$). It is equal to the 
the direct sum $\oplus_{i=1}^{N} A_i$ of the standard operators $A_i$ corresponding to the single particle Hilbert space
of the $i$-th free field. Thus, first we need to construct the standard $A_i$ and the standard Gelfand triples for each of the free fields
of the theory.  $A= \oplus_{i=1}^{N} A_i$ is also standard.  It serves for the construction of the standard Gelfand
triple over the full single particle Hilbert space $L^2(\sqcup \mathbb{R}^3; \mathbb{C}) \cong \mathcal{H}$. 
Namely, with the help of the operator $A$ we construct $E$ as the following projective limit
\[
E = \underset{k\in \mathbb{N}}{\bigcap} \textrm{Dom} \, A^k
\]
with the standard realization of $E$ as the countably Hilbert nuclear (in the sense of Grothendieck) test space and its strong (also
nuclear) dual as the inductive limit
\[
E^* = \underset{k\in \mathbb{N}}{\bigcup} \textrm{Dom} A^{-k}
\]
with the Hilbertian defining norms
\[
\big| \cdot \big|_{{}_{k}} \overset{\textrm{df}}{=} \big|A^k \cdot \big|_{{}_{L^2}}, \,\,\,\,\,\,
\big| \cdot \big|_{{}_{-k}} \overset{\textrm{df}}{=} \big|A^{-k} \cdot \big|_{{}_{L^2}}, \,\,\, k = 0,1,2,3, \ldots.
\]

For any $\Phi$ in $(E)$ or in $(E)^*$ let
\[
\Phi = \sum\limits_{n=0}^{\infty} \Phi_{n} \,\,\,\,
\textrm{with} \,\,\, \Phi_n \in E^{\hat{\otimes} \, n} \,\, \textrm{or, respectively}, \,\,\, \Phi_n \in E^{*\hat{\otimes} \, n}
\]
be its decomposition into $n$-particle states of an element $\Phi$ of the test Hida space $(E)$
or in its strong dual $(E)^*$, convergent, respectively, in $(E)$ or in $(E)^*$. We define
\[
\begin{split}
a(w) \Phi_0 = 0, \,\,\,\, a(w) \Phi_n = n \, \overline{w} \hat{\otimes}_1 \Phi_n
\\
a(w)^+ \Phi_n = w \hat{\otimes} \Phi_n, \,\,\,\, \textrm{for each fixed} \,\, w \in E^*.
\end{split}
\]
\begin{defin}
The Hida operators are obtained when we put here the Dirac delta functional $\delta_{{}_{s,\boldsymbol{p}}}$ for $w$
\[
\partial_{{}_{s,\boldsymbol{p}}} = a_s(\boldsymbol{p}) = a(\delta_{{}_{s,\boldsymbol{p}}}),
\,\,\,\,
\partial_{{}_{s,\boldsymbol{p}}}^+ = a_s(\boldsymbol{p})^+ =  a(\delta_{{}_{s,\boldsymbol{p}}})^+.
\]
\label{Def:HidaOperators}
\end{defin}

Let $\mathscr{L}(E_1,E_2)$ be the linear space of linear continuous operators $E_1\longrightarrow E_2$ endowed with the natural topology of uniform
convergence on bounded sets. 

For each fixed spin-momentum point $(s,\boldsymbol{p})$ the Hida operators are well-defined
(generalized) operators 
\[
\begin{split}
a_s(\boldsymbol{p}) \in  \mathscr{L} \big((E), (E)\big) \subset  \mathscr{L} \big((E), (E)^*\big),
\\
a_s(\boldsymbol{p})^+ \in  \mathscr{L} \big((E)^*, (E)^*\big) \subset  \mathscr{L} \big((E), (E)^*\big),
\end{split}
\]
with the last inclusion coming from the natural topological inclusion $(E) \subset (E)^*$. 
Let $\phi \in \mathscr{E}$ (here $\mathscr{E}$ is the space-time test space $\mathcal{S}$ or $\mathcal{S}^{00}$) 
and let $\kappa_{l, m}$ be any $L(\mathscr{E}, \mathbb{C}) = \mathscr{E}^*$-valued distribution
\[
\kappa_{l, m} \in  \mathscr{L}(E^{\hat{\otimes} (l+m)}, \mathscr{E}^*)  
= L(\mathscr{E}, E^{*\hat{\otimes} (l+m)}) = E^{* \hat{\otimes} (l+m)} \otimes \mathscr{E}^*, 
\]
then we put
\begin{defin}
\[
\left. \begin{array}{ccc} 
\Xi_{{}_{l,m}}(\kappa_{{}_{l,m}})(\Phi \otimes \phi) & \overset{\textrm{df}}{=}  &
\sum\limits_{n=0}^{\infty} \kappa_{l,m} \otimes_m (\Phi_{n+m} \otimes \phi)  \\
\parallel & & \parallel   \\
\Xi_{{}_{l,m}}\big(\kappa_{{}_{l,m}}(\phi)\big) \, \Phi &  &  \sum\limits_{n=0}^{\infty} \kappa_{l,m}(\phi) \otimes_m \Phi_{n+m},
\end{array}\right.
\]
\label{Def:Xi}
\end{defin}
which for any $\mathscr{E}^*$-valued distribution $\kappa_{l,m}$ is a well-defined (generalized) operator 
\[
\Xi_{{}_{l,m}}(\kappa_{lm}) = 
\int \kappa_{lm}\big(\boldsymbol{p}_1, \ldots, \boldsymbol{p}_l,
\boldsymbol{q}_1, \ldots, \boldsymbol{q}_m\big) \,\,
\partial_{\boldsymbol{p}_1}^{*} \ldots \partial_{\boldsymbol{p}_l}^{*} 
\partial_{\boldsymbol{q}_1} \ldots \partial_{\boldsymbol{q}_m} 
d\boldsymbol{p}_1 \ldots d\boldsymbol{p}_l
d\boldsymbol{q}_1 \ldots d\boldsymbol{q}_m,
\]
\[
\Xi_{{}_{l,m}}(\kappa_{{}_{l,m}}) \in \mathscr{L}\big((E)\otimes \mathscr{E}, (E)^* \big) \cong
\mathscr{L}\big(\mathscr{E}, \mathscr{L}((E),(E)^*)\big).
\]
For brevity of notation $\boldsymbol{p}_i$ denote in this formula 
spin-momentum variables  $(s_i,\boldsymbol{p}_i)$ and integrations include summations with respect to spin components
$s_i$.

$\Xi_{{}_{l,m}}(\kappa_{{}_{l,m}})$ defines integral kernel operator $\Xi_{{}_{l,m}}(\kappa_{{}_{l,m}})$ which is uniquely determined by the condition
\[
\langle\langle \Xi_{{}_{l,m}}(\kappa_{{}_{l,m}})(\Phi \otimes \phi), \Psi \rangle\rangle
= \langle \kappa_{{}_{l,m}}(\eta_{{}_{\Phi,\Psi}}), \phi \rangle,
\,\,\,
\Phi, \Psi \in (E), \phi \in \mathscr{E}
\] 
or, respectively,
\[
\langle\langle \Xi_{{}_{l,m}}(\kappa_{{}_{l,m}})(\Phi \otimes \phi), \Psi \rangle\rangle
= \langle \kappa_{{}_{l,m}}(\phi), \eta_{{}_{\Phi,\Psi}}\rangle,
\,\,\,
\Phi, \Psi \in (E), \phi \in \mathscr{E},
\] 
depending on $\kappa_{{}_{l,m}}$ is regarded as an element of
\[
\mathscr{L}(E^{\hat{\otimes} (l+m)}, \mathscr{E}^*)  
\,\,\,
\textrm{or, respectively, of} \,\,\,
 \mathscr{L}(\mathscr{E}, E^{*\hat{\otimes} (l+m)}).
\]
Here 
\begin{multline*}
\eta_{{}_{\Phi,\Psi}}(s_1, \boldsymbol{p}_1, \ldots, s_l, \boldsymbol{p}_l,
s_{l+1}, \boldsymbol{p}_{l+1}, \ldots, s_{l+m}, \boldsymbol{p}_{l+m})
\\
\overset{\textrm{df}}{=}
\langle\langle a_{s_1}(\boldsymbol{p}_{1})^+ \ldots a_{s_l}(\boldsymbol{p}_{l})^+ a_{s_{l+1}}(\boldsymbol{p}_{l+1})
\ldots a_{s_{l+m}}(\boldsymbol{p}_{l+m})\Phi, \Psi \rangle\rangle
\end{multline*}
is the function which always belongs to $E^{\hat{\otimes}(l+m)}$, c.f. \cite{obataJFA}. 

For example the free fields $\mathbb{A}$ are always equal to sums of the two 
integral kernel operators 
\[
\mathbb{A}(\phi) = \mathbb{A}^{(-)}(\phi) + \mathbb{A}^{(+)}(\phi) = \Xi(\kappa_{0,1}(\phi)) +
\Xi(\kappa_{1,0}(\phi))
\]
with the integral kernels $\kappa_{l,m}$ represented by ordinary functions, and giving
the``negative'' and, respectively, the ``positive frequency parts''. For concrete examples compare the following 
Sections.

\begin{lem}
The massless free field operators $\mathbb{A}=A$, and the massive free field operators $\mathbb{A}=\boldsymbol{\psi}$ are operator-valued distributions, 
\emph{i.e.} belong 
to $\mathscr{L}\big(\mathscr{E}, \mathscr{L}((E),(E))\big)$, \emph{i.e.}
\[
\kappa_{0,1}, \kappa_{1,0} \in \mathscr{L}(E^*, \mathscr{E}^{*}) = \mathscr{L}(\mathscr{E}, E) \subset \mathscr{L}(E, \mathscr{E}^{*})
\cong E^{*} \otimes \mathscr{E}^{*},
\]
if and only if
\[
\begin{split}
E= \mathcal{S}^{0}(\mathbb{R}^3; \mathbb{C}^{d'}) = \mathcal{S}_{A_{(3)}}(\mathbb{R}^3;
\mathbb{C}^{d'}), \,\, \mathscr{E} = \mathcal{S}^{00}(\mathbb{R}^4; \mathbb{C}^{d'}) = \mathcal{S}_{\widetilde{A}_{(4)}}(\mathbb{R}^4;
\mathbb{C}^{d'})
\,\,\,\, \textrm{for} \,\,\, A
\\
E= \mathcal{S}(\mathbb{R}^3; \mathbb{C}^{d''}) = \mathcal{S}_{\widetilde{H}_{(3)}}(\mathbb{R}^3;
\mathbb{C}^{d''}), \,\, \mathscr{E} = \mathcal{S}(\mathbb{R}^4; \mathbb{C}^{d''}) = \mathcal{S}_{H_{(4)}}(\mathbb{R}^4; \mathbb{C}^{d''})
\,\,\,\, \textrm{for} \,\,\, \boldsymbol{\psi}.
\end{split}
\]
In this case, moreover,
\[
\kappa_{0,1}, \kappa_{1,0} \in \mathscr{L}(E, \mathcal{O}_M)
\]
and for each $\xi \in E$, $\kappa_{0,1}(\xi), \kappa_{1,0}(\xi)$ are smooth having all derivatives bounded.
\qed
\label{Cont.free.field.kernels}
\end{lem}

Here $\mathcal{S}^{0}(\mathbb{R}^n;\mathbb{C})=\mathcal{S}_{A_{(n)}}(\mathbb{R}^n;\mathbb{C})$
is the closed subspace of the Schwartz space $\mathcal{S}(\mathbb{R}^n;\mathbb{C})$ of all functions
whose all derivatives vanish at zero. $\mathcal{S}^{00}(\mathbb{R}^n;\mathbb{C})$ is the Fourier transform
inverse image of $\mathcal{S}^{0}(\mathbb{R}^n;\mathbb{C})$. The space $\mathcal{S}^{0}(\mathbb{R}^n;\mathbb{C})$ can be  
realized as a countably Hilbert nuclear space $\mathcal{S}_{A_{(n)}}(\mathbb{R}^n;\mathbb{C})$ associated, in the sense of \cite{GelfandIV},
\cite{obata}, with a positive self adjoint operator $A_{(n)}$ in $L^2(\mathbb{R}^n; d^np)$, with $\textrm{Inf Spec} \, A_{(n)} >1$
whose some negative power $\big[A_{(n)} \big]^{-r}$, $r>0$, is of Hilbert-Schmidt class. 
In order to  construct an example of a series
of operators $A_{(n)}$, $n=2,3, \ldots$, let us consider the Hamiltonian operators 
\[
H_{(n)} = - \Delta_{{}_{\mathbb{R}^n}} + r^2 +1,
\,\,\,\,\,\,\,\, r^2= (p_{1})^2 + \ldots +(p_{n})^2,
\,\,\,\,\,\,\,
\textrm{in}
\,\,
L^2(\mathbb{R}^n, d^np),
\] 
and the following $L^2(\mathbb{R}\times \mathbb{S}^{n-1}; dt\times d\mu_{{}_{\mathbb{S}^{n-1}}})$,
$L^2(\mathbb{R}\times \mathbb{S}^{n-1}; \nu_{{}_{n}}(t) dt\times d\mu_{{}_{\mathbb{S}^{n-1}}})$ -spaces
on $\mathbb{R}\times \mathbb{S}^{n-1}$ with the weights
\[
\nu_{{}_{n}}(t) = {\textstyle\frac{(t+\sqrt{t^2+4})^{n-1}}{2^{n-2}(t^2+4-t\sqrt{t^2+4})}},
\] 
and the following unitary operators $U_2: L^2(\mathbb{R}\times \mathbb{S}^{n-1}; \nu_{{}_{n}}(t) dt\times d\mu_{{}_{\mathbb{S}^{n-1}}})
\rightarrow L^2(\mathbb{R}^n; d^n p)$, $U_1: L^2(\mathbb{R}\times \mathbb{S}^{n-1}; dt\times d\mu_{{}_{\mathbb{S}^{n-1}}})
\rightarrow L^2(\mathbb{R}\times \mathbb{S}^{n-1}; \nu_{{}_{n}}(t) dt\times d\mu_{{}_{\mathbb{S}^{n-1}}})$,
$U= U_2U_1: L^2(\mathbb{R}\times \mathbb{S}^{n-1}; dt\times d\mu_{{}_{\mathbb{S}^{n-1}}}) \rightarrow L^2(\mathbb{R}^n; d^n p)$
given by the following formulas
\begin{gather*}
U_1 f(t,\omega) = {\textstyle\frac{1}{\sqrt{\nu_{{}_{n}}(t)}}}f(t,\omega), 
\,\,\,\,\,\,\,\,\,\,\, f \in L^2(\mathbb{R}\times \mathbb{S}^{n-1}; dt\times d\mu_{{}_{\mathbb{S}^{n-1}}}),
\\
U_2 f(r,\omega) = f(t(r),\omega), 
\,\,\,\,\,\,\,\,\,
t(r) = r-r^{-1},
\,\,\,\,\,\,\,\,\,
f\in L^2(\mathbb{R}\times \mathbb{S}^{n-1}; \nu_{{}_{n}}(t) dt \times d\mu_{{}_{\mathbb{S}^{n-1}}}),
\end{gather*}
where $U_2 f$ is expressed in spherical coordinates $(r,\omega)$ in $\mathbb{R}^n$. We can put
\[
A_{(n)} = U \big(H_{(1)}\otimes \boldsymbol{1} + \boldsymbol{1} \otimes \Delta_{{}_{\mathbb{S}^{n-1}}} \big) U^{-1}.
\]
Here $\Delta_{{}_{\mathbb{R}^n}}, \Delta_{{}_{\mathbb{S}^{n-1}}}$ are the standard Laplace operators on $\mathbb{R}^n$
and $\mathbb{S}^{n-1}$, with the standard invariant measures $d^np, d\mu_{{}_{\mathbb{S}^{n-1}}}$. In particular
\[
A_{(3)} = - {\textstyle\frac{r^2}{r^2+1}} \partial_{r}^{2} - {\textstyle\frac{r^3(r^2+4)}{(r^2+1)^3}} \partial_{r}
+ \Big[{\textstyle\frac{r^2(r^2+4)(r^2-2)}{4(r^2+1)^4}} + r^2+r^{-2} \Big]  
\]
in spherical coordinates. 

Let $\kappa^{1}_{0,1}, \kappa_{1,0}$, $\kappa^{2}_{0,1}, \kappa^{2}_{1,0}$, $\ldots, \kappa^{n}_{0,1}, \kappa^{n}_{1,0}$
be the plain wave kernels of the free fields $\mathbb{A}^{(1)}, \ldots, \mathbb{A}^{(n)}$. Then 
\[
\kappa_{lm} = \kappa^{1}_{l_1m_1} \dot{\otimes} \ldots \dot{\otimes} \kappa^{n}_{l_n m_n},
\,\,\,\, l = l_1+ \ldots + l_n, \,\, m = m_1+ \ldots + m_n 
\]
are the kernels of the Wick (pointwise) product operator
\begin{equation}\label{PointWiseWick}
{:}\mathbb{A}^{(1)}(x) \ldots \mathbb{A}^{(n)}(x) {:},
\end{equation} 
where $\dot{\otimes}$ denotes ordinary product $\kappa^{1}_{l_1m_1}(s_1, \boldsymbol{\p}_1;x) \ldots \kappa^{n}_{l_n m_n}(s_n, \boldsymbol{\p}_n;x)$ 
in which all factors $\kappa^{i}_{l_im_i}(s_i, \boldsymbol{\p}_i;x)$
are taken at the same space-time coordinate $x$ and the spin-momenta coordinates $(s_i, \boldsymbol{\p}_i)$ are regarded as independent. The kernels
$\kappa_{lm}$ so defined  should be symmetrized in boson spin-momentum
variables and antisymmetrized in the fermion spin-momentum variables
in order to keep one-to-one correspondence between the kernels and operators. For the (tensor) Wick product operator
\[
{:}\mathbb{A}^{(1)}(x_1) \ldots \mathbb{A}^{(n)}(x_n) {:},
\] 
with $n$ independent space-time variables  $x_j$, we have analogous formula for its kernels $\kappa_{lm}$, but with the
pointwise product $\dot{\otimes}$ replaced with ordinary tensor product $\otimes$, \emph{i.e.} with the product
\begin{multline*}
\kappa_{lm}(s_1, \boldsymbol{\p}_1, \ldots s_n, \boldsymbol{\p}_n, ;x_1, \ldots x_n) = \kappa^{1}_{l_1m_1} \otimes \ldots \otimes \kappa^{n}_{l_n m_n}
\big(s_1, \boldsymbol{\p}_1, \ldots s_n, \boldsymbol{\p}_n, ;x_1, \ldots x_n\big) 
\\
=\kappa^{1}_{l_1m_1}(s_1, \boldsymbol{\p}_1;x_1) \ldots \kappa^{n}_{l_n m_n}(s_n, \boldsymbol{\p}_n;x_n)
\,\,\,\, l = l_1+ \ldots + l_n, \,\, m = m_1+ \ldots + m_n 
\end{multline*}
in which the coordinates
$(s_i, \boldsymbol{\p}_i;x_i)$ of $\kappa^{i}_{l_im_i}$, including the space-time coordinates $x_i$, are independent. In this case the (tensor)
Wick product operator belongs to the class $\mathscr{L}\big(\mathscr{E}^{\otimes \, n}, \, \mathscr{L}((E),(E))\big)$, irrespectively if the 
free fields $\mathbb{A}^{(j)}$ are massive or massless. Again: the kernels
$\kappa_{lm}$ so defined  should be symmetrized in boson spin-momentum
variables and antisymmetrized in the fermion spin-momentum variables
in order to keep one-to-one correspondence between the kernels and operators.

We have the following easily verified lemma for the Wick (pointwise) product operator
\begin{lem}
\[
{:}\mathbb{A}^{(1)} \ldots \mathbb{A}^{(n)} {:} \in
\begin{cases}
\mathscr{L}\big(\mathscr{E}, \, \mathscr{L}((E),(E))\big), 
& \text{if all fields $\mathbb{A}^{(j)}$ are massive},\\
\mathscr{L}\big(\mathscr{E}, \, \mathscr{L}((E),(E)^*)\big), & \text{if some $\mathbb{A}^{(j)}$ are massless fields}.\\
\end{cases}
\]
\qed
\label{WickProduct}
\end{lem}

\begin{twr}
The standard Wick theorem decomposition holds for the (tensor) product operator
\[
{:}\mathbb{A}^{(1)}(x) \ldots \mathbb{A}^{(n)}(x) {:} {:}\mathbb{A}^{(n+1)}(y) \ldots \mathbb{A}^{(n+k)}(y) {:}
\]
with the kernels of the decomposition given by the contractions
\[
\kappa_{{}_{l,m}}(\phi\otimes \varphi) = \sum\limits_{\kappa'_{{}_{l',m'}},\kappa''_{{}_{l'',m''}}, k}
\kappa'_{{}_{l',m'}}(\phi) \otimes_{k} \kappa''_{{}_{l'',m''}}(\varphi)
\]
where in this sum $\kappa'_{{}_{l',m'}}$ and $\kappa''_{{}_{l'',m''}}$ range over the kernels respectively
of the operators
\[
{:}\mathbb{A}^{(1)}(x) \ldots \mathbb{A}^{(n)}(x) {:} \,\,\, \textrm{and} \,\,\, {:}\mathbb{A}^{(n+1)}(y) \ldots \mathbb{A}^{(n+k)}(y) {:}
\]
and
\[
l'+l''-k=l, \,\,\, m'+m''-k=m
\]
and where the contractios $\otimes_k$ are performed upon all $k$ pairs of spin-momenta variables in which the first variable in the pair
corresponds to an annihilation operator variable and the second one to the creation operator variable
or \emph{vice versa}. All these contractions are given by absolutely convergent sums/integrals with respect 
to the contracted variables.  After the contraction, 
the kernels should be symmetrized in boson spin-momentum
variables and antisymmetrized in the fermion spin-momentum variables
in order to keep one-to-one correspondence between the kernels and operators.
\qed
\label{WickDecomposition}
\end{twr}

In the proof of lemma \ref{WickProduct} and theorem \ref{WickDecomposition} we are using the characterization of the integral kernel
operators given by Hida, Obata and Sait\^o, c.f. \cite{obataJFA}. In particular, in the proof of lemma \ref{WickProduct}
we are using lemma \ref{Cont.free.field.kernels}, the analogue of lemma \ref{Cont.free.field.kernels}
for the products of the kernels of the free fields, and finally theorem \ref{Hida-ObataTheorem} given below. 
Difference between the products of massive and massless
free field kernels of the free fields comes from the singular nature of the light cone $\{p: p\cdot p=0, p_0>0\}$, 
which is not smooth at the origin $p=0$, contrary to the massive orbit $\{p: p\cdot p = m, p_0 >0\}$ which is everywhere a smooth
submanifond in the momentum space $\mathbb{R}^4$. The proof of theorem \ref{WickDecomposition} is slightly more laborious, compare
\cite{WN}, but the general idea is simple. The factors ${:}\mathbb{A}^{(1)} \ldots \mathbb{A}^{(n)} {:}(\phi)$
and ${:}\mathbb{A}^{(n+1)} \ldots \mathbb{A}^{(n+k)} {:}(\varphi)$, evaluated respectively at $\phi$ and $\varphi \in \mathscr{E}$,
are ordninary operators and the composition 
${:}\mathbb{A}^{(1)} \ldots \mathbb{A}^{(n)} {:}(\phi) \circ {:}\mathbb{A}^{(n+1)} \ldots \mathbb{A}^{(n+k)} {:}(\varphi)$ 
is meaningful, if all free fields  $\mathbb{A}^{(j)}$ are massive. This composition gives, in this case, the (tensor)
product of theorem \ref{WickDecomposition}, evaluated at the test function $\phi \otimes \varphi$, with the kernels
$\kappa_{{}_{l,m}}(\phi\otimes \varphi)$ given by the ordinary contractions. But if some  $\mathbb{A}^{(j)}$ are massless, the situation is more
subtle, as in this case ${:}\mathbb{A}^{(n+1)} \ldots \mathbb{A}^{(n+k)} {:}(\varphi)$ is a generalized operator transforming
the Hida space $(E)$ into its dual $(E)^*$, and the above stated composition has no immediate meaning. 
In this case, the four-momentum $p$ in the exponent $e^{\mp ip\cdot x}$ defining the plane wave kernels
$\kappa_{0,1}, \kappa_{1,0}$ of the massless free field $\mathbb{A}^{(j)}$, ranges over the corresponding massless orbit 
$\mathscr{O} = \{p: p \cdot p =0 \}$ with zero component $p_0$ of $p$ being the following
function $p_0(\boldsymbol{p}) = | \boldsymbol{p}|$ of the spatial momentum $\boldsymbol{p}$. We replace all the massless
exponents $e^{\mp ip\cdot x}$ in $\kappa_{0,1}, \kappa_{1,0}$ of the massless fields $\mathbb{A}^{(j)}$ 
by the massive exponents with mass $\epsilon$ and $p_0(\boldsymbol{p}) = \sqrt{| \boldsymbol{p}|^2 + \epsilon^2}$. 
After this replacement, the above stated composition is well-defined, as both composed operators map continuously $(E)$
into $(E)$. The limit $\epsilon \rightarrow 0$ 
of the said composition defines the evaluation of the (tensor) product operator of theorem \ref{WickDecomposition} 
at the test function $\phi \otimes \varphi$ in case where some $\mathbb{A}^{(j)}$ are massless. In fact the class of generalized
operators which is closed under the operation of (tensor) product is quite wide. In particular, using the same method of replacing the
exponents of the massless kernels with the exponents of the massive counterparts and passing to the limit, it can be proved the following
\begin{twr}
The class within which the (tensor) product of generalized operators (understood as finite sums of integral kernel operators
with vector valued kernels) is well-defined as a finite sum of integral kernel operators, includes all operators of the form   
\begin{equation}\label{ProductClassGenOp}
t(x_1, \ldots, x_n) \, {:}W_1(x_1)W_2(x_2) \ldots W(x_k){:}, \,\,\, t \in  \mathcal{S}(\mathbb{R}^{4k})^* = \mathcal{S}(\mathbb{R}^{4})^{* \, \otimes \, k},
\end{equation}
with translationally invariant $t$, and $W_i$ being Wick pointwise products of massless or massive free fields. 
\qed
\label{ClassWithProductThm}
\end{twr}
For a detailed proof, compare \cite{WN}. The proof is based on the following two theorems \ref{Hida-ObataTheorem} 
and \ref{ExtensionThm}, applied to 
$V = \mathscr{E}_{{}_{'}} = \mathcal{S}(\mathbb{R}^{4k'}), \mathscr{E}_{{}_{''}}= \mathcal{S}(\mathbb{R}^{4k''})$, 
or $V=\mathscr{E}_{{}_{'}}\otimes \mathscr{E}_{{}_{''}}$.

\begin{twr}[Hida, Obata\cite{obataJFA}]
\[
\Xi(\kappa_{lm}) \in
\begin{cases}
\mathscr{L}\big(V, \, \mathscr{L}((E),(E)^*)\big), 
\\
\mathscr{L}\big(V, \, \mathscr{L}((E),(E))\big),\\
\end{cases}
\]
\begin{center}
if and only if
\end{center}
\[
\begin{cases}
\kappa_{lm} \in 
\mathscr{L}\big(E^{\widehat{\otimes} \, (l+m)}, V^*\big), 
\\
\text{$\kappa_{lm}$ can be extended to a separately cont. map}:  E^{*\widehat{\otimes} l} \times  E^{\widehat{\otimes} m} \longrightarrow V^*,\\
\end{cases}
\]
\qed
\label{Hida-ObataTheorem}
\end{twr}

Let $E, E_1 , F, F_1$  be t.v.s. such that $E, F$ are, respectively, dense in $E_1, F_1$. 
Suppose that $\mathfrak{S}$ is a family of bounded subsets of $E$ with
the property that $\mathfrak{S}_1$ covers $E_1$ , where $\mathfrak{S}_1$ denotes the family of the closures,
taken in $E_1$ , of all subsets in $\mathfrak{S}$; analogously let $\mathfrak{F},\mathfrak{F}_1$ be such families
in  $F, F_1$; finally, let $G$ be a quasi-complete Hausdorff t.v.s. Under these assumptions,
the following extension theorem holds (compare e.g. the proposition 5.4, Chap. III.5.4 in the book\cite{Schaefer} of H. H. Schaefer):

\begin{twr}
Every $(\mathfrak{S}, \mathfrak{F})$-hypocontinuous bilinear mapping of $E \times F$ into $G$ has a unique
extension to $E_1 \times F_1$ (and into $G$) which is bilinear and$(\mathfrak{S}_1, \mathfrak{F}_1)$-hypocontinuous.
\qed
\label{ExtensionThm}
\end{twr}

Theorem \ref{ClassWithProductThm} allows the application of the white noise method to the Bogoliubov formulation of perturbative
QFT, based on the causality axioms. We can understand the free field operators, their Wick products, interaction Lagrangian generalized operators
$\mathcal{L}$ given by Wick polynomials,
and higher order contributions $S_n$ to the scattering operator, 
as finite sums of generalized integral kernel
operators in the sense of \cite{obataJFA}. 
Recall, please, that the causality axioms (I)-(V) for $S_{n}$ read:
\begin{gather*}
\textrm{(I)} \,\,\,\,\,\,\,\,\,\,\,\,\,\,\,\,\,\,\,\,   
              S_{n}(x_1, \ldots, x_{n}) = S_{k}(x_1, \ldots, x_{k})S_{n-k}(x_{k+1}, \ldots, x_{n}),
\,\,\,\, \textrm{whenever $\{x_{k+1}, \ldots, x_{n} \} \preceq \{x_1, \ldots, x_k\}$},   \\
\textrm{(II)}  \,\,\,\,\,\,\,\,\,\,\,\,\,\,\,\,\,\,\,\,      
U_{b,\Lambda} S_n(x_1, ..,x_n) U_{b, \Lambda}^{+} = S_n(\Lambda^{-1}x_1 - b, .., \Lambda^{-1}x_n - b),\,\,\,\,\,\,\,
\,\,\,\,\,\,\,\,\,\,\,\,\,\,\,\,\,\,\,\,\,\,\,\,\,\,\,\,\,\,\,\,\,\,\,\,\,\,\,\,\,\,\,\,\,\,\,\,\,\,\,\,\,\,\,\,\,\,
\,\,\,\,\,\,\,\,\,\,\,\,\,\,\,\,\,\,\,\,\,\,\,\,\,\,\,\,\,\,\,\,\,\,\,\,\,\,\,\,\, \\
\textrm{(III)}  \,\,\,\,\,\,\,\,\,\,\,\,\,\,\,\,\,\,\,\,        
\overline{S}_{n}(x_1, \ldots, x_{n}) = \eta S_n(x_1, \ldots, x_{n})^{+} \eta,\,\,\,\,\,\,\,\,\,\,\,\,\,\,\,\,\,\,\,\,\,\,\,\,\,\,
\,\,\,\,\,\,\,\,\,\,\,\,\,\,\,\,\,\,\,\,\,\,\,\,\,\,\,\,\,\,\,\,\,\,\,\,\,\,\,\,\,\,\,\,\,\,\,\,\,\,\,\,\,\,\,\,\,\,\,\,\,\,\,\,\,
\,\,\,\,\,\,\,\,\,\,\,\,\,\,\,\,\,\,\,\,\,\,\,\,\,\,\,\,\,\,\,\,\,\,\,\,\,\,\,\,\,\,\,\,\,\,\,\,\,\,\,\,\,\,\,\,\,\,\,\,\, \\
\textrm{(IV)}  \,\,\,\,\,\,\,\,\,\,\,\,\,\,\,\,\,\,\,\,                    
S_{1}(x_1) = i \mathcal{L}(x_1), \,\,\,\,\,\,\,\,\,\,\,\,\,\,\,\,\,\,\,\,\,\,\,\,\,\,\,\,\,\,\,\,\,\,\,\,\,\,\,\,\,\,\,\,\,\,\,\,\,\,\,\,\,\,\,\,\,
\,\,\,\,\,\,\,\,\,\,\,\,\,\,\,\,\,\,\,\,\,\,\,\,\,\,\,\,\,\,\,\,\,\,\,\,\,\,\,\,\,\,\,\,\,\,\,\,\,\,\,\,\,\,\,\,\,\,\,\,\,\,\,\,\,\,\,\,\,\,\,\,\,
\,\,\,\,\,\,\,\,\,\,\,\,\,\,\,\,\,\,\,\,\,\,\,\,\,\,\,\,\,\,\,\,\,\,\,\,\,\,\,\,\,\,\,\,\,\,\,\,\,\,\,\,\,\,\,\,\,\,\,\,\,\,\,\,\,\,\,\,\,\,\,\,\, 
\end{gather*}
\begin{enumerate}
\item[(V)] \,\,\,\,\,\,\,\,\,\,\,\,
The singularity degree of the retarded part of a kernel
should coincide with the singularity degree of this kernel, for the kernels of the generalized integral kernel 
and causal operators $D_{(n)}$ which are equal to linear combinations of products of the generalized operators $S_k$,
\,\,\,\,\,\,\,\,\,\,\,\,\,\,\,\,\,\,\,\,\,\,\,\,\,\,\,\,\,\,\,\,\,\,\,\,\,\,\,\,\,\,\,\,\,\,\,\,\,\,\,\,\,\,\,\,\,\,\,\,
\end{enumerate}

It is immediately seen that the axioms involve the (tensor) products
\begin{equation}
S_{k}(x_1, \ldots, x_{k})S_{n-k}(x_{k+1}, \ldots, x_{n}),
\end{equation}
and, assuming the first order contribution $S_{1}(x_1)$ in a form $i \mathcal{L}(x_1)$ of a Wick (pointwise) product
of free fields or in a form of a Wick polynomial, we can see that indeed $S_n$ can be understood as finite sums of 
generalized integral kernel operators, compare \cite{WN}. This follows in fact from 
theorem \ref{ClassWithProductThm} and
the Epstein-Glaser inductive construction, giving the $n$-th order contribution $S_n$ in terms of the contributions $S_k$, $k\leq n-1$,
which uses solely the (tensor) products $S_{k}^{+} S_{n-k}$, and $S_{n-k}S_{k}^{+}$, $1 \leq k <n$, 
compare the following Sections, in which we present the inductive step in detail. 
These products are indeed meaningful also in case the interaction Lagrangian density operator $\mathcal{L}$ contains massless fields, as e.g. in QED.

This converts the free fields $\mathbb{A}$ and the $n$-th order contributions, $S_n(g^{\otimes \, n})$ and 
$\mathbb{A}_{{}_{\textrm{int}}}^{(n)}(g^{\otimes \, n},\phi)$,
(written frequently as $S_n(g)$ and $\mathbb{A}_{{}_{\textrm{int}}}^{(n)}(g,\phi)$)
to the scattering operator
into the finite sums of generalized integral kernel operators
\[
\mathbb{A}(\phi) = \int \kappa_{0,1}(\boldsymbol{\p};\phi) \, \partial_{\boldsymbol{\p}} \, d\boldsymbol{\p} 
+ \int \kappa_{1,0}(\boldsymbol{\p};\phi) \, \partial_{\boldsymbol{\p}}^{*} \, d\boldsymbol{\p}, 
\]
\begin{multline*}
S_n(g^{\otimes \, n}) = \sum\limits_{\mathpzc{l},\mathpzc{m}}
\int \kappa_{\mathpzc{l}\mathpzc{m}}\big(\boldsymbol{\p}_1, \ldots, \boldsymbol{\p}_\mathpzc{l},
\boldsymbol{\q}_1, \ldots, \boldsymbol{\q}_\mathpzc{m};  g^{\otimes \, n} \big) \,\,
\partial_{\boldsymbol{\p}_1}^{*} \ldots \partial_{\boldsymbol{\p}_\mathpzc{l}}^{*} 
\partial_{\boldsymbol{\q}_1} \ldots \partial_{\boldsymbol{\q}_\mathpzc{m}} 
d\boldsymbol{\p}_1 \ldots d\boldsymbol{\p}_\mathpzc{l}
d\boldsymbol{\q}_1 \ldots d\boldsymbol{\q}_\mathpzc{m}
\\
=
\int \ud^4 x_1 \ldots \ud^4x_n \, S_n(x_1, \ldots, x_n) \, g(x_1) \ldots g(x_n),
\end{multline*}
and
\begin{multline*}
\mathbb{A}_{{}_{\textrm{int}}}^{(n)}(g^{\otimes \, n},\phi) = \sum\limits_{\mathpzc{l},\mathpzc{m}}
\int \kappa_{\mathpzc{l}\mathpzc{m}}\big(\boldsymbol{\p}_1, \ldots, \boldsymbol{\p}_\mathpzc{l},
\boldsymbol{\q}_1, \ldots, \boldsymbol{\q}_\mathpzc{m};  g^{\otimes \, n}, \phi \big) \,\,
\partial_{\boldsymbol{\p}_1}^{*} \ldots \partial_{\boldsymbol{\p}_\mathpzc{l}}^{*} 
\partial_{\boldsymbol{\q}_1} \ldots \partial_{\boldsymbol{\q}_\mathpzc{m}} 
d\boldsymbol{\p}_1 \ldots d\boldsymbol{\p}_\mathpzc{l}
d\boldsymbol{\q}_1 \ldots d\boldsymbol{\q}_\mathpzc{m}
\\
=
\int \ud^4 x_1 \ldots \ud^4x_n \ud^4 x \, \mathbb{A}_{{}_{\textrm{int}}}^{(n)}(x_1, \ldots, x_n; x) \, g(x_1) \ldots g(x_n) \, \phi(x),
\end{multline*}
with vector-valued distributional kernels $\kappa_{\mathpzc{l}\mathpzc{m}}$ in the sense of \cite{obataJFA}, with the values in the distributions
over the test nuclear space
\begin{gather*}
\mathscr{E} \ni \phi,
\,\,\,\,\,\,\,\,\,
\textrm{or}
\,\,\,\,\,
\mathscr{E}^{\otimes \, n} \ni g^{\otimes \, n} 
\,\,\,\,\,\,\,\,\,
\textrm{or, respectively,}
\,\,\,\,\,\,\,\,\,
\mathscr{E}^{\otimes \, n} \otimes (\oplus_{1}^{d}\mathscr{E}) \ni g^{\otimes \, n} \otimes \phi.
\end{gather*}
Here $\mathscr{E} = \mathcal{S}(\mathbb{R}^4;\mathbb{C})$.

Each of the $3$-dim Euclidean integration $d\boldsymbol{\p}_i$ with respect to the spatial momenta $\boldsymbol{\p}_i$ components
$\boldsymbol{\p}_{i1}, \boldsymbol{\p}_{i2}, \boldsymbol{\p}_{i3}$,
also includes here summation over the corresponding discrete spin components $s_i\in(1,2,\ldots)$ hidden under the symbol $\boldsymbol{\p}_i$. 

The class to which the operators $S_n$ and $\mathbb{A}_{{}_{\textrm{int}}}^{(n)}$ belong, expressed in terms of the Hida test space,
depend on the fact if there are massless free fields present in the interaction Lagrange density operator $\mathcal{L}$ or not.
Namely, \cite{WN}: 
\[
S_n \in
\begin{cases}
\mathscr{L}\big(\mathscr{E}^{\otimes \, n}, \, \mathscr{L}((E),(E))\big), 
& \text{if all fields in $\mathcal{L}$ are massive},\\
\mathscr{L}\big(\mathscr{E}^{\otimes \, n}, \, \mathscr{L}((E),(E)^*)\big), 
& \text{if there are massless fields in $\mathcal{L}$},\\
\end{cases}
\]
\[
\mathbb{A}_{{}_{\textrm{int}}}^{(n)} \in
\begin{cases}
\mathscr{L}\big(\mathscr{E}^{\otimes \, n}\otimes (\oplus_{1}^{d}\mathscr{E}), \, \mathscr{L}((E),(E))\big), 
& \text{if all fields in $\mathcal{L}$ are massive},\\
\mathscr{L}\big(\mathscr{E}^{\otimes \, n}\otimes (\oplus_{1}^{d}\mathscr{E}), \, \mathscr{L}((E),(E)^*)\big), 
& \text{if there are massless fields in $\mathcal{L}$}.\\
\end{cases}
\]

Now we are ready to point out one important benefit of using Hida operators in place of the Wightman's operator valued distributions. 
It turns out that if in theories with massless fields in $\mathcal{L}$, as e.g. in QED's with the minimal $U(1)$-coupling,
the higher order contributions $\mathbb{A}_{{}_{\textrm{int}}}^{(n)}$ to interacting fields are understood 
as finite sums of generalized integral kernel operators, then 
the adiabatic limit $g\rightarrow 1$ exists for them as a well-defined  finite sum of generalized integral kernel operators. 
In case of QED's, this limit exists if and only if the charged field is massive and, moreover, 
if the splitting into retarded and advanced parts of causal distribution in computation of $S_n$ is ``natural''. For the
definition of the ``natural'' normalization in the splitting, compare \cite{IF}. Thus, we have \cite{IF}:

\begin{twr}
For QED with massive charged field, the higher order contributions $\psi_{{}_{\textrm{int}}}^{(n)}(g^{\otimes \, n})$ 
and $A_{{}_{\textrm{int}}}^{(n)}(g^{\otimes \, n})$
 to interacting fields $\psi_{{}_{\textrm{int}}}$ and $A_{{}_{\textrm{int}}}$ 
in the adiabatic limit $g\rightarrow 1$ are well-defined as sums of generalized integral kernel operators
with vector valued kernels in the sense of Obata \cite{obataJFA},
\[
\underset{g\rightarrow 1}{\textrm{lim}} \psi_{{}_{\textrm{int}}}^{(n)}(g^{\otimes \, n}),
\underset{g\rightarrow 1}{\textrm{lim}} A_{{}_{\textrm{int}}}^{(n)}(g^{\otimes \, n})
\in \mathscr{L}\big( \oplus_{1}^{d}\mathscr{E}, \, L((E),(E)^*)\big), 
\] 
and this is the case only for the ``natural''
choice in the Epstein-Glaser splitting in the construction of the scattering operator.
\label{g->1IntFields}
\end{twr}

But:

\begin{twr}
For causal perturbative QED on the Minkowski space-time with the Hida operators as 
the creation-annihilation operators and with massless charged field, the higher order contributions to
interacting fields  in the adiabatic limit $g\rightarrow 1$ are not well-defined, 
even as sums of generalized integral kernel operators in the sense of Obata, and for
no choice in the Epstein-Glaser splitting in the construction of the scattering operator.
\label{g->1IntFieldsm=0}
\end{twr}

In the limit $g=1$, the higher order contributions, regarded as finite sums of generalized integral kernel operators,
map continuously space-time test functions $\phi$ into the continuous operators trasforming continuously $(E)$ into $(E)^*$: 
\[
\oplus_{1}^{4} \mathscr{E} \ni \phi \xrightarrow{\textrm{continously}} \mathbb{A}_{{}_{\textrm{int}}}^{(n)}(g=1, \phi)
\in \mathscr{L}\big((E), (E)^*\big). \,\,\,\,\,\,\,\,\,\,\,\, \textrm{case A)}
\] 
Only for some exceptional higher order subcontributions their limits $g=1$ are more regular and map continuously space-time test functions $\phi$ 
into the continuous operators, trasforming continously $(E)$ into $(E)$:
\[
\oplus_{1}^{4} \mathscr{E} \ni \phi \xrightarrow{\textrm{continously}} \mathbb{A}_{{}_{\textrm{int}}}^{(n)}(g=1, \phi)
\in \mathscr{L}\big((E), (E)\big). \,\,\,\,\,\,\,\,\,\,\,\, \textrm{case B)}
\] 
Only the more regular case B) can be understood as operator-valued distribution, but not the case A), which
cannot be subsumed by the approach based on the Wightman's operator valued distributions. The essential point lies in the fact, that
the integral kernels $\kappa_{{}_{l,m}}$ of the higher order contributions, when evaluated at the space-time test functions $\phi$, although
they do not represent any tensor products of the single particle Schwartz test functions (in the momenta variables), are still equal to well-defined
$\big(\oplus_{1}^{4} \mathscr{E}\big)^*$-valued distributions, and, by theorem \ref{Hida-ObataTheorem}, represent well-defined
integral kernel operators. For a more detailed analysis of these phenomena, compare \cite{IF}.

In this paper, we do not enter into the proof of theorem \ref{ClassWithProductThm} in order to 
show that $S$ is a Fock series of generalized integral kernel operators. 
Nor are we entering the study of the adiabatic limit here. These kinds of study
has been provided in \cite{WN} and \cite{IF} for the general case including the Grassmann valued test functions. 
Here we give a proof of existence of the higher order contributions $S_n$ to $S$, regarded as finite sums of integral kernel operators, 
with $\mathcal{L}$ containing massless fields.
We do it solely by the immediate application of the inductive step, and by application of 
lemma \ref{Cont.free.field.kernels} and  theorem \ref{Hida-ObataTheorem}. 
We assume that the reader knows the white noise construction of the general free field according to the line 
shortly summarized above. Hida himself has given only the white noise construction of the simplest example of the scalar Bose massive field, 
which is of course not sufficient for our purposes. A detailed general white nose construction of the free fields on the Minkowski space-time, 
including massless and Fermi case, can be found in \cite{QFTSG}. 
In Section \ref{InductiveStep} we remind the inductive
step of Epstein-Glaser. In Section \ref{1to2step} we present in details the inductive step from order one to order two
for general interaction $\mathcal{L}$ including massless fields, and show that existence of $S_n$ is an immediate consequence
of lemma \ref{Cont.free.field.kernels} and  theorem \ref{Hida-ObataTheorem} and from existence of the splitting of causal 
distribution into advanced and retarded part. 
In this Section, we are using the relation
between the Wick products of free fields and tensor products of the kernels corresponding to the free fields, and between
the pointwise Wick products of free fields and the pointwise products of the kernels of the free field operators. We express
there are the kernels of $S_2$ in terms of the advanced and retarded parts of the $q$-contractions $\otimes_q$ of the tensor products 
of kernels of the free fields, entering $\mathcal{L}$.  In Section \ref{SplittingOf(x)q} we give a proof of existence and give an 
explicit formula for the splitting into retarded and advanced parts of the said contractions, which in turn give the kernels of 
the operator $S_2$. There is nothing special about the second order $S_2$, as we have analogous formulas for the kernels of 
$S_n$ in terms of the retarded and advanced parts
of the linear combinations of the contractions of the tensor products of the kernels of $S_k$ for $k<n$. We have chosen
$S_2$ in order to explain the computation in terms of the kernels of the generalized integral kernel operators.

\section{Inductive step.
The key point in the removal of UV-divergence}\label{InductiveStep}

The inductive step which we apply was invented by Epstein and Glaser, \cite{Epstein-Glaser}, \cite{Scharf}, and it is based solely
on the Bogoliubov's  \cite{Bogoliubov_Shirkov} axioms (I)-(IV) and the preservation of the singularity degree axiom (V),
\cite{Epstein-Glaser}, given in Introduction. It replaces the naive multiplication by the step theta function procedure
in the construction of the chronological product, which caused ultraviolet divergences.
This inductive step construction converts the chronological product, or each $S_n$, into a mathematically
rigorous construction. Namely, suppose all 
$\{S_k\}_{{}_{k\leq n-1}}$ are already constructed. In order to construct  
$S_n(Z, x_n) = S(Z, x_n)$ \,\,\,\, (here $X, Y, Z$ denote the sets of space-time variables 
such that  $X \cup Y = \{x_1, \ldots, x_{n-1}\} = Z$, $X \cap Y = \emptyset$) we construct the following
generalized operators
\[
A'_{(n)}(Z, x_n)  
= \sum \limits_{X\sqcup Y=Z, X\neq \emptyset} \overline{S}(X)S(Y,x_n), \,\,\,
R'_{(n)}(Z, x_n)  
= \sum \limits_{X\sqcup Y=Z, X\neq \emptyset} S(Y,x_n)\overline{S}(X),
\]
where the sums run over all divisions $X\sqcup Y=Z$ of the set $Z$ of variables $\{x_1, \ldots, x_{n-1} \}$
into two disjoint subsets $X$ and $Y$ with $X\neq \emptyset$,
and which, by the inductive assumption are known. Next we construct the following generaized operators
\[
\begin{split}
A_{(n)}(Z, x_n) = \sum \limits_{X\sqcup Y=Z} \overline{S}(X)S(Y,x_n)
= A'_{(n)}(Z, x_n)  + S(x_1, \ldots, x_n), \\
R_{(n)}(Z, x_n) = \sum \limits_{X\sqcup Y=Z} S(Y,x_n)\overline{S}(X)
= R'_{(n)}(Z, x_n) + S(x_1, \ldots, x_n),
\end{split}
\]
where the sums run over all divisions $X\sqcup Y=Z$ of the set $Z$ of variables $\{x_1, \ldots, x_{n-1} \}$
into two disjoint subsets $X$ and $Y$ including the empty set $X = \emptyset$.
The unprimed generalized operator $R_{(n)}$ has retarded support, meaning that it is supported within the range of each
of the variables $x_1, \ldots, x_{n-1}$ which lie in the future closed light cone emerging from $x_n$, and the 
unprimed generalized operator $A_{(n)}$ has advanced support, meaning that it is supported within the range of each
of the variables $x_1, \ldots, x_{n-1}$ which lie in the past closed light cone emerging from $x_n$, for the proof compare
\cite{Epstein-Glaser} or \cite{Scharf}. Therefore, 
\[
D_{(n)} = R'_{(n)} - A'_{(n)} = R_{(n)} - A_{(n)} 
\,\,\,\,\,\,\,\,\,\,\,\,\,\,\,\, \textrm{is causally supported},
\]
with
\[
R_{(n)} \,\,\, \textrm{-- being the retarded part of $D_{(n)}$}
\,\,\,\,\,\,\,\,\,\,\,\,\,\,\,\,\,\,\,\,
A_{(n)} \,\,\, \textrm{-- being the advanced part of $D_{(n)}$}.
\]
Next we compute the retarded $R_{(n)}$  and the advanced part $A_{(n)}$ of the given generaliized operator $D_{(n)}$ 
independently of the axioms (I)-(V), just observing that each of the numerical tempered distributional coefficient in each of the Wick monomials
in the Wick decomposition of $D_{(n)}$ is a numerical causal distribition, and applying the ordinary theory of distribution
splitting. This splitting is in general not unique with the number of arbitrary consctants in the splitting 
depending on the singularity degree of the splitted distribution. Thus, we arrive at the formula 
\begin{equation}\label{Sn}
 S_n(x_1, \ldots, x_n) =  R_{(n)}(x_1, \ldots, x_n)-R'_{(n)}(x_1, \ldots, x_n) 
\end{equation}
which is determined up to a finite number of constants, with the number of constants depending on the singularity degree od 
the scalar causal coefficients in the Wick decomposition of $D_{(n)}$.

We replace the step theta function $\theta$ in the heuristic definition
of the chronological product by a one-parameter family $\theta_\varepsilon$ of smooth functions. Then we apply the
Wick Theorem (using integral kernel calculus, given in \cite{WN}, \cite{obataJFA}) with the kernels depending on $\theta_\varepsilon$,
and with the kernels in the Wick decomposition containing various contractions $\otimes_q$. By the Epstein-Glaser
splitting we know that all terms with contractions $\otimes_q$ with $q>1$ can be convergent, when $\theta_\varepsilon
\rightarrow \theta$, only on a closed
subspace of the space-time test space of finite codimension. Indeed, we show that families
$\theta_\varepsilon$ exist for which such convergence holds, and with the limit independent of the choice of the family.
This allows to determine the limit kernels of $S_n$ up to a finite number of constants
in each order contribution. Further extension of such constructed kernels 
all over the whole space-time test space
cannot be determined uniquely by the causality axioms (I)-(V) for the scattering operator.
In case of QED, the arbitrary constants, which arise in such extension, can then be determined by further and natural
condition, independent of the causality axioms (I)-(V) for the scattering operator itself, namely, the condition
of existence of the adiabatic limit for the higher order contributions to interacting fields, well-defined 
as the integral kernel operators \cite{IF}. This choice of normalization in the splitting we call ``natural''. 
It is important to emphasize that this is possible if we are using the Hida operators 
as the creation-annihilation operators and the integral kernel calculus of Hida, Obata and Sait\^o.

As we did for the construction of the ``product'' of Wick ordered factors including massless fields (using integral kernel calculus),
we give here a similar construction of a '``natural'' chronological product, which is essentially based on the step theta
function $\theta(x) \overset{\textrm{df}}{=} \theta(x_0)=\theta(t)$, where the last $\theta$ is the ordinary step $\theta$-function on $\mathbb{R}$,
and is immediately motivated by the heuristic definition:
\begin{multline}\label{thetaS_n}
S_n(x_1, \ldots, x_n) = i^n \, T\big(\mathcal{L}(x_1) \ldots \mathcal{L}(x_n) \big) \\ = i^{n}
\sum\limits_{\pi} \theta(t_{\pi(1)} -t_{\pi(2)}) \theta(t_{\pi(2)} -t_{\pi(3)}) \ldots \theta(t_{\pi(n-1)} -t_{\pi(n)}) \,
\mathcal{L}(x_{\pi(1)}) \ldots \mathcal{L}(x_{\pi(n)}), \\
\,\,\,\,\,\,\,\,\,\,\,\,\,\,\,\,\,\,\,\,\,\,\,\,\,\,\,\,\,\,\,\,
x_{\pi(k)} = (t_{\pi(k)}, \boldsymbol{\x}_{\pi(k)}), \,\,\, \pi \in \textrm{Permutations of} \,\{1, \ldots, n\}.
\end{multline}
We give to the expression (\ref{thetaS_n}) a rigorous meaning in case the Wick monomial, or each of the Wick monomials
in the interaction Lagrange density Wick polynomial $\mathcal{L}$ contains massless free field factors or not. Let for example
\[
\mathcal{L}(x) = \boldsymbol{{:}} \boldsymbol{\psi}(x)^{+}\gamma_{0} \gamma^\mu \boldsymbol{\psi}(x) A_\mu(x) \boldsymbol{{:}},
\]
as in spinor QED, but our analysis is general, and we can replace $\mathcal{L}$ with any Wick polynomial of free fields
with each of the Wick monomials possibly containing massless free field factors.

We apply the causal method of Bogoliubov-Epstein-Glaser \cite{Epstein-Glaser}, \cite{Scharf}. 
In particular for the second order contribution the formula
reduces to
\begin{equation}\label{S2}
S_2(x,y) = R_{(2)}(x,y) - R'_{(2)}(x,y) =  \textrm{ret} \, \big[\mathcal{L}(x)\mathcal{L}(y) -\mathcal{L}(y)\mathcal{L}(x) \big] 
- \mathcal{L}(y)\mathcal{L}(x),
\end{equation} 
where $R_{(2)}(x,y)$ is the retarded part of a causally supported
distribution $D_{(2)}(x,y)$. In particular all scalar factors in terms of the Wick decomposition of 
\[
D_{(2)}(x,y)= S(y)\overline{S}(x) - \overline{S}(x)S(y) = i^2\big[\mathcal{L}(y)\mathcal{L}(x) - \mathcal{L}(x)\mathcal{L}(y)\big]
\]
(those with pairings in the Wick decomposition of
$D_{(2)}(x,y)$) are indeed causal distributions, \emph{i.e.} distributions of one space-time
variable $x-y$, which is supported within the light cone. Therefore the method of splitting of causal distributions
into the retarded and advanced part (due to Epstein-Glaser \cite{Epstein-Glaser} or \cite{Scharf})
can indeed be used for the computation of $S_2$.

We observe that the general splitting method, to which Epstein and Glaser or Scharf refer in \cite{Epstein-Glaser} or \cite{Scharf},
works not only for the strictly causally supported distributions. In fact it is the local property around $x-y=0$
of a distribution $C(x-y)$ which is to be splitted, which is important for the splitting, and is governed
by a single number $\omega$ pertinent to the distribution $C$, and called \emph{singular order $\omega$ of} $C$ at zero
(or at infinity for the Fourier transformed $\widetilde{C}$). 
Namely, for the plane wave kernels 
\begin{equation}\label{FFkernels}
\kappa^{(1)}_{0,1}, \kappa^{(1)}_{1,0}, \ldots, \kappa^{(q)}_{0,1}, \kappa^{(q)}_{1,0},
\end{equation}
of the free fields we show in this paper that all contraction distributions (products of pairings) 
\begin{equation}\label{ProductsOfPairings}
\Big(\kappa^{(1)}_{0,1} \dot{\otimes} \ldots \dot{\otimes} \kappa^{(q)}_{0,1}\Big)
\otimes_q
\Big(\kappa^{(1)}_{0,1} \dot{\otimes} \ldots \dot{\otimes} \kappa^{(q)}_{0,1}\Big)(x,y) = \kappa^{(-)}_q(x-y)
\end{equation}
have finite order $\omega$ of singularity at zero and can be splitted as in the cited works. 
Contractions $\otimes_q$ are always understood for the spin momentum variables
in the tensor products of single particle test spaces $E_1$, $E_2$, $\ldots$ and their duals, and expressed
through sum/integration with respect to the $q$ pairs of the contracted spin momentum variables.
In general 
 (\ref{ProductsOfPairings}) is not causally suported. 
This splitting of (\ref{ProductsOfPairings}) is possible because the following differences or sums 
of negative frequency $\otimes_q$-contractions (\ref{ProductsOfPairings}) and their positive frequency counterparts
(being equal to products of pairings)
\begin{multline}\label{CausalDifferencesOfProductsOfPairings}
\Big(\kappa^{(1)}_{0,1} \dot{\otimes} \ldots \dot{\otimes} \kappa^{(q)}_{0,1}\Big)
\otimes_q
\Big(\kappa^{(1)}_{1,0} \dot{\otimes} \ldots \dot{\otimes} \kappa^{(q)}_{1,0}\Big)(x,y)
-(-1)^{f(q)}
\Big(\kappa^{(1)}_{1,0} \dot{\otimes} \ldots \dot{\otimes} \kappa^{(q)}_{1,0}\Big)
\otimes_q
\Big(\kappa^{(1)}_{0,1} \dot{\otimes} \ldots \dot{\otimes} \kappa^{(q)}_{0,1}\Big)(x,y)
\\
= \kappa^{(-)}_q(x-y) -(-1)^{f(q)} \kappa^{(+)}_q(x-y)
\end{multline}
have always causal support, to which the Epstein-Glaser splitting can be applied. 
Here $f(q)$ is the number
of Fermi field kernels among the contracted kernels $\kappa^{(1)}_{0,1}, \ldots, \kappa^{(q)}_{0,1}$.
The condition
for the convergence of the retarded part of the difference or sum (\ref{CausalDifferencesOfProductsOfPairings}) 
of the contractions (\ref{ProductsOfPairings}) implies convergence for the retarded part of  (\ref{ProductsOfPairings}).
The only difference is that now the retarded part of each $\kappa^{(-)}_q(x-y)$ and $\kappa^{(+)}_{q}(x-y)$ in (\ref{CausalDifferencesOfProductsOfPairings})
taken separately, is not Lorentz invariant with the time-like unit versor $v$ of the reference frame in the theta function
$\theta(v\cdot x)$ giving the support $\textrm{supp} \theta$ of the retarded part. Only in the sum/diffrene of the retarded 
parts of $\kappa^{(-)}_q(x-y)$ and $\kappa^{(+)}_{q}(x-y)$ in (\ref{CausalDifferencesOfProductsOfPairings}) the dependence on $v$ drops out, with the retarded part
of the whole distribution (\ref{CausalDifferencesOfProductsOfPairings}) being Lorentz invariant.

Using this fact, to the second order contribution and heuristic formula
\begin{equation}\label{S2(x1,x2)}
S_2(x_1,x_2) = i^2\theta(x_1-x_2)\mathcal{L}(x_1)\mathcal{L}(x_2)+i^2\theta(x_2-x_1)\mathcal{L}(x_2)\mathcal{L}(x_1)
\end{equation}
can be given a rigorous sense, as the application of the Wick formula (using integral kernel calculus) to the operators
\[
\mathcal{L}(x_1)\mathcal{L}(x_2) \,\,\,\, \textrm{and} \,\,\,\,
\mathcal{L}(x_2)\mathcal{L}(x_1)
\]
gives the normal order product of free fields plus the terms with pairings (contractions) which can be splitted
into the retarded and advanced parts by the Epstein-Glaser method. In particular we can compute the retarded part of 
all the pairing distributions involved into the Wick decomposition of 
\[
\mathcal{L}(x_1)\mathcal{L}(x_2) \,\,\,\, \textrm{and} \,\,\,\,
\mathcal{L}(x_2)\mathcal{L}(x_1).
\]
Thus, in the Wick decomposition formula for
\[
\mathcal{L}(x_1)\mathcal{L}(x_2) \,\,\,\, \textrm{and} \,\,\,\,
\mathcal{L}(x_2)\mathcal{L}(x_1)
\]
each term, containing the scalar causal distribution factor
(coming from the pairing, \emph{i.e.} $\otimes_q$-contraction) multiplied with $\theta(x_1-x_2)$ or $\theta(x_2-x_1)$
we interpret as the replacement of the pairing (contraction) with the retarded part of the scalar distribution
or as the retarded part of the $\otimes_q$-contraction.
The term without pairings is trivial
as $\theta(x_1-x_2)$ is a well-defined translationally invariant distribution and the normal product
of free field operators in the variables $x_1$ and $x_2$ multiplied by $\theta(x_1-x_2)$ is well-defined.
The only difference in comparison to the standard formal approach is that we use the Epstein-Glaser splitting
into the retarded and advanced part of contractions (pairings) instead of the naive multiplication by $\theta$. 
In fact the splitting is non-trivial
only in case of $\otimes_q$ contractions with $q>1$. For $\otimes_0 = \otimes$-contraction or $\otimes_1$-contraction
the splitting is trivial and can be realized through the ordinary multiplication by $\theta$. The contributions
with $\otimes_q$ contractions with $q>1$ correspond to the contributions represented by the loop-graphs insertions,
which in the standard renormalization approach are divergent. At the second order level there are 
only three types of such divergent insertions in case of QED.

Moreover, using this method of computation of $S_2$ we have the same singularity orders $\omega$ in the splitting as in 
the Epstein-Glaser formula for $S_2$, which uses $R_{(2)}$ and $D_{(2)}$ , with the same range of freedom in the splitting.
We obtain the same $S_2$ as in the Epstein-Glaser method, using the same standard normalization conditions
determining the splitting, e.g. by imposing the standard normalization conditions. Having the contractions (which we encounter
in the computation of $S_2$) and their splitting into retarded and advanced parts fixed, we determine the higher order
contributions $S_n$.

For the computation of the higher order contribution $S_n(x_1, \ldots, x_n)$ we apply the Epstein-Glaser causal method,
repeating the same computation as for $S_2$, but instead of (\ref{S2}) we use (\ref{Sn}), and instead of computing
the retarded/advanced part of causal distributions of single space-time variable, we need to apply the analogous dispersion
integral for a retarded/advanced part of causal distributions of $n-1$ space-time variables.
This computation method of $S_n(x_1, \ldots, x_n)$ is reduced to the application
of the Wick theorem \cite{WN}  
to the product of at most two normally ordered operators (which we treat as the generalized integral kernel operators),
as the operator $D_{(n)}(x_1, \ldots, x_n)$ is equal to the finite sum of products of at most two normally ordered
operators:
\[
D_{(n)}(Z,x_n) = R'_{(n)}(Z,x_n)-A'_{(n)}(Z,x_n) 
 = 
\sum \limits_{X\sqcup Y=Z, X\neq \emptyset} S(Y,x_n)\overline{S}(X)
-
\sum \limits_{X\sqcup Y=Z, X\neq \emptyset} \overline{S}(X)S(Y,x_n), 
\]
where the sums run over all divisions $X\sqcup Y=Z$ of the set $Z$ of variables $\{x_1, \ldots, x_{n-1} \}$
into two disjoint subsets $X$ and $Y$ with $X\neq \emptyset$. Moreover, the operator $D_{(n)}(x_1, \ldots, x_n)$
has causal support in the subspace where all $x_i$, $1\leq i \leq n-1$, lie in the closed future light cone emerging from $x_n$.
Computation is reduced to the computation of the contractions $\kappa_{{}_{lm}}\otimes_{{}_{q}}\kappa'_{{}_{l'm'}}$ of the kernels of $S_k$, for $1 \leq k < n$
and retarded/advanced parts of the causal sums of the contractions $\kappa_{{}_{lm}}\otimes_{{}_{q}}\kappa'_{{}_{l'm'}}$.
Indeed, the computation of the retarded part (symbolically)
\begin{multline*}
R_{(n)}(x_1, \ldots, x_n) = \theta(x_1-x_n)\theta(x_2-x_n) \ldots \theta(x_{n-1}-x_n) \, D_{(n)}(x_1, \ldots, x_n)
= \sum \limits_{X\sqcup Y=Z} S(Y,x_n)\overline{S}(X)
\end{multline*}
of $D_{(n)}$ is reduced to the computation of the retarded/advanced parts of the 
causal sums of the contractions $\kappa_{{}_{lm}}\otimes_{{}_{q}}\kappa'_{{}_{l'm'}}$. In principle, dispersion integrals
giving a retarded/advanced parts make sense for each simple contraction summand taken separately, but the computation
is simplified much by the Lorentz invariance and causality properties, so that dispersion integrals
are more easily handled for the whole causal sums. 

Therefore, we may concentrate on the computation of the second order $S_2$ in the rest part of the paper, as the computation
of higher orders is a repeated application of the computation of the second order.

The basic distributions in one space-time variable $x_1-x_2$ are the products of
non-zero pairings (contractions $\otimes_q$) of the free fields entering into the Wick formula \cite{WN}
for the (tensor) product generalized operator $\mathcal{L}(x_1)\mathcal{L}(x_2)$, where $\mathcal{L}(x)$
is the interaction Lagrange density of the QFT in question, together with the particular choice of their splitting
into the retarded and advanced parts. The particular choice of the splitting will have to be used already in the computation
of the second order contribution $S_2$ outlined above, so we will especially concentrate on the analysis of the computation
of $S_2$. The computation of the higher order terms brings no new ingredients from the point of view of analysis of distributions
reduced to computation of dispersion integrals of the same kind as in the computation of $S_2$.
In particular the computation of the singularity degree of the scalar coefficients in $D_{(n)}$ can be as easily computed as in the second order case
provided we have Fourier transforms of all coefficients in $S_k$, $k<n$ already computed. In fact the singularity degree of these coefficients
can be relatively easily computed by induction without the knowledge of the explicit form of the said Fourier transforms, using the additive
behavior of the singularity degree under tensor product operation. If the interaction Lagrangian $\mathcal{L}$ contains derivatives,
we are usng in addition the fact that the space-time derivation $D^\alpha$ of degree $|\alpha|$ raises the singularity degree by $|\alpha|$.

\section{Second order contribution $S_2$ with general interaction Lagrangian}\label{1to2step}

So let us concentrate on the general mathematical analysis of $S_2$, making only general remarks concerning the
higher order generalized operators $D_{(n)}$, $R'_{(n)}$, $A'_{(n)}$ and $S_n$ which all can be reduced to the application of the Wick theorem
for (finite sum of) products of at most two normally ordered (generalized integral kernel) operators and in addition the factor-by-factor
application of the splitting of the basic pairings in one space-time variable in case of $S_n$. 

Therefore we consider the product $\mathcal{L}(x_1)\mathcal{L}(x_2)$ of two normally ordered generalized integral kernel
operators $\mathcal{L}(x_1)$, $\mathcal{L}(x_2)$, with vector valued distributional kernels 
\begin{equation}\label{kernelsOfL(x)L(y)}
\kappa_{l,m} = \kappa'_{l',m'} \otimes_{q'} \kappa''_{l'',m''}
\end{equation}
of the product $\mathcal{L}(x_1)\mathcal{L}(x_2)$,
and the kernels $\kappa'_{l',m'}$, $\kappa''_{l'',m''}$ ranging over the kernels of the operator
$\mathcal{L}(x)$,  
and giving the Wick decomposition of products of integral kernel operators
in terms of kernels of the integral kernel operators. The possible values of $q'$ range from zero up to the degree
of the Wick polynomial $\mathcal{L}$. 
Recall, that contractions $\otimes_q$ are always understood for the spin momentum variables
in the tensor products of single particle test spaces $E_1$, $E_2$, $\ldots$ and their duals.
Denoting the kernels
of the free fields which are involved in the Wick polynomial $\mathcal{L}$, respectively, by
\[
\kappa'_{0,1}, \kappa'_{1,0},\kappa''_{0,1}, \kappa''_{1,0}, \ldots 
\] 
we easily see that the kernels  (\ref{kernelsOfL(x)L(y)}) have the general form
\begin{multline}\label{ProductForm}
\Big[
\big(\kappa'_{0,1} \dot{\otimes}  \kappa''_{0,1} \dot{\otimes} \ldots \kappa^{(q)}_{0,1} \big)
\, \otimes_{q'} \,\, \big(\kappa'_{1,0} \dot{\otimes} \kappa''_{1,0}
\dot{\otimes} \ldots \kappa^{(q)}_{1,0} \big)\Big]
\\
=
\Big[
\big(\kappa^{(k_1)}_{0,1} \dot{\otimes} \kappa^{(k_2)}_{0,1} \ldots \dot{\otimes} \kappa^{(k_{q'})}_{0,1} \big)
\, \otimes_{q'} \,\, \big(\kappa^{(k_1)}_{1,0} \dot{\otimes} \kappa^{(k_2)}_{1,0} \ldots \dot{\otimes} \kappa^{(k_{q'})}_{1,0}  \big)\Big] \, 
\Big[
\big(\kappa'_{0,1} \dot{\otimes} \kappa''_{0,1} \dot{\otimes} \widehat{\ldots} \kappa^{(q)}_{0,1} \big)
\, \otimes \,\, \big(\kappa'_{1,0} \dot{\otimes} \kappa''_{1,0} \dot{\otimes} \widehat{\ldots} \kappa^{(q)}_{1,0}  \big)
\Big]
\end{multline}
Hat means that the contracted kernels are deleted. The product $\mathcal{L}(x_1)\mathcal{L}(x_2)$ still
makes sense and is a well-defined element of
\[
\mathscr{L} \big(\mathscr{E}\otimes \mathscr{E}, \mathscr{L}((E), (E)^*) \big)
\]
whenever massless fields are present in $\mathcal{L}$, even though  in this case the product belongs to
\[
 \mathscr{L} \big(\mathscr{E}\otimes \mathscr{E}, \mathscr{L}((E), (E)^*) \big), 
\,\,\, \textrm{but, in general, not to} \,\,\, 
\mathscr{L} \big(\mathscr{E}\otimes \mathscr{E}, \mathscr{L}((E), (E)) \big);
\]
for a proof, compare \cite{WN}. Here $(E)$ is the test Hida space in the total Fock space of all free fields underlying
the actual QFT.
The product operator $\mathcal{L}(x_1)\mathcal{L}(x_2)$ is defined by a limit 
process on replacing the massless fields by their massive counterparts and then by passing with the auxiliary mass to the zero limit.
If all free fields are massive in $\mathcal{L}$, then 
\[
\mathcal{L} \in \mathscr{L} \big(\mathscr{E}, \mathscr{L}((E), (E) \big),
\]
and the product $\mathcal{L}(x_1)\mathcal{L}(x_2)$ represents a finite sum of generalized integral kernel operators with vector
valued kernels, and belonging to
\[
\mathscr{L} \big(\mathscr{E}\otimes \mathscr{E}, \mathscr{L}((E), (E)) \big),
\]
for a proof, compare \cite{WN}. We we will see it below by explicit computation.

Next we consider the retarded part
\begin{equation}\label{retL(x)L(y)}
\textrm{ret} \, \mathcal{L}(x_1)\mathcal{L}(x_2)
\end{equation}
which we symbolically write as
\[
\theta(x_1-x_2)\mathcal{L}(x_1)\mathcal{L}(x_2),
\]
of the product $\mathcal{L}(x_1)\mathcal{L}(x_2)$, expressed through the Wick formula.
This will allow us to compute $S_2$ given by (\ref{S2(x1,x2)}).  The same Wick formula can be applied 
to the sum $D_{(n)}$ of products of at most two normally ordered operators. Becuse $D_{(n)}$ is causally
supported (compare \cite{Epstein-Glaser} or \cite{Scharf}) then we can repeat
analogue computation of the retarded part\footnote{Here with the multidimensional theta
$\theta(x_1, \ldots, x_n) \overset{\textrm{df}}{=}\theta(x_1-x_n)\ldots\theta(x_{n-1}-x_n)$.} $R_{(n)} = \theta D_{(n)}$ of $D_{(n)}$,
with the analogue analysis also valid for $S_n = R_{(n)}-R'_{(n)}$, $n>2$. 

The kernels of (\ref{retL(x)L(y)}) have the general form
\[
\textrm{ret} \,\,
\Big[
\big(\kappa'_{0,1} \dot{\otimes}  \kappa''_{0,1} \dot{\otimes} \ldots \kappa^{(q)}_{0,1} \big)
\, \otimes_{q'} \,\, \big(\kappa'_{1,0} \dot{\otimes} \kappa''_{1,0}
\dot{\otimes} \ldots \kappa^{(q)}_{1,0} \big)\Big]
\]
\begin{multline}\label{specialRetkappa'timesq'kappa''}
=
\textrm{ret} \,\,
\Big[
\big(\kappa^{(k_1)}_{0,1} \dot{\otimes} \kappa^{(k_2)}_{0,1} \ldots \dot{\otimes} \kappa^{(k_{q'})}_{0,1} \big)
\, \otimes_{q'} \,\, \big(\kappa^{(k_1)}_{1,0} \dot{\otimes} \kappa^{(k_2)}_{1,0} \ldots \dot{\otimes} \kappa^{(k_{q'})}_{1,0}  \big)\Big] \, \times
\\
\times
 \,
\Big[
\big(\kappa'_{0,1} \dot{\otimes} \kappa''_{0,1} \dot{\otimes} \widehat{\ldots} \kappa^{(q)}_{0,1} \big)
\, \otimes \,\, \big(\kappa'_{1,0} \dot{\otimes} \kappa''_{1,0} \dot{\otimes} \widehat{\ldots} \kappa^{(q)}_{1,0}  \big),
\Big]
\end{multline}
with the integral kernel operators $\Xi(\kappa_{l,m})$, corresponding to the kernels $\kappa_{l,m}$ equal (\ref{specialRetkappa'timesq'kappa''}),
which belong to
\[
\mathscr{L} \big(\mathscr{E}\otimes \mathscr{E}, \mathscr{L}((E), (E)) \big),
\]
or, respecively, to
\[
\mathscr{L} \big(\mathscr{E}\otimes \mathscr{E}, \mathscr{L}((E), (E)^*) \big),
\]
depending on whether all non contracted kernels are massive in (\ref{specialRetkappa'timesq'kappa''}) or not.

The scalar distribution (scalar $\otimes_{q'}$ contraction) in (\ref{specialRetkappa'timesq'kappa''}) 
\begin{equation}\label{specialscalarq'contraction}
\textrm{ret} \,\,
\Big[
\big(\kappa^{(k_1)}_{0,1} \dot{\otimes} \kappa^{(k_2)}_{0,1} \ldots \dot{\otimes} \kappa^{(k_{q'})}_{0,1} \big)
\, \otimes_{q'} \,\, \big(\kappa^{(k_1)}_{1,0} \dot{\otimes} \kappa^{(k_2)}_{1,0} \ldots \dot{\otimes} \kappa^{(k_{q'})}_{1,0}  \big)\Big](x_1,x_2)
\end{equation}
we compute by the ordinary multiplication by the step theta function
\begin{multline*}
\textrm{ret} \,\,
\Big[
\big(\kappa^{(k_1)}_{0,1} \dot{\otimes} \kappa^{(k_2)}_{0,1} \ldots \dot{\otimes} \kappa^{(k_{q'})}_{0,1} \big)
\, \otimes_{q'} \,\, \big(\kappa^{(k_1)}_{1,0} \dot{\otimes} \kappa^{(k_2)}_{1,0} \ldots \dot{\otimes} \kappa^{(k_{q'})}_{1,0}  \big)\Big]
\\
=
\theta(x_1-x_2) \,\,
\Big[
\big(\kappa^{(k_1)}_{0,1} \dot{\otimes} \kappa^{(k_2)}_{0,1} \ldots \dot{\otimes} \kappa^{(k_{q'})}_{0,1} \big)
\, \otimes_{q'} \,\, \big(\kappa^{(k_1)}_{1,0} \dot{\otimes} \kappa^{(k_2)}_{1,0} \ldots \dot{\otimes} \kappa^{(k_{q'})}_{1,0}  \big)\Big](x_1,x_2)
\\
\textrm{if} \,\, q'=0, \,\, \textrm{or} \,\, q'=1,
\end{multline*}
\emph{i.e.} if the singularity degree $\omega$ at zero of the distribution (scalar $\otimes_{q'=0}$, $\otimes_{q'=1}$-contraction)
\[
\big(\kappa^{(k_1)}_{0,1} \dot{\otimes} \kappa^{(k_2)}_{0,1} \ldots \dot{\otimes} \kappa^{(k_{q'})}_{0,1} \big)
\, \otimes_{q'} \,\, \big(\kappa^{(k_1)}_{1,0} \dot{\otimes} \kappa^{(k_2)}_{1,0} \ldots \dot{\otimes} \kappa^{(k_{q'})}_{1,0}  \big)
\]
is negative, and can be mulitplied by the step theta function.
If the singularity degree $\omega$ of 
\[
\kappa_{q'}(x_1-x_2)
=
\big(\kappa^{(k_1)}_{0,1} \dot{\otimes} \kappa^{(k_2)}_{0,1} \ldots \dot{\otimes} \kappa^{(k_{q'})}_{0,1} \big)
\, \otimes_{q'} \,\, \big(\kappa^{(k_1)}_{1,0} \dot{\otimes} \kappa^{(k_2)}_{1,0} \ldots \dot{\otimes} \kappa^{(k_{q'})}_{1,0}  \big)(x_1,x_2)
\]
is zero, or positive, which is the case for $q'>1$, then we define (\ref{specialscalarq'contraction})
on the subspace of finite codimension of all space-time test functions $\chi\in \mathscr{E}^{\otimes \, 2}$, $\mathscr{E} = \mathcal{S}(\mathbb{R}^4)$,
$\chi(x_1,x_2) = \phi(x_1-x_2)\varphi(x_2)$, $\phi, \varphi \in \mathscr{E}$,  for which all derivatives of 
$\phi$, up to order $\omega$, vanish at zero. Denoting the continuous idemponent operator projecting
$\mathscr{E}$ on the subspace of functions with vanishing derivatives at zero up to order $\omega$, by $\Omega'$,
we define (\ref{specialscalarq'contraction}), first by projecting  $\Omega'$ and then by multiplication by $\theta$
function 
\[
\langle \textrm{ret} \, \kappa_{q'}, \phi\rangle = \langle \kappa_{q'}, \theta.\Omega'\phi \rangle,
\]
and putting $\textrm{ret} \, \kappa_{q'}(x_1-x_2)$ for (\ref{specialscalarq'contraction}). This definition is
not unique and we can add to  $\textrm{ret} \, \kappa_{q'}$ a distribution 
\[
\sum\limits_{|\alpha|=0}^{\omega} C_\alpha \delta^{(\alpha)},
\]
with arbitrary (in principle) $C_\alpha$, which is zero on the $\textrm{Im} \, \Omega'$
and is most general on the finite dimensional subspace $\textrm{Ker} \, \Omega'$, giving the most general 
retarded part of $\kappa_{q'}$ on the  whole test space
\[
\mathscr{E} = \textrm{Im} \, \Omega' \, \oplus \, \textrm{Ker} \, \Omega',
\]
and correspondingly, giving the most general (\ref{specialscalarq'contraction}) which together with
\begin{multline*}
\textrm{av} \,\,
\Big[
\big(\kappa^{(k_1)}_{0,1} \dot{\otimes} \kappa^{(k_2)}_{0,1} \ldots \dot{\otimes} \kappa^{(k_{q'})}_{0,1} \big)
\, \otimes_{q'} \,\, \big(\kappa^{(k_1)}_{1,0} \dot{\otimes} \kappa^{(k_2)}_{1,0} \ldots \dot{\otimes} \kappa^{(k_{q'})}_{1,0}  \big)
\Big]
\\
=
-\big(\kappa^{(k_1)}_{0,1} \dot{\otimes} \kappa^{(k_2)}_{0,1} \ldots \dot{\otimes} \kappa^{(k_{q'})}_{0,1} \big)
\, \otimes_{q'} \,\, \big(\kappa^{(k_1)}_{1,0} \dot{\otimes} \kappa^{(k_2)}_{1,0} \ldots \dot{\otimes} \kappa^{(k_{q'})}_{1,0}  \big)
\\
+
\textrm{ret} \,\,
\Big[
\big(\kappa^{(k_1)}_{0,1} \dot{\otimes} \kappa^{(k_2)}_{0,1} \ldots \dot{\otimes} \kappa^{(k_{q'})}_{0,1} \big)
\, \otimes_{q'} \,\, \big(\kappa^{(k_1)}_{1,0} \dot{\otimes} \kappa^{(k_2)}_{1,0} \ldots \dot{\otimes} \kappa^{(k_{q'})}_{1,0}  \big)
\Big]
\end{multline*}
gives the most general splitting of 
\[
\big(\kappa^{(k_1)}_{0,1} \dot{\otimes} \kappa^{(k_2)}_{0,1} \ldots \dot{\otimes} \kappa^{(k_{q'})}_{0,1} \big)
\, \otimes_{q'} \,\, \big(\kappa^{(k_1)}_{1,0} \dot{\otimes} \kappa^{(k_2)}_{1,0} \ldots \dot{\otimes} \kappa^{(k_{q'})}_{1,0}  \big)
\]
into the difference of the retarded and advanced parts. 

Introducing the continuous idempotent operator $\Omega$: 
\[
\Omega\chi(x_1,x_2) = \phi(x_1-x_2)\varphi(x_2),
\]
which projects the test functions $\chi(x_1,x_2) = \phi(x_1-x_2)\varphi(x_2)$ onto the subspace
of $\chi(x_1,x_2) = \phi'(x_1-x_2)\varphi(x_2)$ in which all derivatives of $\phi'$ vanish at zero
up to order $\omega$, we can write 
\begin{equation}\label{DoubleLimitContraction}
\textrm{ret} \, \kappa'_{l',m'} \otimes_{q'} \kappa''_{l'',m''} = \theta \kappa'_{l',m'} \otimes_{q'} \kappa''_{l'',m''} \circ \Omega
= \theta \kappa'_{l',m'} \otimes||_{q'} \kappa''_{l'',m''},  
\end{equation}
where we have introduced the ``limit contraction''
\[
\kappa'_{l',m'} \otimes||_{q'} \kappa''_{l'',m''}(\chi) =
\kappa'_{l',m'} \otimes_{q'} \kappa''_{l'',m''}(\Omega \chi),
\]
on the test functions $\chi(x_1,x_2) = \phi(x_1-x_2)\varphi(x_2)$, $\phi,\varphi \in \mathscr{E}$ and where 
the multiplication by $\theta$ in (\ref{DoubleLimitContraction}) is inderstood as the multiplication by the 
following functon $(x_1,x_2) \mapsto\theta(x_1-x_2)$ 
of two space-time variables $x_2,x_2$.

Similarly, we have for the kernels of the operator $\theta D_{(n)}$, $n>2$
which are given by the analogue expression. In the last case $\kappa'_{l',m'}$
and $\kappa''_{l'',m''}$ run independently over the kernels, 
respectively, of the operators $\overline{S_{l}}$ and $S_{k}$, $l+k=n$, $l>0$. But in this case $\theta D_{(n) \, \epsilon}$ 
the function $\theta$
is understand as equal to the product of the one-dimensional $\theta$-functions
evaluated at $x_1-x_n, \ldots, x_{n-1}-x_n$:
\[
\theta(x_1-x_n, x_2-x_n, \ldots, x_{n-1}-x_n) =  \theta(x_1-x_n) \ldots \theta(x_{n-1}-x_n).
\]
In this case we consider elements $\chi \in \mathscr{E}^{\otimes \, n}$  of the form
\begin{multline}\label{FormOfchi}
\chi(x_1, \ldots, x_n) = \phi(x_1 -x_n, x_2-x_n, \ldots, x_{n-1} -x_{n})\varphi(x_n), 
\\
= (\phi \otimes \varphi) \circ L^{-1}(x_1, \ldots, x_n), \,\,\,\,\,\,
\phi \in \mathscr{E}^{n-1}, \varphi\in \mathscr{E}
\end{multline}
which respect the condition
\begin{equation}\label{Dalphaphi|0=0,dim=k-1}
D^\alpha\phi(0) = 0, \,\,\, 0 \leq |\alpha| \leq \omega, 
\end{equation}
or, equivalently, the condition
\begin{equation}\label{Der|0chi=0,dim=n}
D^\alpha_{{}_{x}}\chi(x_1=x_n, \ldots, x_{n-1} = x_n, y) = 0, \,\,\, 0 \leq |\alpha| \leq \omega, 
\end{equation}
\[
x= (x_1, \ldots, x_{n-1}) \in \big[\mathbb{R}^4\big]^{n-1}, \,\,\, y = x_n \in \mathbb{R}^4.
\]
Here $L^{-1}$ is the invertible linear map on $\big[\mathbb{R}^{4}\big]^{\times \, n}$ given by:
\[
L^{-1}: (x_1, \ldots, x_n) \longmapsto (x_1-x_n, x_2 - x_n, \ldots, x_{n-1}-x_{n}, x_{n}).
\]
Here $\omega$ depends only on the \emph{singularity degree} at zero of the vector-valued distribution
\begin{equation}\label{chronologicalkappal,m}
\kappa'_{l',m'} \otimes|_{{}_{q}} \, \kappa''_{l'',m''} \,\,
\end{equation}
$\phi \in \mathscr{E}^{\otimes \, n}$ in the first $n-1$  space-time variables
$(x_1-x_n, x_2-x_n, \ldots, x_{n-1}-x_n)$. 
This singularity degree is in fact equal to the singularity degree (in the sense of \cite{Epstein-Glaser})
of a scalar-valued translationally invariant (with
causally supported singular part) distribution canonically related to (\ref{chronologicalkappal,m}), which follows from the canonical
factorization of (\ref{chronologicalkappal,m}) into a scalar-valued factor, which captures all contractions,
and a vector-valued factor, without any contractions. Below we also explain how this factorization appears.

Let, for each multi-index $\alpha$, such that $0 \leq |\alpha| \leq \omega$, 
$\omega_{{}_{o \,\, \alpha}} \in \mathscr{E}^{\otimes \, (n-1)}$ on $\mathbb{R}^{4(n-1)}$ be such 
functions that\footnote{Such functions $\omega_{{}_{o \,\, \alpha}} \in \mathscr{E}^{\otimes \, (n-1)}$, $0 \leq |\alpha| \leq \omega$, do exist. 
Indeed let
\[
f_{{}_{\alpha}}(x) = 
{\textstyle\frac{x^\alpha}{\alpha!}}, \,\,\, x=(x_1, \ldots, x_{n-1}) \in \mathbb{R}^{4(n-1)}, 0 \leq |\alpha| \leq \omega,
\]
\[
\alpha ! = \prod \limits_{i=1}^{n-1} \prod \limits_{\mu=0}^{3} \alpha_{i\mu} !, 
\,\,\, x^\alpha = \prod \limits_{i=1}^{n-1} \prod \limits_{\mu=0}^{3} (x_{1\mu})^{\alpha_{1\mu}}, \,\,\, \mu = 0,1,2,3.
\]
It is not difficult to see that there exists $w \in \mathscr{E}^{n-1} = \mathcal{S}(\mathbb{R}^4)^{\otimes(n-1)}$,
which is equal $1$ on some neighborhood of zero.
Indeed,  let $w \in \mathscr{C}^\infty(\mathbb{R}^{4(n-1)})$, which is equal $1$ on some neighborhood of zero and zero outside some larger
neighborhood of zero (such a function does exist, compare  e.g. \cite{Rudin}).
Then we can put
\[
\omega_{{}_{o \,\, \alpha}} \overset{\textrm{df}}{=} f_{{}_{\alpha}}.w.
\]}
\[
D^\beta \omega_{{}_{o \,\, \alpha}} (0) = \delta^{\beta}_{\alpha}, \,\,\,\, 0 \leq |\alpha|, |\beta| \leq \omega.
\]
Let for any $\phi \in \mathscr{E}^{\otimes \, (n-1)}$ 
\begin{equation}\label{OmegaphiR(k-1)}
\Omega' \phi = \phi - \sum \limits_{0\leq |\alpha| \leq \omega} D^\alpha \phi(0) \, \omega_{{}_{o \,\, \alpha}} 
\end{equation}

Of course $\Omega'$ is a continuous operator $\mathscr{E}^{\otimes \, (n-1)} \rightarrow \mathscr{E}^{\otimes \, (n-1)}$,
whose image has finite codimension in $\mathscr{E}^{\otimes \, (n-1)}$ and consists of all functions $\Omega' \phi$
such that
\[
\big(D^\alpha \Omega' \phi\big)(0) = 0, \,\,\,  0 \leq |\alpha| \leq \omega.
\]
Next, let $\phi \in \mathscr{E}^{\otimes \, (n-1)}$ and $\varphi \in \mathscr{E}$, and 
let 
\[
\chi(x_1, \ldots, x_n) \overset{\textrm{df}}{=} \phi(x_1-x_n, x_2 - x_n, \ldots, x_{n-1} - x_n) \varphi(x_n)
= (\phi \otimes \varphi) \circ L^{-1}(x_1, \ldots, x_n).
\]

Let, for any such $\chi$, the following function be defined
\begin{equation}\label{SimpleFormulaForOmega}
\Omega\chi(x_1, \ldots, x_n) \overset{\textrm{df}}{=} \Omega' \phi(x_1-x_n, x_2 - x_n, \ldots, x_{n-1} - x_n) \varphi(x_n)
= (\Omega' \phi \otimes \varphi) \circ L^{-1}(x_1, \ldots, x_n),
\end{equation}
which by construction respects the condition (\ref{Der|0chi=0,dim=n}). 
More generally for any 
\[
\chi \in \mathscr{E}^{\otimes \, n}
\]
we define the function
\begin{multline*}
\chi^\natural(x_1, \ldots, x_n) = \chi(x_1+ x_n, x_2+ x_n, \ldots , x_{n-1} + x_n, x_n)
\\
=
\chi \circ L(x_1, \ldots, x_n),
\end{multline*}
where $L$ is the linear map
\[
(x_1, \ldots, x_n) \longmapsto (x_1+ x_n, \ldots, x_{n-1} +x_n, x_n).
\]

It is immediately seen that
\[
\chi^\natural \in \mathscr{E}^{\otimes \, n}
\]
and there exists a series 
\[
\chi^\natural = \sum \limits_{j} \phi_j \otimes \varphi_j, \,\,\,\,\,\,\, \chi = \sum \limits_{j} (\phi_j \otimes \varphi_j ) \circ L^{-1},
\]
of simple tensors $\phi_j \otimes \varphi_j$, 
$\phi_j \in \mathscr{E}^{\otimes \, (n-1)}$ and $\varphi_j \in \mathscr{E}$,
converging in $\mathscr{E}^{\otimes \, n}$. We define 
\begin{multline*}
\Omega \chi(x_1, \ldots, x_n) \overset{\textrm{df}}{=} \sum \limits_{j} \big(\Omega' \phi_j \big)(x_1-x_n, x_2-x_n, \ldots, x_{n-1}-x_n) 
\varphi_j(x_n)
\\
\overset{\textrm{df}}{=} \sum \limits_{j} \big(\Omega' \phi_j \otimes \varphi_j \big) \circ L^{-1}(x_1, \ldots, x_n),
\end{multline*}
so that
\[
\Omega(\chi) \circ L(x_1, \ldots, x_n) = \sum \limits_{j} \big(\Omega' \phi_j \big)(x_1, \ldots, x_{n-1}) 
\varphi_j(x_n)
=
\sum \limits_{j} \big[(\Omega'\phi_j) \otimes \varphi_j \big](x_1, \ldots, x_n),
\]
with $\Omega' \phi_j$
given by (\ref{OmegaphiR(k-1)}). Writing the same otherwise
\begin{multline*}
\Omega(\chi) \circ L (x_1, \ldots, x_n) = \chi \circ L(x_1, \ldots, x_n)
- \sum\limits_{|\beta|=0}^{\omega}  \omega_{{}_{0 \,\, \beta}}(x_1, \ldots, x_{n-1}) \,\, \times
\\
\times \,\,
D^{\beta}_{{}_{x_1, \ldots, x_{n-1}}} (\chi \circ L)(x_1=0, \ldots, x_{n-1}=0,x_n).
\end{multline*}

Equivalently
\begin{multline}\label{Omega-n-variables}
\Omega \chi(x_1, \ldots, x_n) 
= \chi(x_1, \ldots, x_n) -  \sum \limits_{|\beta|=0}^{\omega}
 \omega_{{}_{0 \,\, \beta}}(x_1-x_n, x_2-x_n, \ldots, x_{n-1}-x_n) \,\, \times
\\
\times \,\,
D^{\beta}_{{}_{x_1, \ldots, x_{n-1}}} \chi(x_1=x_n, x_2=x_n, \ldots, x_{n-1}=x_{n}, x_n).
\end{multline}

Existence of the kernels (\ref{specialRetkappa'timesq'kappa''}) of $S_2$, and more generally of $S_n$,
 and their corresponding continuity, which assures
\begin{gather*}
S_n \in \mathscr{L} \big(\mathscr{E}^{\otimes \, n}, \mathscr{L}((E), (E)) \big)
\,\,\, 
\textrm{with no massless fields in $\mathcal{L}$}
\\
S_n \in  \mathscr{L} \big(\mathscr{E}^{\otimes \, n}, \mathscr{L}((E), (E)^*) \big)
\,\,\, 
\textrm{with massless fields in $\mathcal{L}$},
\end{gather*}
 we have already proved in \cite{WN}, where we have used the fact that there exists well-defined
splitting of the scalar $\otimes_{q'}$-contractions (\ref{specialscalarq'contraction}) into retarded and advanced parts, 
and which was shown earlier for the causal symmetric or antisymmetric combinations of the products (\ref{CausalDifferencesOfProductsOfPairings}) 
of pairings in \cite{Epstein-Glaser}, and which moreover is unique if $q\leq 1$, and non unique if $q>1$, with the non uniqueness
depending on a finite number $N(\omega)$ of constants, with $N(\omega)$ depending on the singularity degree $\omega$ at zero of (\ref{specialscalarq'contraction}). 
Although this is already a well-established result, we present the splitting of the scalar contractions (products of pairings)
(\ref{specialscalarq'contraction}) or their causal symmetric or antisymmetric parts (\ref{CausalDifferencesOfProductsOfPairings}),
and then, using this fact, we show once again existence and respective continuity of the kernels (\ref{specialRetkappa'timesq'kappa''}).

Having in view applications to realistic QFT, like QED, we are especially concentrated here on the case in which the
interaction Lagrangian $\mathcal{L}$ contains massless free fields.

Let us remind that in QED we have the free fields $\mathbb{A} = A,\boldsymbol{\psi}$ equal to the free e.m. potential field $A$ or the free
Dirac bispinor field $\boldsymbol{\psi}$, regarded as sums of two integral kernel operators
\[
\mathbb{A}(\phi) = \mathbb{A}^{(-)}(\phi) + \mathbb{A}^{(+)}(\phi) = \Xi(\kappa_{0,1}(\phi)) +
\Xi(\kappa_{1,0}(\phi))
\]
with the integral kernels $\kappa_{l,m}$ represented by ordinary functions:
\[
\begin{split}
\kappa_{0,1}(\nu, \boldsymbol{\p}; \mu, x) =
{\textstyle\frac{\delta_{\nu \mu}}{(2\pi)^{3/2}\sqrt{2p^0(\boldsymbol{\p})}}}
e^{-ip\cdot x}, \,\,\,\,\,\,
p = (|p_0(\boldsymbol{\p})|, \boldsymbol{\p}), \, p\cdot p=0, \\
\kappa_{1,0}(\nu, \boldsymbol{\p}; \mu, x) =
{\textstyle\frac{-g_{\nu \mu}}{(2\pi)^{3/2}\sqrt{2p^0(\boldsymbol{\p})}}}
e^{ip\cdot x},
\,\,\,\,\,\,
p \cdot p = 0,
\end{split}
\]
for the free e.m.potential field $\mathbb{A}=A$ (in the Gupta-Bleuler gauge) and
\[
\kappa_{0,1}(s, \boldsymbol{\p}; a,x) = \left\{ \begin{array}{ll}
(2\pi)^{-3/2}u_{s}^{a}(\boldsymbol{\p})e^{-ip\cdot x}, \,\,\, \textrm{$p = (|p_0(\boldsymbol{\p})|, \boldsymbol{\p}), \, p \cdot p = m^2$} & \textrm{if $s=1,2$}
\\
0 & \textrm{if $s=3,4$}
\end{array} \right.,
\]
\[
\kappa_{1,0}(s, \boldsymbol{\p}; a,x) = \left\{ \begin{array}{ll}
0 & \textrm{if $s=1,2$}
\\
(2\pi)^{-3/2} v_{s-2}^{a}(\boldsymbol{\p})e^{ip\cdot x}, \,\,\, \textrm{$p \cdot p = m^2$} & \textrm{if $s=3,4$}
\end{array} \right.
\]
for the free Dirac spinor field $\mathbb{A}=\boldsymbol{\psi}$, and
which are in fact the respective plane wave solutions of d'Alembert and of Dirac equation, which span the corresponding generalized eigen-solution sub spaces. 
Here $g_{\nu\mu}$ are the components of the space-time Minkowski metric tensor with signature $(1,-1,-1,-1)$, and 
\[
u_s(\boldsymbol{\p}) =  \frac{1}{\sqrt{2}} \sqrt{\frac{E(\boldsymbol{\p}) + m}{2 E(\boldsymbol{\p})}}
\left( \begin{array}{c}   \chi_s + \frac{\boldsymbol{\p} \cdot \boldsymbol{\sigma}}{E(\boldsymbol{\p}) + m} \chi_s
\\                                           
              \chi_s - \frac{\boldsymbol{\p} \cdot \boldsymbol{\sigma}}{E(\boldsymbol{\p}) + m} \chi_s                         \end{array}\right),
\,\,\,\,\,\,\,\,  
v_s(\boldsymbol{\p}) =  \frac{1}{\sqrt{2}} \sqrt{\frac{E(\boldsymbol{\p}) + m}{2 E(\boldsymbol{\p})}}
\left( \begin{array}{c}   \chi_s + \frac{\boldsymbol{\p} \cdot \boldsymbol{\sigma}}{E(\boldsymbol{\p}) + m} \chi_s
\\                                           
              -\big(\chi_s - \frac{\boldsymbol{\p} \cdot \boldsymbol{\sigma}}{E(\boldsymbol{\p}) + m}\chi_s \big)                          \end{array}\right) 
\]
where
\[
\chi_1 = \left( \begin{array}{c} 1  \\
                                                  0 \end{array}\right), \,\,\,\,\,
\chi_2 = \left( \begin{array}{c} 0  \\
                                                  1 \end{array}\right),
\]
are the Fourier transforms of the complete system of the free Dirac equation in the chiral represenation in which 
\[
\gamma^0 = \left( \begin{array}{cc}   0 &  \bold{1}_2  \\
                                           
                                                   \bold{1}_2              & 0 \end{array}\right), \,\,\,\,
\gamma^k = \left( \begin{array}{cc}   0 &  -\sigma_k  \\
                                           
                                                   \sigma_k             & 0 \end{array}\right),
\]
with the Pauli matrices
\[
\boldsymbol{\sigma} = (\sigma_1,\sigma_2,\sigma_3) = 
\left( \,\, \left( \begin{array}{cc} 0 & 1 \\

1 & 0 \end{array}\right), 
\,\,\,\,\,
\left( \begin{array}{cc} 0 & -i \\

i & 0 \end{array}\right),
\,\,\,
\left( \begin{array}{cc} 1 & 0 \\

0 & -1 \end{array}\right)
\,\,
\right).
\]
Given the kernels $\kappa_{0,1},\kappa_{1,0}$ of the Dirac field, the kernels of the Dirac conjugated
field $\boldsymbol{\psi}^{\sharp}$ are equal
\[
\kappa_{0,1}^{\sharp}(s, \boldsymbol{\p}; a,x) = \sum\limits_{b}\left[ \gamma^0 \right]_{ab}\overline{\kappa_{1,0}(s, \boldsymbol{\p}; b,x)},
\,\,\,\,\,
\kappa_{1,0}^{\sharp}(s, \boldsymbol{\p}; a,x) = \sum\limits_{b}\left[ \gamma^0 \right]_{ab}\overline{\kappa_{0,1}(s, \boldsymbol{\p}; b,x)}.
\]
Recall that in the Fock space of the e.m.potential field we have the Gupta-Bleuler operator $\eta$
respecting the commutation law $\eta a_{{}_{\nu}}(\boldsymbol{\p}) = - g_{{}_{\nu \nu}} \eta a_{{}_{\nu}}(\boldsymbol{\p})$. The e.m. potential field is
not self-adjoint but Krein self adjoint: $\eta A_{\mu}(x)^{+}\eta = A_{\mu}(x)$.

The single particle momentum test space $E_1$ of the Dirac field is equal $\mathcal{S}(\mathbb{R}^3; \mathbb{C}^4)$ 
(provided it is massive of mass $m\neq 0$),  and the single particle
momentum test space $E_2$ of the free e.m. potential massless field is equal to $\mathcal{S}^{0}(\mathbb{R}^3;\mathbb{C})$ -- the closed subspace
of  $\mathcal{S}(\mathbb{R}^3; \mathbb{C}^4)$ consisting of all those functions whose all derivaives vanish at zero.

Recall that the products $\otimes$ and $\overline{\dot{\otimes}}$ 
are respectively symmetrized in Bose momentum and spin variables and antisymmetrized in Fermi spin and momentum variables, according to the ordinary relation 
between the kernel and the integral kernel operator, in order to keep one-to-one relation between integral kernel operator and its kernel.

However, in practical computations it is convenient to use non-symmetrized products in the intermediate steps, and perform
the symmetrization and antisymmetrization at the very end, in order to keep the one-to-one relation between the kernel and the operator.
We will do so from now on.

We should emphasize here that the Wick theorem of \cite{WN} assures the contraction kernel (\ref{ProductForm}) 
to be a well-defined element of
\[
\mathscr{L}\big(E_{1} \otimes \widehat{\ldots} \otimes E_{q},  \mathscr{E}^{* \, \otimes \, 2}\big)
\cong \mathscr{L}\big(\mathscr{E}^{\otimes \, 2}, \,  E_{1}^{*}  \otimes \widehat{\ldots} \otimes E_{q}^{*} \big)
\]
(here $\widehat{\ldots}$ means that $q'$ pairs of single particle spaces $E_i$, corresponding to the $q'$ pairs of the paired
spin-momentum variables are removed from the tensor product). 
Let us show it more directly now. Indeed, for the dot product 
\[
\kappa'_{0,1} \dot{\otimes} \ldots  \dot{\otimes} \kappa^{(q)}_{0,1} \dot{\otimes}
\kappa'_{1,0} \dot{\otimes} \ldots \dot{\otimes} \kappa^{(q)}_{1,0} 
\]
containing massless plane wave kernels
\[
\kappa'_{0,1}, \ldots, \kappa^{(q)}_{0,1},
\kappa'_{1,0}, \ldots, \kappa^{(q)}_{1,0}, 
\]
corresponding to free fields with single particle Gelfand triples
\[
E_{{}_{(1)}} \subset \mathcal{H}_{{}_{(1)}} \subset E_{{}_{(1)}}^{*}, \,\,\,\, \ldots,
E_{{}_{(q)}} \subset \mathcal{H}_{{}_{(q)}} \subset E_{{}_{(q)}}^{*},
\]
the map (for each fixed component of the dot product)
\begin{multline*}
E_{{}_{(1)}} \otimes \ldots \otimes E_{{}_{(q)}}\otimes E_{{}_{(1)}} \otimes \ldots \otimes E_{{}_{(q)}} \ni
\xi_{{}_{(1)}} \otimes \ldots \otimes \xi_{{}_{(q)}} \longmapsto 
\\
\longmapsto 
\big(\kappa'_{0,1} \dot{\otimes} \ldots  \dot{\otimes} \kappa^{(q)}_{0,1} \dot{\otimes}
\kappa'_{1,0} \dot{\otimes} \ldots \dot{\otimes} \kappa^{(q)}_{1,0}\big)
(\xi_{{}_{(1)}} \otimes \ldots \otimes \xi_{{}_{(q)}}) \in \mathcal{O}_{M}(\mathbb{R}^4; \mathbb{C})
\end{multline*}
is continuous for the Schwartz' operator topology in $\mathcal{O}_{M}$.
This means that the map 
\begin{multline*}
E_{{}_{(1)}} \otimes \ldots \otimes E_{{}_{(q)}} \otimes E_{{}_{(1)}} \otimes \ldots \otimes E_{{}_{(q)}} \ni
\xi_{{}_{(1)}} \otimes \ldots \otimes \xi_{{}_{(2q)}} \longmapsto 
\\
\longmapsto 
\big(\kappa'_{0,1} \dot{\otimes} \ldots  \dot{\otimes} \kappa^{(q)}_{0,1} \dot{\otimes}
\kappa'_{1,0} \dot{\otimes} \ldots  \dot{\otimes} \kappa^{(q)}_{1,0}\big)
(\xi_{{}_{(1)}} \otimes \ldots \otimes \xi_{{}_{(2q)}}) 
\\
\in \textrm{Algebra of multipliers of }\mathscr{E}^{*}
\end{multline*}
is continuous for $\mathscr{E} = \mathcal{S}(\mathbb{R}^4)$, 
with the Schwartz' operator topology in the algebra of multipliers of $\mathscr{E}^{*}$.

Indeed, by lemma \ref{Cont.free.field.kernels}, we have the continuity \cite{WN}:
\[
E_{{}_{(i)}} \ni \xi_{{}_{(i)}} \longmapsto \kappa^{(i)}_{\ell,m}(\xi_{{}_{(i)}}) \in \mathcal{O}_{M}(\mathbb{R}^4; \mathbb{C}), \,\,\, (\ell,m) = (0,1), (1,0).
\]
The general case for dot products  of more plane wave kernels follows from the product formula
\[
\big(\kappa'_{0,1} \dot{\otimes} \ldots  \dot{\otimes} \kappa^{(q)}_{0,1} \dot{\otimes}
\kappa'_{1,0} \dot{\otimes} \ldots \dot{\otimes} \kappa^{(q)}_{1,0}\big)
(\xi_{{}_{(1)}} \otimes \ldots \otimes \xi_{{}_{(2q)}}) 
= 
\kappa'_{0,1}(\xi_{{}_{(1)}})\cdot \ldots \cdot \kappa^{(q)}_{0,1}(\xi_{{}_{(q)}}) \cdot
\kappa'_{1,0}(\xi_{{}_{(q+1)}}) \cdot \ldots \cdot \kappa^{(2q)}_{1,0}(\xi_{{}_{(q)}})
\]
and the fact that each factor
\[
\kappa^{(i)}_{\ell,m}(\xi_{{}_{(i)}}),  \,\,\,\,\,\, (\ell,m) = (0,1), (1,0),
\]
is a multiplier of $\mathscr{E}^*$ continuously depending on $\xi_{{}_{(i)}}$.  

In particular, writing $\kappa_{\ell,m}$ for the vector valued factor 
\[
\big(\kappa'_{0,1} \dot{\otimes} \widehat{\ldots} \dot{\otimes} \kappa^{(q)}_{0,1} \big)
\, \otimes \,\, \big(\kappa'_{1,0} \dot{\otimes} \widehat{\ldots} \dot{\otimes} \kappa^{(q)}_{1,0}  \big)
\]
of the right hand side of (\ref{ProductForm})
we see that it defines a continuous map
\[
E_{1} \otimes \widehat{\ldots} \otimes E_{q} \ni
\xi \longmapsto \kappa_{\ell,m}(\xi) \in \mathcal{O}_M\big([\mathbb{R}^4]^{\times \, 2}\big)
\]
into the Schwartz' algebra $\mathcal{O}_M\big([\mathbb{R}^4]^{\times \, 2}\big)$ of multipliers of
$\mathcal{S}\big([\mathbb{R}^4]^{\times \, 2}\big)^*$, with the Schwartz operator topology on
$\mathcal{O}_M\big([\mathbb{R}^4]^{\times \, 2}\big)$. Because the scalar factor 
\begin{equation}\label{ScalarCotractionFactor}
\Big[
\big(\kappa^{(k_1)}_{0,1} \dot{\otimes} \kappa^{(k_2)}_{0,1} \ldots \dot{\otimes} \kappa^{(k_{q'})}_{0,1} \big)
\, \otimes_{q'} \,\, \big(\kappa^{(k_1)}_{1,0} \dot{\otimes} \kappa^{(k_2)}_{1,0} \ldots \dot{\otimes} \kappa^{(k_{q'})}_{1,0}  
\big)\Big] \, 
\end{equation}
of the right hand side of (\ref{ProductForm}) 
is a well-defined element of $\mathscr{E}^{* \otimes \, 2}$
then it follows that the whole product (\ref{ProductForm})
defines a continuous map
\begin{equation}\label{productkernelsmultipliers'}
E_{1} \otimes \widehat{\ldots} \otimes E_{q}  \ni
\xi 
\longmapsto
\Big[
\big(\kappa^{(k_1)}_{0,1} \dot{\otimes} \kappa^{(k_2)}_{0,1} \ldots \dot{\otimes} \kappa^{(k_{q'})}_{0,1} \big)
\, \otimes_{q'} \,\, \big(\kappa^{(k_1)}_{1,0} \dot{\otimes} \kappa^{(k_2)}_{1,0} \ldots \dot{\otimes} \kappa^{(k_{q'})}_{1,0}  
\big)\Big] \,
 \cdot
\kappa_{\ell,m}(\xi) \in \mathscr{E}^{* \otimes \, 2}
\end{equation}
and the product (\ref{ProductForm}) is a well-defined element of 
\[
\mathscr{L}\big(E_{1} \otimes \widehat{\ldots} \otimes E_{q},  \mathscr{E}^{* \, \otimes \, 2}\big),
\]
because $\kappa_{\ell,m}(\xi)$ is a multiplier of $\mathscr{E}^{* \otimes \, 2}$.
In particular the kernel function (\ref{ProductForm}) can be written as the
ordinary product of the scalar kernel (depending only on space-time variables) and of a vector-valued
kernel depending on space-time and momentum variables. 

From this it also immediately follows that the multiplication of a Wick product of free fields, say respectively of
$x_1, \ldots, x_n$ space-time variables, by the product $\theta(x_1-x_n)\theta(x_2-x_n) \ldots \theta(x_{n-1}-x_n)$
of $\theta$-functions is a well-defined integral kernel operator with vector-valued kernel.
The same argument shows that also the kernel (\ref{specialRetkappa'timesq'kappa''}) is a well-defined element
of 
\begin{equation}\label{inclusionRet}
\mathscr{L}\big(E_{1} \otimes \widehat{\ldots} \otimes E_{q},  \mathscr{E}^{* \, \otimes \, 2}\big)
\end{equation}
provided the retarded part
\[
\textrm{ret} \,\,
\Big[
\big(\kappa^{(k_1)}_{0,1} \dot{\otimes} \kappa^{(k_2)}_{0,1} \ldots \dot{\otimes} \kappa^{(k_{q'})}_{0,1} \big)
\, \otimes_{q'} \,\, \big(\kappa^{(k_1)}_{1,0} \dot{\otimes} \kappa^{(k_2)}_{1,0} \ldots \dot{\otimes} \kappa^{(k_{q'})}_{1,0}  \big)\Big] 
\]
of the scalar contraction factor (\ref{ScalarCotractionFactor}) of (\ref{ProductForm}) is a well-defined ditribution
in $\mathscr{E}^{* \otimes \, 2}$. Therefore our analysis we can restrict to the spliting into retarded and advanced parts
of scalar factors of the contractions, in which all spin-momentum variables are contracted. It is sufficient to show that the retarded and advanced
parts of the scalar factors of the contractions of the kernels are well-defined
tempered distributions. Indeed, inclusion (\ref{inclusionRet}) shows that the kernels of $S_2$  are well-defined vector-valed distributions and, 
by theorem \ref{Hida-ObataTheorem}, $S_2$ is a finite sum of generalized integral kernel operators. The same method applies to
to $S_n$, $n>2$: retarded and advanced parts of contractions of tensor products of the kernels of $S_k$, $k<n$, define the kernels
of $S_n$. Because retarded and advanced parts of causal distributions are well-defined (and up to the freedom
depending on the singularity degree), then by identical argument, it follows that $S_n$ are finite sums of well-defined integral kernel
operators.  In case of the second order we therefore have to show that the scalar contractions $\otimes_q$ of the kernels having the form
of pointwise (in space-time) products of the kernels of the free fields are well-defined tempered distributions. The conraction
is scalar in the sense that all spin-momentum variables are contracted, and thus we consider only such tensor product kernels which contains 
the spin momenta variables which all can be arrangend in pairs which can be mutually contracted.

The (negative frequency) pairing functions
\[
\big[\mathbb{A}^{(i)(-)}, \mathbb{A}^{(j)(+)} \big]_\mp = \kappa^{(i)}_{0,1} \otimes_1 \kappa^{(j)}_{1,0} = \pm iD^{(-)}
\]
of the free fields $\mathbb{A}^{ (i)}, \mathbb{A}^{ (j)}$
are equal to the $1$-contractions 
\[
\kappa^{(i)}_{0,1} \otimes_1 \kappa^{(j)}_{1,0}
\]
of the corresponding kernels, have the following general form (for the free scalar field of mass $m$, compare \cite{Bogoliubov_Shirkov})
\begin{multline}\label{D(+)}
D^{(+)}(x) = {\textstyle\frac{1}{4\pi}}\varepsilon(x_0)\delta(x\cdot x)
- {\textstyle\frac{m i}{8\pi \sqrt{x\cdot x}}} \theta(x\cdot x)
\big[
N_1(m \sqrt{x\cdot x}) - i\varepsilon(x_0) J_1(m\sqrt{x\cdot x}) 
\big]
\\
-\theta(-x\cdot x) {\textstyle\frac{m i}{4\pi^2 \sqrt{-x\cdot x}}} K_1(m\sqrt{- x\cdot x}), 
\end{multline}
\begin{multline}\label{D(-)}
D^{(-)}(x) = {\textstyle\frac{1}{4\pi}}\varepsilon(x_0)\delta(x\cdot x)
+ {\textstyle\frac{m i}{8\pi \sqrt{x\cdot x}}} \theta(x\cdot x)
\big[
N_1(m \sqrt{x\cdot x}) + i\varepsilon(x_0) J_1(m\sqrt{x\cdot x}) 
\big]
\\
+\theta(-x\cdot x) {\textstyle\frac{m i}{4\pi^2 \sqrt{-x\cdot x}}} K_1(m\sqrt{- x\cdot x}).
\end{multline}
Here
\[
\varepsilon(x_0) = \textrm{sgn}(x_0) =  \theta(x_0) - \theta(-x_0),
\]
$J_\nu$ are the Bessel functions of the first kind, $N_\nu$ -- the Neumann functions 
(or the Bessel functions of the second kind) and $K_\nu$ -- the Hankel functions (or the Bessel functions
of the third kind). In order to obtain the pairings for other massive fields we only need to apply a corresponding invariant differential
operator to (\ref{D(+)}) and (\ref{D(-)}), e.g. the operator $i\gamma^\mu \partial_\mu +m$ in case of the Dirac field,
or pass to the limit $m\rightarrow 0$ in case of a massless field. Therefore in each case the singular part of  $D^{(+)}$ and of $D^{(-)}$
is located at the light cone. 

From (\ref{D(+)}) and (\ref{D(-)}) we see that the commutation functions, \emph{i.e.} the sums $D^{(+)}+D^{(-)}$ of the pairing functions, 
have causal support lying within the closure of the interiors of the past
and forward light cones. Using the support property of (\ref{D(+)}) and (\ref{D(-)}), and thus of the pairings, or the $1$-contractions, 
one can rather easily prove that the scalar factors in the Wick monomials of the Wick decomposition of
\[
D_{(2)}(x_1,x_2) = \mathcal{L}(x_1)\mathcal{L}(x_2) - \mathcal{L}(x_2)\mathcal{L}(x_1),
\]
are causally supported. These causal scalar factors
are linear combinations (\ref{CausalDifferencesOfProductsOfPairings}) of the scalar contractions of the form
(\ref{ProductsOfPairings}). More generally, using (\ref{D(+)}) and (\ref{D(-)}) one can show that the
the difference of the products of even number of pairings (or sum of products of odd number of pairings), has causal support.

Because the support of the operator $D_{(n)}$ is causal, then, both, in case of the computation of the retarded part
\[
\theta(x_1-x_2)\mathcal{L}(x_1)\mathcal{L}(x_2) 
\]
of $\mathcal{L}(x_1)\mathcal{L}(x_2)$ in
\[
S_2(x_1,x_2) = \theta(x_1-x_2)\mathcal{L}(x_1)\mathcal{L}(x_2) + \theta(x_2-x_1)\mathcal{L}(x_2)\mathcal{L}(x_1), 
\]
as well as in the computation
of the retarded part $R_{(n)}=\theta D_{(n)}$ of $D_{(n)}$ in $S_n = R_{(n)} - R'_{(n)}$ we can use
the splitting method of \cite{Scharf} (inspired by \cite{Epstein-Glaser}) applied to
the scalar factors (\ref{ScalarCotractionFactor}) of the kernels of 
$\mathcal{L}(x_1)\mathcal{L}(x_2)$ or, respectively, of 
$D_{(n)}$, into the retarded and advanced parts, although each of these factors (\ref{ScalarCotractionFactor}) taken separately is not causal.
Recall, please, that only full scalar factors of the contributions collecting all terms in  $D_{(n)}$ which multiply 
a fixed Wick monomial, are causal, by the causality of $D_{(n)}$.
The terms without contractions in the Wick product expansion of $\mathcal{L}(x_1)\mathcal{L}(x_2)$
or, respectively, of $D_{(n)}(x_1, \ldots, x_n)$, can be multiplied by $\theta(x_1-x_2)$, or respectively, by
$\theta(x_1-x_n) \ldots \theta(x_{n-1}-x_n)$, as we have already seen. 

The singularity degree $\omega$ of (\ref{ScalarCotractionFactor}) with $k=2$ (for the kernels of $\mathcal{L}(x_1)\mathcal{L}(x_2)$), 
in the sense of \cite{Epstein-Glaser}, 
determines canonically the subspace of $\mathscr{E}^k$ (here with $k=2$ for the kernels of
$\mathcal{L}(x_1)\mathcal{L}(x_2)$, or $k>2$ for the kernels of $D_{(k)}$) 
on which the scalar distribution 
(\ref{ScalarCotractionFactor}), multiplied by the step theta function $\theta(x_1-x_2)$ (or, respectively, by 
$\theta(x_1-x_k) \ldots \theta(x_{k-1}-x_k)$ for  $R_{(k)}=\theta D_{(k)}$),
is given  by an ordinary contraction integral which is absolutely convergent, and defines a retarded part of 
(\ref{ScalarCotractionFactor}) on this subspace. Moreover, by translational
invariance it determines
the retarded part of  (\ref{ScalarCotractionFactor}) up to 
\begin{equation}\label{SimpleEpsteinGlaserReminder}
\delta^{\omega}_{{}_{1, \ldots, k-1; \,\, k}}(x_1, \ldots, x_k)
\overset{\textrm{df}}{=}
\sum \limits_{|\alpha|=0}^{\omega} C_\alpha D^\alpha \big[ \delta(x_1 - x_k) \delta(x_2-x_k) \ldots \delta(x_{k-1} -x_k)\big]
\end{equation}
with a finite sequence of arbitrary constants
$C_0, \ldots, C_\omega$, compare \cite{Epstein-Glaser}.

Let us analyze in more details the particular case of the \emph{limit contraction} $\otimes||_{{}_{q}}$
(\ref{DoubleLimitContraction})  in which we have only two kernels $\theta \kappa'_{\ell',m'}$
and $\kappa''_{\ell'',m''}$. Let $\kappa'_{\ell',m'}$
and $\kappa''_{\ell'',m''}$ be kernels of the Fock expansion, respectively, of $\mathcal{L}(x)$ and $\mathcal{L}(y)$.  
Recall that we replace them with $\theta_\varepsilon \kappa'_{\epsilon \,\, \ell',m'}$,
$\kappa''_{\epsilon'' \,\, \ell'',m''}$,
then make the ordinarily $q$-contraction $\otimes_q$, $\theta_\varepsilon \kappa'_{\epsilon \,\, \ell',m'} \otimes_q \,
\kappa''_{\epsilon'' \,\, \ell'',m''}$, next project on the closed subspace of the space-time test space $\mathscr{E}^{\otimes \, 2}$ with the projection $\Omega$, 
depending on the singularity degree of the kernel\footnote{Recall that we use the limit contraction sign $\otimes|_{{}_{q}}$ exchangeably with
the ordinary contraction sign $\otimes_q$ because the contraction integral is convergent even for kernels being equal to pointed products of massless kernels,
compare our previous paper.}
\[
\kappa'_{\ell',m'} \otimes|_q  \,  \kappa''_{\ell'',m''} = \kappa'_{\ell',m'} \otimes_q  \,  \kappa''_{\ell'',m''},
\]
and finally integrate.

Namely, for the kernels $\kappa'_{\ell',m'}, \kappa''_{\ell'',m''}$  of the Fock 
expansion of the operator $\Xi' = \mathcal{L}$,  we have
\begin{equation}\label{kappa'otimes||kappa'}
\theta \kappa'_{\ell',m'} \otimes||_{{}_{q}}  \,  \kappa''_{\ell'',m''} \overset{\textrm{df}}{=}
\theta \kappa'_{\epsilon \,\, \ell',m'} \otimes_q \, \kappa''_{\epsilon \,\, \ell'',m''} \circ \Omega
\end{equation}
in
\[
\mathscr{L}\big(\mathscr{E} \otimes \mathscr{E} , E_{1}^{* \otimes 2} \otimes  \widehat{\ldots} \otimes E_{n}^{* \otimes 2} \big)
\]
\[
\cong \mathscr{L}\big(E_{1}^{\otimes 2} \otimes \widehat{\ldots} \otimes E_{n}^{\otimes 2}, \, \mathscr{E}^* \otimes \mathscr{E}^*   \big),
\]
which, can also be undertand as a limit, in which all exponents of the non contracted massless free field kernels
are replaced with the exponents of massive kernels, next integration and finally by passing to the limit zero with
the auxiliary mass.  This limit is convergent, as we have shown in the previous Subsection, where existence of the retarded part 
of the scalar contractions (\emph{i.e.} products of pairings, in which no non contracted varaibles are present) was assumed. This 
assumption was not arbitrary, as it was proved to hold earlier by Epstein and Glaser, compare also \cite{Scharf}.

The limit (\ref{kappa'otimes||kappa'}) does exist because of the presence of the operator $\Omega$ which projects on the closed subspace
on which the convergence holds. If the operator $\Omega$ was removed from (\ref{kappa'otimes||kappa'}), then the limit
(\ref{kappa'otimes||kappa'}) would not be existing in general for the kernel 
\[
\kappa'_{\ell',m'} \otimes_q  \,  \kappa''_{\ell'',m''}
\]
with $q>1$, whose singularity degree $\omega>0$.

Let us explain this fact in more details in case of particular but generic examples. In these examples we consider 
$\kappa'_{\ell',m'}, \kappa''_{\ell'',m''}$ being equal not only to simple dot products of just three massless or massive plane wave kernels 
(as in QED), but a more general situation of simple products of arbitrary large number of plane wave kernels, massless or massive.
We give the analysis of the limit (\ref{kappa'otimes||kappa'}) in case $q>1$ in which the projection operator $\Omega$ is essential,
and omit the simpler case $q=0,1$, in which $\Omega$ can be removed and the retarded part can be computed through the ordinary 
multiplication by the step theta function.

\subsection{Splitting of general $\otimes_{q}$ contractions for $q>1$}\label{SplittingOf(x)q}

Let
\begin{gather}\label{freekernels}
\kappa'_{0,1}(\boldsymbol{\p}'; x) = u'(\boldsymbol{\p}') e^{-ip'\cdot x},
\, \kappa'_{1,0}(\boldsymbol{\p}'; x) = v'(\boldsymbol{\p}') e^{ip'\cdot x}, 
\\
\kappa''_{0,1}(\boldsymbol{\p}''; x) = u''(\boldsymbol{\p}'') e^{-ip''\cdot x},
\, \kappa''_{1,0}(\boldsymbol{\p}'; x) = v'(\boldsymbol{\p}'') e^{ip'''\cdot x}, 
\\
\ldots
\end{gather}
be the plane wave kernels of a free, say massless, fields $\mathbb{A}', \mathbb{A}'', \ldots$.  

We consider the value of the contraction integral ($\theta_y(x)=\theta(x-y)$)
\begin{equation}\label{theta(x-y)masslesskappa01.masslesskappa10contractionmasslesskappa10.masslesskappa10}
\Big[
\theta_y\big(\kappa'_{0,1} \dot{\otimes} \kappa''_{0,1} \dot{\otimes} \dots \dot{\otimes} \kappa^{(q)}_{0,1}\big)
 \otimes_q \big(\kappa'_{1,0} \dot{\otimes} \kappa''_{1,0} \dot{\otimes} \ldots 
\dot{\otimes} \kappa^{(q)}_{1,0}\big)\Big](\chi)
\end{equation}
for a general element
\begin{equation}\label{xiInS00timesS00}
\chi  \in \mathcal{S}(\mathbb{R}^4; \mathbb{C}^{d'd''}) \otimes \mathcal{S}(\mathbb{R}^4; \mathbb{C}^{d'd''}) =
\mathcal{S}(\mathbb{R}^4\times \mathbb{R}^4; \mathbb{C}^{d'd''d'd''}),
\end{equation}
in which the simple tensor $\chi = \phi \otimes \varphi$ fulfils
\begin{equation}\label{Dalphaphi(x,x)=0,alpha=<omega}
\big(\chi \circ L\big) (x=0,y) = 0, \,\,\,
D_{{}_{x}}^{\alpha} \big(\chi \circ L\big)(x=0,y) = 0, \,\, 1 \leq |\alpha| \leq \omega,
\end{equation}
in the Schwartz multi-index notation for partial derivatives in the variables $x = (x_0, \ldots x_3)$,
where $L$ is the following linear invertible transformation
\[
\mathbb{R}^4 \times \mathbb{R}^4 \ni (x,y) \longmapsto L(x,y) = (x+y, y)
\in \mathbb{R}^4 \times \mathbb{R}^4,
\]
and where $\omega$ is the singularity degree of the distribution
\[
\big(\kappa'_{0,1} \dot{\otimes} \kappa''_{0,1} \big)
\, \otimes_2 \,\, \big(\kappa'_{1,0} \dot{\otimes} \kappa''_{1,0} \big) \subset
\mathcal{S}(\mathbb{R}^4\times \mathbb{R}^4; \mathbb{C}^{d'd''d'd''})^*,
\]
(in the sense of \cite{Epstein-Glaser}). Then the integral
(\ref{theta(x-y)masslesskappa01.masslesskappa10contractionmasslesskappa10.masslesskappa10}) with $q=2$
\begin{equation}\label{[A'(-),A'(+)]^2}
\int
u'(\boldsymbol{\p}') v'(\boldsymbol{\p}') u''(\boldsymbol{\p}'') v''(\boldsymbol{\p}'')
e^{-i(|\boldsymbol{p}'| +|\boldsymbol{p}''|)(x_0-y_0) +i(\boldsymbol{p}' + \boldsymbol{p}'')\cdot (\boldsymbol{\x} - \boldsymbol{\y})}
\,\, 
\theta(x-y) \, \chi(x,y)
\, \ud^3 \boldsymbol{p}' \, \ud^3 \boldsymbol{p}'' \, \ud^4 x \ud^4 y
\end{equation}
becomes absolutely convergent. Analogous statement holds for arbitrary $\otimes_q$-contraction integrals, with the respective
order $\omega$ depending on $q$ and on the plane wave kernels $\kappa'_{0,1}, \kappa''_{0,1}, \ldots$
in the two contracted kernels, one being equal to dot product $\dot{\otimes}$ of $q$
plane waves $\kappa'_{0,1}, \kappa''_{0,1}, \ldots$ and multiplied by $\theta(x-y)$ and the second kernel being equal to the
dot product $\dot{\otimes}$ of $q$
plane waves $\kappa'_{1,0}, \kappa''_{1,0}, \ldots$.
This can be seen relatively simply for the causal symmetric/antisymmetric parts of the contractions if we use the translation 
invariance of the respective contraction distributions and causal support, on using their ultraviolet quasiasymptotis. 
By translational invariance the integral (\ref{[A'(-),A'(+)]^2}) 
and for $\chi$ which respect (\ref{Dalphaphi(x,x)=0,alpha=<omega}), is equal to the integral
\begin{equation}\label{[A'(-),A'(+)]^2(x)}
\int
u'(\boldsymbol{\p}') v'(\boldsymbol{\p}') u''(\boldsymbol{\p}'') v''(\boldsymbol{\p}'')
e^{-i(|\boldsymbol{p}'| +|\boldsymbol{p}''|)x_0 +i(\boldsymbol{p}' + \boldsymbol{p}'')\cdot \boldsymbol{\x}}
\,\, 
\theta(x) \, \phi(x)
\, \ud^3 \boldsymbol{p}' \, \ud^3 \boldsymbol{p}'' \, \ud^4 x \, \widetilde{\varphi}(0)
\end{equation}
\[
=
\int
u'(\boldsymbol{\p}') v'(\boldsymbol{\p}') u''(\boldsymbol{\p}'') v''(\boldsymbol{\p}'')
e^{-i(|\boldsymbol{p}'| +|\boldsymbol{p}''|)x_0 +i(\boldsymbol{p}' + \boldsymbol{p}'')\cdot \boldsymbol{\x}}
\,\, 
\widetilde{\theta \phi}(-\boldsymbol{\p}'-\boldsymbol{\p}'', -|\boldsymbol{\p}'| - |\boldsymbol{\p}''|)
\, \ud^3 \boldsymbol{\p}' \, \ud^3 \boldsymbol{\p}'' \,\, \widetilde{\varphi}(0)
\]
for any
\[
\phi \in 
\mathcal{S}(\mathbb{R}^4; \mathbb{C}^{d'd''}),
\]
which fulfills
\begin{equation}\label{up-to-omegaorder-derivatives-of-phi=0}
\phi(0) = 0, \,\,\,
D^{\alpha} \phi(0) = 0, \,\, 1 \leq |\alpha| \leq \omega,
\end{equation}
in the Schwartz multi-index notation for partial derivatives, and where
$\omega$ is the singularity degree of the distribution $\kappa_2$, such that
\begin{equation}\label{kappa2(x-y)}
\kappa_2(x-y) = \big(\kappa'_{0,1} \dot{\otimes} \kappa''_{0,1} \big)
\, \otimes_2 \,\, \big(\kappa'_{1,0} \dot{\otimes} \kappa''_{1,0} \big)(x,y)
= \big(\kappa'_{0,1} \dot{\otimes} \kappa''_{0,1} \big)
\, \otimes_2 \,\, \big(\kappa'_{1,0} \dot{\otimes} \kappa''_{1,0} \big)(x-y),
\end{equation}
and which, by definition, coincides with the singularity degree of the distribution
\[
\big(\kappa'_{0,1} \dot{\otimes} \kappa''_{0,1} \big)
\, \otimes_2 \,\, \big(\kappa'_{1,0} \dot{\otimes} \kappa''_{1,0} \big)(x,0)
\]
at $x=0$.

Instead of the negative frequency $\otimes_q$-contractions themselves, we concentrate our attention first
on the negative and positive frequency combinations of the $\otimes_q$-contractions. Using (\ref{D(+)}) and (\ref{D(-)}), we can prove that
\begin{multline*}
 \big(\kappa'_{0,1} \dot{\otimes} \kappa''_{0,1} \dot{\otimes} \kappa'''_{0,1} \dot{\otimes} \ldots \dot{\otimes} \kappa^{(q)}_{0,1} \big)
\, \otimes_{q} \,\, \big(\kappa'_{1,0} \dot{\otimes} \kappa''_{1,0}  \dot{\otimes} \kappa'''_{1,0} \dot{\otimes} 
\ldots \dot{\otimes} \kappa^{(q)}_{1,0} \big)(x,y)
\\
- (-1)^{f(q)}
\big(\kappa'_{1,0} \dot{\otimes} \kappa''_{1,0} \dot{\otimes} \kappa'''_{1,0} \dot{\otimes} \ldots \dot{\otimes} \kappa^{(q)}_{1,0} \big)
\, \otimes_{q} \,\, \big(\kappa'_{0,1} \dot{\otimes} \kappa''_{0,1}  \dot{\otimes} \kappa'''_{0,1} \dot{\otimes} 
\ldots \dot{\otimes} \kappa^{(q)}_{0,1} \big)(x,y)
\\
= \kappa^{(-)}_q(x-y) - (-1)^{f(q)} \kappa^{(+)}_q(x-y) = \big[\kappa^{(-)}_q - (-1)^q \kappa^{(+)}_q\big](x-y)
\end{multline*}
is always causally supported.

Then, for 
\[
\chi(x,y) = \phi(x-y)\varphi(y), \,\,\,\, \phi,\varphi \in \mathscr{E} = \mathcal{S}(\mathbb{R}^4),
\]
and with 
\[
\phi \in \mathscr{E}= \mathcal{S}(\mathbb{R}^4) 
\]
which fulfil (\ref{up-to-omegaorder-derivatives-of-phi=0}), the valuation  integral
\begin{multline}\label{thetaASkappa2(chi)}
\Big[
\theta_y\big(\kappa'_{0,1} \dot{\otimes} \kappa''_{0,1} \dot{\otimes} \dots \dot{\otimes} \kappa^{(q)}_{0,1}\big)
 \otimes_q \big(\kappa'_{1,0} \dot{\otimes} \kappa''_{1,0} \dot{\otimes} \ldots 
\dot{\otimes} \kappa^{(q)}_{1,0}\big)
\\
- 
(-1)^{f(q)}
\theta_y\big(\kappa'_{1,0} \dot{\otimes} \kappa''_{1,0} \dot{\otimes} \dots \dot{\otimes} \kappa^{(q)}_{1,0} \big) \otimes_2\big(\kappa'_{0,1} \dot{\otimes} \kappa''_{0,1} \dot{\otimes} \dots \dot{\otimes} \kappa^{(q)}_{0,1} \big)
\Big] (\chi)
\end{multline}
becomes equal
\begin{multline*}
\int
[\kappa^{(-)}_{q}(x-y) -(-1)^{f(q)}\kappa^{(+)}_q(x-y)]
\theta(x-y) \, \phi(x-y)\varphi(y) \, \ud^4 x \, \ud^4 y
\\
=
\widetilde{\varphi}(0) \,\, \Big\langle \theta \big[\kappa^{(-)}_q -(-1)^{f(q)}\kappa^{(+)}_q\big], \phi \Big\rangle 
= \widetilde{\varphi}(0) \,\, \Big\langle \big[\kappa^{(-)}_q -(-1)^{f(q)}\kappa^{(+)}_q\big], \theta \phi \Big\rangle
\end{multline*}
\begin{multline}\label{Convergent[A',A''][A'', A'']theta}
=
\widetilde{\varphi}(0) \,\,
\int
u'(\boldsymbol{\p}') v'(\boldsymbol{\p}') u''(\boldsymbol{\p}'') v''(\boldsymbol{\p}'') \ldots 
\,
\widetilde{\theta\phi}(-\boldsymbol{p}' - \boldsymbol{p}'' - \ldots, \,-|\boldsymbol{p}'| -|\boldsymbol{p}''|-\ldots) 
\, \ud^3 \boldsymbol{p}' \, \ud^3 \boldsymbol{p}'' \ldots  \ud^3 \boldsymbol{p}^{(q)}
\\
-
 (-1)^{f(q)} \widetilde{\varphi}(0) \,\,
\int
u'(\boldsymbol{\p}') v'(\boldsymbol{\p}') u''(\boldsymbol{\p}'') v''(\boldsymbol{\p}'') \ldots 
\,
\widetilde{\theta\phi}(\boldsymbol{p}' + \boldsymbol{p}''+ \ldots, \,|\boldsymbol{p}'| +|\boldsymbol{p}''| + \ldots) 
\, \ud^3 \boldsymbol{p}' \, \ud^3 \boldsymbol{p}''\ldots \ud^3 \boldsymbol{p}^{(q)}. 
\end{multline}
which can also be understood (and which is non-trivial) as a limit $\lambda \rightarrow +\infty$ of
\begin{equation}\label{Convergent[A',A''][A'', A'']Slambda(chi0)}
\widetilde{\varphi}(0) \,\, \Big\langle (S_{\lambda}\chi_0) \big[\kappa_q - (-1)^q \check{\kappa_q}\big], \phi \Big\rangle 
\end{equation}
where for any function $h$ on $\mathbb{R}^4$ and real $\lambda$, $(S_\lambda h)(x) = h(\lambda x)$, 
and where $\chi_0(x) = f(x_0)$
with an everywhere smooth non decreasing function $f$, equal zero for $x_0 \leq 0$, $0\leq f \leq 1$ for $0\leq x_0 \leq 1$,
and $f=1$ for $x_0\geq 1$. Thus $S_\lambda \chi_0$ converges to $\theta(x)$ in $\mathcal{S}(\mathbb{R}^4)^*$ if $\lambda \rightarrow +\infty$,
with the fixed choice of the timelike unit versor $v$ of the reference frame equal $v =(1,0,0,0)$.  
More generally, for any other unit timelike $v\cdot v=1$ versor $v$ of the reference frame, we can consider
$\chi_0(x) = f(v\cdot x)$, with $S_\lambda \chi_0$ converging to $\theta(v\cdot x)$ in $\mathcal{S}(\mathbb{R}^4)^*$ if $\lambda \rightarrow +\infty$.

First we show convergence of the limit $\lambda \rightarrow \infty$
of (\ref{Convergent[A',A''][A'', A'']Slambda(chi0)}),
for each $\phi\in \mathscr{E}$ fulfilling (\ref{up-to-omegaorder-derivatives-of-phi=0}), with $\omega$ not less than the singularity degree
of $\kappa_q - (-1)^q \check{\kappa_q}$. For this purpose, it is sufficient to determine
the quasi-asymptotics together with its homogeneity of the distribution $\kappa_q$ 
at zero (in $x$ or at infinity in $p$ for its Fourier transform).
Let for any test function $\phi \in \mathcal{S}(\mathbb{R}^4)$ and $\lambda>0$, $S_\lambda\phi(x) = \phi(\lambda x)$.
Let us remind that the distribution $\kappa_{q \,0}$ is equal to the quasi-asymptotic part $\kappa_{q \,0}$
of $\kappa_q \in \mathcal{S}(\mathbb{R}^4)^*$ at zero if there exists a number $\omega$, 
called singularity degree at zero (in $x$ and at $\infty$
in $p$), such that the limit
\[
\underset{\lambda \rightarrow +\infty}{\textrm{lim}} \,\,
\big({\textstyle\frac{1}{\lambda}}\big)^\omega \big\langle \kappa_q, S_{\lambda}\phi \big\rangle
= 
\underset{\lambda \rightarrow +\infty}{\textrm{lim}} \,\,
\big({\textstyle\frac{1}{\lambda}}\big)^\omega \big({\textstyle\frac{1}{\lambda}}\big)^4
 \big\langle \widetilde{\kappa_q}, S_{1/\lambda} \widetilde{\phi} \big\rangle
=
\big\langle \kappa_{q \,0},\phi\big\rangle
\]
exists for each $\phi\in \mathcal{S}(\mathbb{R}^4)$ and is not equal zero for all $\phi\in \mathcal{S}(\mathbb{R}^4)$;
$\kappa_{q \,0}$ represents in fact the asymtoticaly homogeneous
part of $\kappa_{q} $, which is of ultraviolet homogeneity degree $\omega$. 
It is easily seen that $\kappa_{q \,0}$ is homogeneous of homogeneity degree $\omega$.

The singularity degrees $s', s'', \ldots$,  of the functions $u',v'$, $u'',v'',\ldots$ of $\boldsymbol{\p}'$,
$\boldsymbol{\p}'', \ldots$, defining the plane wave kernels (\ref{freekernels})
 of the free fields $\mathbb{A}', \mathbb{A}'', \ldots$ 
and regarded as Fourier transform 
of distributions in $\mathbb{R}^3$, can be easily computed.  By definition $s',s'', \ldots$, is the number for which,
respectively,
\begin{gather*}
\underset{\lambda \rightarrow +\infty}{\textrm{lim}} \,\,
\big({\textstyle\frac{1}{\lambda}}\big)^{s'}u'(\boldsymbol{\lambda \p}'),  
\,\,\,\,\,\,\,\,\,\,\,\,
\underset{\lambda \rightarrow +\infty}{\textrm{lim}} \,\,
\big({\textstyle\frac{1}{\lambda}}\big)^{s'}v'(\boldsymbol{\lambda\p}'),
\\
\underset{\lambda \rightarrow +\infty}{\textrm{lim}} \,\,
\big({\textstyle\frac{1}{\lambda}}\big)^{s''}u''(\lambda\boldsymbol{\p}''), 
\,\,\,\,\,\,\,\,\,\,\,\,
\underset{\lambda \rightarrow +\infty}{\textrm{lim}} \,\,
\big({\textstyle\frac{1}{\lambda}}\big)^{s''}v''(\lambda\boldsymbol{\p}''),
\\
\ldots
\end{gather*}
exists in distributional sense and is non zero.
Becuse the functions $u',v'$, $u'',v'',\ldots$, respectively, of $\boldsymbol{\p}'$, $\boldsymbol{\p}', \ldots$, are ordinary functions, analytic,
except eventually at zero in the massless case, then in practice it is sufficient to check for what choice of the number
$s',s'', \ldots$, the above limit exists pointwisely and represents a
nonzero function, respectively, of $\boldsymbol{\p}'$, $\boldsymbol{\p}'', \ldots$. It is easily seen
that the numbers $s',s'', \ldots$, respectively, are the same, for $u'$ and $v'$, then for $u''$ and $v''$, 
$\dots$, and the same for the massive field and for its massless counterpart. In particular for the
free spinor Dirac field $s = 0$. For the free electromagnetic potential field $s=-1/2$. For the scalar field $s = -1/2$.

Because
\begin{multline*}
\big\langle \kappa_q, \phi \big\rangle = \big\langle \widetilde{\kappa_q}, \widetilde{\phi} \big\rangle  =
\int 
u'(\boldsymbol{\p}') v'(\boldsymbol{\p}') u''(\boldsymbol{\p}'') v''(\boldsymbol{\p}'') \ldots 
u^{(q)}(\boldsymbol{\p}^{(q)}) v^{(q)}(\boldsymbol{\p}^{(q)}) \, \times 
\\
\times
\,
\widetilde{\phi}(-\boldsymbol{p}' - \boldsymbol{p}'' - \ldots, \,-p_0(\boldsymbol{p}') -p_0(\boldsymbol{p}'') - \ldots)
\, \ud^3 \boldsymbol{p}' \, \ud^3 \boldsymbol{p}''  \ldots \boldsymbol{p}^{(q)}
\end{multline*}
then the singularity degree $\omega$ at zero of the scalar $\otimes_q$-contraction, or product 
\begin{equation}\label{q-contraction}
\kappa_q(x-y) =
\big(\kappa'_{0,1} \dot{\otimes} \kappa''_{0,1} \dot{\otimes} \kappa'''_{0,1} \dot{\otimes} \ldots \dot{\otimes} \kappa^{(q)}_{0,1} \big)
\, \otimes_{q} \,\, \big(\kappa'_{1,0} \dot{\otimes} \kappa''_{1,0}  \dot{\otimes} \kappa'''_{1,0} \dot{\otimes} 
\ldots \dot{\otimes} \kappa^{(q)}_{1,0} \big)(x,y)
\end{equation}
of $q$ pairings, respectively, of $q$ free
fields $\mathbb{A}', \mathbb{A}'', \ldots$, is equal
\begin{equation}\label{omegaOfq-contraction}
\omega = 2(s'+s'' + \ldots +s^{(q)}) +3q - 4. 
\end{equation}

In each QFT we have the finite set of basic $\otimes_1$, $\otimes_2, \ldots$, $\otimes_q$-contractions 
$\kappa_1$, $\kappa_2, \ldots$, $\kappa^{(q)}$,
which enter into the Wick product decomposition of the product $\mathcal{L}(x)\mathcal{L}(y)$  of the
interaction Lagrangian of the theory. The whole theory can be constructed with the help of the advanced and retarded 
parts of the basic $\kappa_1$, $\kappa_2, \ldots$, $\kappa^{(q)}$. The maximal number $q$ depends on the degree of the Wick
polynomial $\mathcal{L}(x)$, which in case of QED is equal $3$.   
In spinor QED we have the following basic $\otimes_1$,$\otimes_2$ and $\otimes_3$-contractions: 
\begin{enumerate}
\item[1)]
the one-contraction $\kappa_1$ of the Dirac field
with the conjugated Dirac field, with $s' = 0$, $q=1$ and with $\omega = 2(0) +3-4 = -1$, 
\item[2)]
the one-contraction $\kappa_1$ of the electromagnetic potential with itself, with $s'=-1/2$, $s' = +1/2$ , 
$q=1$, and with  $\omega= 2(-1/2) +3 -4= -2$, 

\item[4)]
the two-contraction $\kappa_2$ containing two contractions of the Dirac field
with the conjugated Dirac field with $s'=s''=0$, $q=2$ and $\omega = 2(0+0) + 3\cdot 2 - 4 = 2$,

\item[5)]
the two-contraction $\kappa_2$ containing one contraction of the Dirac field
with the conjugated Dirac field and one contraction of electromagnetic potential with itself,
with $s'=0$, $s''=-1/2$, $q=2$,
and with $\omega = 2(0-1/2) +3\cdot 2 - 4 = 1$,

\item[6)]
the three-contraction $\kappa_3$ containing two contractions of the Dirac field
with the conjugated Dirac field and one contracion 
of the electromagnetic potential with itself, with $s'=s''=0$, $s'''=-1/2$, 
$q=3$, and $\omega = 2(0+0-1/2) +3\cdot 3 -4 = 4$. 
\end{enumerate}

The singularity degree $\omega$ of the causal combination
\[
\kappa^{(-)}_q -(-1)^{f(q)}\kappa^{(+)}_q \in \mathscr{E}^*,
\]
of the product $\kappa_q$ of pairings,  is the same as the singularity degree of the respective $\kappa_q$. 

Now we observe that in order to show the convergence of (\ref{Convergent[A',A''][A'', A'']Slambda(chi0)})
of the limit $\lambda\rightarrow +\infty$ for each $\phi\in\mathscr{E}$ fulfilling 
(\ref{up-to-omegaorder-derivatives-of-phi=0}), it is sufficient to show convergence of the limit
$n \rightarrow +\infty$ of
\begin{equation}\label{Convergent[A',A''][A'', A'']Slambda^n(chi0)}
\Big\langle (S_{\lambda^n}\chi_0) \big[\kappa^{(-)}_q -(-1)^{f(q)}\kappa^{(+)}_q\big], \phi \Big\rangle 
\end{equation}
 for each fixed $\lambda >1$ and $\phi\in\mathscr{E}$ fulfilling 
(\ref{up-to-omegaorder-derivatives-of-phi=0}), or, equivalently, that (\ref{Convergent[A',A''][A'', A'']Slambda^n(chi0)})
is a Cauchy sequence in $n$, for each fixed $\lambda >1$ and $\phi\in\mathscr{E}$ fulfilling 
(\ref{up-to-omegaorder-derivatives-of-phi=0}). We present here a proof due to \cite{Scharf}, pp. 175-177, and going back to \cite{Epstein-Glaser},
that (\ref{Convergent[A',A''][A'', A'']Slambda^n(chi0)}) is a Cauchy sequence. 

Because 
\[
\underset{\lambda \rightarrow +\infty}{\textrm{lim}} \,\,
\big({\textstyle\frac{1}{\lambda}}\big)^\omega \Big\langle [\kappa^{(-)}_q -(-1)^{f(q)}\kappa^{(+)}_q], S_{\lambda}\phi \Big\rangle
=
\Big\langle [\kappa^{(-)}_q -(-1)^{f(q)}\kappa^{(+)}_q],\phi\big\rangle
\]
exists and is finite for each $\phi\in\mathscr{E}$, then for each $h \in \mathscr{E}$
\[
\underset{\lambda \rightarrow +\infty}{\textrm{lim}} \,\,
\big({\textstyle\frac{1}{\lambda}}\big)^\omega \Big\langle h[\kappa^{(-)}_q -(-1)^{f(q)}\kappa^{(+)}_q], S_{\lambda}\phi \Big\rangle
=
\Big\langle [\kappa^{(-)}_q -(-1)^{f(q)}\kappa^{(+)}_q],\phi\big\rangle \, h(0)
\]
exists and is finite for each $\phi\in\mathscr{E}$, and thus 
\begin{multline*}
\underset{\lambda \rightarrow +\infty}{\textrm{lim}} \,\,
\big({\textstyle\frac{1}{\lambda}}\big)^{\omega-|\alpha|} \Big\langle x^\alpha h [\kappa^{(-)}_q -(-1)^{f(q)}\kappa^{(+)}_q], S_{\lambda} \phi \Big\rangle
=
\underset{\lambda \rightarrow +\infty}{\textrm{lim}} \,\,
\big({\textstyle\frac{1}{\lambda}}\big)^{\omega-|\alpha|} \Big\langle h [\kappa^{(-)}_q -(-1)^{f(q)}\kappa^{(+)}_q], 
S_{\lambda} \big (\textstyle{\frac{x^\alpha}{\lambda^{|\alpha|}}} \phi\big) \Big\rangle
\\
=
\underset{\lambda \rightarrow +\infty}{\textrm{lim}} \,\,
\big({\textstyle\frac{1}{\lambda}}\big)^{\omega} \Big\langle h [\kappa^{(-)}_q -(-1)^{f(q)}\kappa^{(+)}_q], 
S_{\lambda} \big (x^\alpha \phi\big) \Big\rangle
=
\Big\langle [\kappa^{(-)}_{q \,0} -(-1)^{f(q)}\kappa^{(+)}_{q \, 0}], x^\alpha\phi \Big\rangle \,\, h(0)
\end{multline*}
exists and is finite for each $h,\phi\in\mathscr{E}$.
Next, for each $\phi\in\mathscr{E}$ fulfilling 
(\ref{up-to-omegaorder-derivatives-of-phi=0}) there exists multiindices $\alpha$, $|\alpha| = \omega+1$, and $\phi_\alpha \in \mathscr{E}$
such that (summation is understood over $\alpha$ with $|\alpha| = \omega+1$)
\[
\phi = x^\alpha \phi_\alpha.
\]
Let $\psi$ be an auxiliary smooth bounded function equal $1$ on the closure of the future and past light cones, 
\emph{i.e.} on the support of $\kappa_q - (-1)^q \check{\kappa_q}$,
equal zero outside an open $\epsilon$-neighborhood of the light cone 
(sum of open $\epsilon$-balls centered at the points of the closed future and past light cones). Note also
that for each  $\lambda>0$, also $S_\lambda\psi$ is smooth, bounded, equal $1$ on the cone
(the support of $\kappa_q - (-1)^q \check{\kappa_q}$), and equal zero outside the  $\epsilon/\lambda$-neighborhood of the cone. 
Then, for each $\lambda \geq 1$ the intersection
\[
\textrm{supp} \, \psi \cap \textrm{supp} \, [S_{\lambda}\chi_0 - \chi_0]
\]
has compact closure and 
\[
[S_{\lambda}\chi_0 - \chi_0] \psi \in \mathcal{D}(\mathbb{R}^4) \subset \mathcal{S}(\mathbb{R}^4).
\]   
Therefore, using the causal support of $\kappa^{(-)}_q -(-1)^{f(q)}\kappa^{(+)}_q$,  for each fixed $\lambda >1$ and $\phi\in\mathscr{E}$ fulfilling 
(\ref{up-to-omegaorder-derivatives-of-phi=0}) we get
\begin{multline*}
\Big\langle (S_{\lambda^{m+n}}\chi_0) \big[\kappa^{(-)}_q -(-1)^{f(q)}\kappa^{(+)}_q\big], \phi \Big\rangle 
-
\Big\langle (S_{\lambda^m}\chi_0) \big[\kappa^{(-)}_q -(-1)^{f(q)}\kappa^{(+)}_q\big], \phi \Big\rangle 
\\
=
\Big\langle (S_{\lambda^{m+n}}\chi_0 -S_{\lambda^m}\chi_0) x^\alpha\big[\kappa^{(-)}_q -(-1)^{f(q)}\kappa^{(+)}_q\big], \phi_\alpha \Big\rangle 
=
\Big\langle (S_{\lambda^{m+n}}\chi_0 -S_{\lambda^m}\chi_0) x^\alpha\big[\kappa^{(-)}_q -(-1)^{f(q)}\kappa^{(+)}_q\big], \psi \phi_\alpha \Big\rangle 
\end{multline*}
\[
= \Big\langle 
\phi_\alpha x^\alpha \big[\kappa^{(-)}_q -(-1)^{f(q)}\kappa^{(+)}_q\big],
\, 
(S_{\lambda^{m+n}}\chi_0 -S_{\lambda^m}\chi_0)
\psi \Big\rangle 
= \Big\langle 
\phi_\alpha x^\alpha \big[\kappa^{(-)}_q -(-1)^{f(q)}\kappa^{(+)}_q\big],
\, 
S_{\lambda^m}(S_{\lambda^{n}}\chi_0 -\chi_0).\psi\big]
\Big\rangle 
\]
\begin{multline*}
=
\sum\limits_{j=0}^{n-1}
\Big\langle 
\phi_\alpha x^\alpha \big[\kappa^{(-)}_q -(-1)^{f(q)}\kappa^{(+)}_q\big],
\, 
S_{\lambda^j}S_{\lambda^m}\big[(S_{\lambda}\chi_0 -\chi_0)\big] \, \psi
\Big\rangle 
\\
=
\sum\limits_{j=0}^{n-1}
\Big\langle 
\phi_\alpha x^\alpha \big[\kappa^{(-)}_q -(-1)^{f(q)}\kappa^{(+)}_q\big],
\, 
S_{\lambda^j}S_{\lambda^m}\big[(S_{\lambda}\chi_0 -\chi_0)\big] \, S_{\lambda^j}S_{\lambda^m} \psi
\Big\rangle 
\end{multline*}
\begin{multline*}
=
\sum\limits_{j=0}^{n-1}
\Big\langle 
\phi_\alpha x^\alpha \big[\kappa^{(-)}_q -(-1)^{f(q)}\kappa^{(+)}_q\big],
\, 
S_{\lambda^j}S_{\lambda^m}\big[(S_{\lambda}\chi_0 -\chi_0)\psi\big]
\Big\rangle 
\\
=
\sum\limits_{j=0}^{n-1}
\big(\lambda^{j+m}\big)^{\omega-|\alpha|}
\big({\textstyle\frac{1}{\lambda^{j+m}}}\big)^{\omega-|\alpha|}
\Big\langle 
\phi_\alpha x^\alpha \big[\kappa^{(-)}_q -(-1)^{f(q)}\kappa^{(+)}_q\big],
\, 
S_{\lambda^j}S_{\lambda^m}\big[(S_{\lambda}\chi_0 -\chi_0)\psi\big]
\Big\rangle 
\end{multline*}
Now, because
\begin{multline*}
\big({\textstyle\frac{1}{\lambda^{j+m}}}\big)^{\omega-|\alpha|}
\Big\langle 
\phi_\alpha x^\alpha \big[\kappa^{(-)}_q -(-1)^{f(q)}\kappa^{(+)}_q\big],
\, 
S_{\lambda^j}S_{\lambda^m}\big[(S_{\lambda}\chi_0 -\chi_0)\psi\big]
\Big\rangle 
\\
\overset{m,j\rightarrow\infty}{\longrightarrow}
\Big\langle 
 \big[\kappa^{(-)}_q -(-1)^{f(q)}\kappa^{(+)}_q\big]_0,
\, 
x^\alpha (S_{\lambda}\chi_0 -\chi_0)\psi
\Big\rangle \,\, \phi_\alpha(0)
\end{multline*}
are convergent, then for each fixed $\lambda>1$, $\phi\in\mathscr{E}$, the set of the numbers
\[
\big({\textstyle\frac{1}{\lambda^{j+m}}}\big)^{\omega-|\alpha|}
\Big\langle 
\big[\phi_\alpha x^\alpha (\kappa^{(-)}_q -(-1)^{f(q)}\kappa^{(+)}_q)\big],
\, 
S_{\lambda^j}S_{\lambda^m}\big[(S_{\lambda}\chi_0 -\chi_0)\psi\big]
\Big\rangle, \,\,\, j,m \in \mathbb{N}
\]
is bounded. Let $B$ denote an upper bound of their absolute values. Thus, putting $\varepsilon = |\alpha|-\omega$,
\begin{multline*}
\Big|
\Big\langle (S_{\lambda^{m+n}}\chi_0) \big[\kappa^{(-)}_q -(-1)^{f(q)}\kappa^{(+)}_q\big], \phi \Big\rangle 
-
\Big\langle (S_{\lambda^m}\chi_0) \big[\kappa^{(-)}_q -(-1)^{f(q)}\kappa^{(+)}_q\big], \phi \Big\rangle 
\Big|
\\
\leq 
B
\sum\limits_{j=0}^{n-1}
\big(\lambda^{j+m}\big)^{\omega-|\alpha|}
= B
\sum\limits_{j=0}^{n-1}
\big(\lambda^{j+m}\big)^{-\varepsilon}
= B \lambda^{-\varepsilon m}
\,
{\textstyle\frac{1-\lambda^{-\varepsilon n}}{1-\lambda^{-\varepsilon}}} \,\,\,
\overset{m,n\rightarrow\infty}{\longrightarrow} 0,
\end{multline*}
and (\ref{Convergent[A',A''][A'', A'']Slambda^n(chi0)}) is a Cauchy sequence in $n$
for each $\phi\in\mathscr{E}$ fulfilling 
(\ref{up-to-omegaorder-derivatives-of-phi=0}). This also proves
convergence of (\ref{Convergent[A',A''][A'', A'']Slambda(chi0)})
if  $\lambda\rightarrow +\infty$ for each $\phi\in\mathscr{E}$ fulfilling 
(\ref{up-to-omegaorder-derivatives-of-phi=0}).

If the integral (\ref{Convergent[A',A''][A'', A'']theta}) is convergent, then
it has to be equal to the limit $\lambda\rightarrow +\infty$ of (\ref{Convergent[A',A''][A'', A'']Slambda(chi0)}),
on the subspace of test functions $\phi$ which respect (\ref{up-to-omegaorder-derivatives-of-phi=0}).
Indeed, by construction the limit has singularity degree $\omega$, the same as $\kappa_q-(-1)^q\check{\kappa_q}$
and the same as $\kappa_q$. It is easily seen that also the singularity degree
of (\ref{Convergent[A',A''][A'', A'']theta}), provided it is convergent, also must be equal to the singularity 
degree $\omega$ of $\kappa_q$. Because (\ref{Convergent[A',A''][A'', A'']theta}) and the 
 limit $\lambda\rightarrow +\infty$ of (\ref{Convergent[A',A''][A'', A'']Slambda(chi0)}) are both retarded parts
of the same causal distribution $\kappa^{(-)}_q -(-1)^{f(q)}\kappa^{(+)}_q$, they differ at most by a distribution supported at zero:
\[
\sum\limits_{\alpha} C_\alpha \delta^{(\alpha)}
\]
and because, moreover, they both have the same singularity degree $\omega$, this sum cannot
involve terms with $|\alpha|>\omega$, thus their difference can at most be equal 
\[
\sum\limits_{|\alpha|=0}^{\omega} C_\alpha \delta^{(\alpha)}.
\] 
But the last distribution is zero on the subspace of test functions $\phi$ respecting (\ref{up-to-omegaorder-derivatives-of-phi=0}).

Convergence of the limit $\lambda\rightarrow +\infty$ of (\ref{Convergent[A',A''][A'', A'']Slambda(chi0)}) only shows existence
of the retarded part of $\kappa^{(-)}_q -(-1)^{f(q)}\kappa^{(+)}_q$ on the subspace of $\phi$ which respect (\ref{up-to-omegaorder-derivatives-of-phi=0}).
For the practical computations this result is not very much interesting yet, as we need a practical method for computation
of the retarded part, or its Fourier transform, explicitly. For this task the formula (\ref{Convergent[A',A''][A'', A'']theta})
is useful, because it can be converted into a practical formula.

Therefore we concentrate now on the proof of convergence of
(\ref{Convergent[A',A''][A'', A'']theta}) or, equivalently, (\ref{thetaASkappa2(chi)}), for
\[
\chi(x,y) = \phi(x-y)\varphi(y), \,\,\,\,\,\,\,\, \phi, \varphi \in \mathscr{E},
\]
with $\phi$ which respects (\ref{up-to-omegaorder-derivatives-of-phi=0}), with $\omega$ in (\ref{up-to-omegaorder-derivatives-of-phi=0})
equal at least to the singularity order of the distribution
\[
\kappa^{(-)}_q -(-1)^{f(q)}\kappa^{(+)}_q.
\]

Presented proof of convergence of the limit $\lambda\rightarrow +\infty$ of (\ref{Convergent[A',A''][A'', A'']Slambda(chi0)}), 
based on the singularity degree, is not very much interesting also because this proof strongly uses causality of 
$\kappa^{(-)}_q -(-1)^{f(q)}\kappa^{(+)}_q$, property which no longer holds for each
term $\kappa_q$ or $\kappa^{(+)}_q$, taken separately. Practical computations, which we need to perform, give
more effective construction of the retarded part, and which is inspired by the integral formula, and allows to prove
that the following integral is convergent in distributional sense and represents well-defined functional of $\chi$ 
\begin{equation}\label{thetakappaq(chi)}
\Big[
\theta_y\big(\kappa'_{0,1} \dot{\otimes} \kappa''_{0,1} \dot{\otimes} \dots \dot{\otimes} \kappa^{(q)}_{0,1}\big)
 \otimes_q \big(\kappa'_{1,0} \dot{\otimes} \kappa''_{1,0} \dot{\otimes} \ldots 
\dot{\otimes} \kappa^{(q)}_{1,0}\big)
\Big] (\chi)
\end{equation}
equal
\[
\int
\kappa_{q}(x-y)
\theta(x-y) \, \phi(x-y)\varphi(y) \, \ud^4 x \, \ud^4 y
=
\widetilde{\varphi}(0) \,\, \big\langle \kappa_q, \phi \big\rangle 
= \widetilde{\varphi}(0) \,\, \big\langle\kappa_q, \theta \phi \big\rangle
\]
\begin{multline}\label{ConvergentProdPairings[A',A''][A'', A'']theta}
=
\widetilde{\varphi}(0) \,\,
\int \big\{
u'(\boldsymbol{\p}') v'(\boldsymbol{\p}') u''(\boldsymbol{\p}'') v''(\boldsymbol{\p}'') \ldots 
\,
\widetilde{\theta\phi}(-\boldsymbol{p}' - \boldsymbol{p}'' - \ldots, \,-|\boldsymbol{p}'| -|\boldsymbol{p}''|-\ldots) 
\\
\big\}
\, \ud^3 \boldsymbol{p}' \, \ud^3 \boldsymbol{p}'' \ldots  \ud^3 \boldsymbol{p}^{(q)},
\end{multline}
for 
\[
\chi(x,y) = \phi(x-y)\varphi(y), \,\,\,\, \phi,\varphi \in \mathscr{E} = \mathcal{S}(\mathbb{R}^4),
\]
and with 
\[
\phi \in \mathscr{E}= \mathcal{S}(\mathbb{R}^4) 
\]
which fulfil (\ref{up-to-omegaorder-derivatives-of-phi=0}). After \cite{Scharf} we briefly report a practical method for calculation which, 
as far as we know, goes back to \cite{Scharf}. Namely, the Fourier transform of the scalar $\otimes_q$-contraction
$\kappa_q$ can be computed explicitly quite easily, on using the completeness relations of the plane wave kernels. 
Below in this Subsection, as an example, we give Fourier transforms $\widetilde{\kappa_q}$ of all basic scalar $\kappa_q$, $q=1,2,3$, $\otimes_q$-contractions 
for QED. It turns out that for $q>1$ (the most interesting case)  $\widetilde{\kappa_q}$ are regular, \emph{i.e.} function-like,
distributions, analytic in $p$ everywhere, except the the finite set of characteristic submanifolds of the type $p\cdot p = const.$, or $p_0=0$,
where they have $\theta(p\cdot p - const.)$, $\theta(p_0)$-type finite jump. Next  we observe
that each $\phi \in \mathscr{E}$, which fulfils (\ref{up-to-omegaorder-derivatives-of-phi=0}), can be written
as an element of the image of a continuous idempotent operator $\Omega'$ acting on $\mathscr{E} = \mathcal{S}(\mathbb{R}^4)$,
and  (\ref{thetakappaq(chi)}) can be understood as defined on $\chi(x,y)=(\Omega'\phi)(x-y)\varphi(y)$,
now for $\phi,\varphi$ ranging over general elements in $\mathscr{E}$. 
Indeed, let, for each multi-index $\alpha$, such that $0 \leq |\alpha| \leq \omega$,
$\omega_{{}_{o \,\, \alpha}} \in \mathscr{E}$ on $\mathbb{R}^{4}$ be such
functions that\footnote{As we have already mentioned such functions $\omega_{{}_{o \,\, \alpha}} \in \mathscr{E}$, $0 \leq |\alpha| \leq \omega$ do exist.
Indeed, let
\[
f_{{}_{\alpha}}(x) =
{\textstyle\frac{x^\alpha}{\alpha!}}, \,\,\, x \in \mathbb{R}^{4}, 0 \leq |\alpha| \leq \omega,
\]
\[
\alpha ! = \prod \limits_{\mu=0}^{3} \alpha_{\mu} !,
\,\,\, x^\alpha = \prod \limits_{\mu=0}^{3} (x_{\mu})^{\alpha_{\mu}}, \,\,\, \mu = 0,1,2,3.
\]
It is well-known fact that there exists $w \in \mathscr{E} = \mathcal{S}(\mathbb{R}^4)$,
which is equal $1$ on some neighborhood of zero.
Then we can put
\[
\omega_{{}_{o \,\, \alpha}} \overset{\textrm{df}}{=} f_{{}_{\alpha}}.w.
\]}
\[
D^\beta \omega_{{}_{o \,\, \alpha}} (0) = \delta^{\beta}_{\alpha}, \,\,\,\, 0 \leq |\alpha|, |\beta| \leq \omega.
\]
Then, for any $\phi \in \mathscr{E}$ we put
\[
\Omega' \phi = \phi - \sum \limits_{0 \leq |\alpha| \leq \omega} D^\alpha \phi(0) \,\, \omega_{{}_{o \,\, \alpha}}
\,\,\,
=
 \phi - \sum \limits_{0 \leq |\alpha| \leq \omega} D^\alpha \phi(0) \,\, {\textstyle\frac{x^\alpha}{\alpha!}}.w
\]
Now, the integral (\ref{thetakappaq(chi)}), with $\chi(x,y)=(\Omega'\phi)(x-y)\varphi(y)$ and $\phi, \varphi \in \mathscr{E}$, becomes equal 
\[
\widetilde{\varphi}(0) \,\, \big\langle\kappa_q, \theta \Omega'\phi \big\rangle = 
\big\langle \widetilde{\kappa_q}, \widetilde{\theta} \ast \widetilde{\Omega'\phi} \big\rangle, 
\] 
and is expected to be a well-defined continuous functional of $\widetilde{\phi},\widetilde{\varphi} \in \mathcal{S}(\mathbb{R}^4)$,
or equivalently
\[
\big\langle \widetilde{\kappa_q}, \widetilde{\theta} \ast \widetilde{\Omega'\phi} \big\rangle, 
\]
should give a well-defined functional of $\phi \in \mathcal{S}(\mathbb{R}^4)$. It defines
(\ref{thetakappaq(chi)}) for all $\chi(x,y)=\phi(x-y)\varphi(y)$ with
$\phi \in \textrm{Im} \, \Omega'$, and because $\textrm{Ker} \, \Omega' \neq\{0\}$ for $q>1$, its definition
is not unique and can be extended by addition of any functional which is zero 
on $\chi(x-y)\varphi(y) = \phi(x-y)\varphi(y)$ with
$\phi \in \textrm{Im} \, \Omega'$. Equivalently the retarded part
\begin{gather}
\textrm{ret} \, \kappa_q \overset{\textrm{df}}{=} \kappa_q \circ (\theta \Omega'),
\label{Def(retkappaq)}
\\
\big\langle \textrm{ret} \, \kappa_q, \phi \big\rangle
= \big\langle \widetilde{\textrm{ret} \, \kappa_q}, \widetilde{\phi} \big\rangle
=\big\langle\kappa_q, \theta \Omega'\phi \big\rangle  
=\big\langle \widetilde{\kappa_q}, \widetilde{\theta} \ast \widetilde{\Omega'\phi} \big\rangle
\label{FT(retkappaq)}
\end{gather}
of $\kappa_q$ is not unique, and we can add to it
\begin{equation}\label{freedom}
\sum\limits_{|\alpha|=0}^{\omega} C_\alpha \delta^{\alpha},
\end{equation} 
which is most general on $\textrm{Ker} \, \Omega'$ and which is zero on $\textrm{Im} \, \Omega'$. 
In order to show (\ref{Def(retkappaq)}) is a well-defined tempered distribution, we proceed after \cite{Scharf},
using the explicit formula for the function $\widetilde{\kappa_q}$ in (\ref{FT(retkappaq)}). 

Since
\[
\widetilde{x^\alpha w}(p) = (iD_{p})^\alpha \widetilde{w}(p),
\]
and
\[
D^\alpha\phi(0) = (-1)^\alpha \big\langle D^\alpha \delta, \phi \big\rangle 
= (-1)^\alpha \big\langle \widetilde{D^\alpha\delta}, \widetilde{\phi} \big\rangle
=
 (2\pi)^{-4/2} \big\langle (ip)^\alpha, \widetilde{\phi}  \big\rangle,
\]
immediately from (\ref{FT(retkappaq)}) we obtain
\begin{multline*}
\big\langle \widetilde{\textrm{ret} \, \kappa_q}, \widetilde{\phi} \big\rangle
= \big\langle \widetilde{\kappa_q}, \widetilde{\theta \Omega'\phi}\big\rangle
= 
(2\pi)^{-4/2} \Bigg\langle \widetilde{\kappa_q},  \,\,\,\, \widetilde{\theta}
\,\,
\ast \,\, \Big[ \widetilde{\phi} - \sum\limits_{|\alpha|=0}^{\omega} {\textstyle\frac{1}{\alpha!}} (iD_{p})^\alpha \widetilde{w}
(2\pi)^{-4/2} \big\langle (ip')^\alpha, \widetilde{\phi} \big\rangle \Big] \Bigg\rangle
\\
=
(2\pi)^{-4/2} \Big\langle \widetilde{\theta} \ast \widetilde{\kappa_q}, \,\,\,
\widetilde{\phi} - \sum\limits_{|\alpha|=0}^{\omega} {\textstyle\frac{1}{\alpha!}} (iD_{p})^\alpha \widetilde{w}
(2\pi)^{-4/2} \big\langle (ip')^\alpha, \widetilde{\phi} \,\, \Big\rangle
\end{multline*}
were the convolution $\widetilde{\theta} \ast \widetilde{\kappa_q}$ is defined only on the subtracted $\widetilde{\phi}$,
\emph{i.e.} on Fourier transforms of test functions with all derivatives vanishing  at zero up to order $\omega$. Interchanging the integration
variables $p'$ and $p$ in the subtracted terms we get
\[
\big\langle \widetilde{\textrm{ret} \, \kappa_q}, \widetilde{\phi} \big\rangle 
= (2\pi)^{-4/2} \int d^4 k \, \widetilde{\theta}(k) \,\, \Big\langle \widetilde{\kappa_q}(p-k)
- (2\pi)^{-4/2} \sum\limits_{|\alpha|=0}^{\omega} {\textstyle\frac{(-1)^\alpha}{\alpha!}}p^\alpha 
\int d^4 p' \,  \widetilde{\kappa_q}(p'-k) D_{p'}^{\alpha}\widetilde{w}, \, \widetilde{\phi} \Big\rangle.
\]
By partial integration in the $p'$ variable we arrive at the formula
\begin{equation}\label{FT(retkappaq)1}
\widetilde{\textrm{ret} \, \kappa_q}(p) = {\textstyle\frac{1}{(2\pi)^2}} \int d^4k \, \widetilde{\theta}(k)
\Bigg[
\widetilde{\kappa_q}(p-k) - {\textstyle\frac{1}{(2\pi)^2}} \sum\limits_{|\alpha|=0}^{\omega}
{\textstyle\frac{p^\alpha}{\alpha!}}\int d^4 p' \, D^\alpha \widetilde{\kappa_q}(p'-k)\widetilde{w}(p')  
\Bigg],
\end{equation}
which, in general should be understood in the distributional sense (converging when
integrated in $p$ variable with a test function of $p$). But in practical computations we are interesting
in the regularity domain, outside the characteristic submanifold $p\cdot p = const.$, where $\widetilde{\textrm{ret} \, \kappa_q}$
is represented by ordinary function of $p$ and where 
this integral should be convergent in the ordinary sense. 
Then we assume that there exists a point $p''$ around which (\ref{thetakappaq(chi)}) is regular, and moreover
possess all derivatives in the usual sense (should not be mixed with distributional) at $p''$, 
up to order $\omega$ equal to the singularity degree. This is not entirely arbitrary assumption, because
if the formula (\ref{thetakappaq(chi)}) makes any sense at all as a functional of $\phi$, it must have support
in the half space (or forward cone in case we replace $\kappa_q$ with the causal $\kappa_q - (-1)^q\check{\kappa_q}$),
so its Fourier transform should be a boundary value of an analytic function, regular all over $\mathbb{R}^4$,
but compare the comment we give below on this topic.
Next, we subtract  the first 
Taylor terms of $\widetilde{\textrm{ret} \, \kappa_q}$ up to order $\omega$ around the regularity point $p''$ 
\begin{equation}\label{FT(retkappa)p''}
\big[\widetilde{\textrm{ret} \, \kappa_q}\big]_{{}_{{}_{p''}}}(p) = \widetilde{\textrm{ret} \, \kappa_q}(p)  
- \sum\limits_{|\beta|=0}^{\omega}
{\textstyle\frac{(p-p'')^\beta}{\beta!}} D^\beta \widetilde{\textrm{ret} \, \kappa_q}(p'') 
\end{equation}
and get another possible retarded part $\big[\widetilde{\textrm{ret} \, \kappa_q}\big]_{{}_{{}_{p''}}}$, 
Fourier transformed, of $\kappa_q$, as the added terms have the form (\ref{freedom}). 
By construction (\ref{FT(retkappa)p''}) is so normalized, that all its derivatives vanish up to order $\omega$
at $p''$. This point is called \emph{normalization point}.
Next we compute $D^\beta \widetilde{\textrm{ret} \, \kappa_q}$ from the formula (\ref{FT(retkappaq)1}), and substitute
into the formula (\ref{FT(retkappa)p''}). Using
\[
\sum\limits_{\beta\leq \alpha} {\textstyle\frac{(p-p'')^\beta}{\beta!}}
 {\textstyle\frac{{p''}^{\alpha-\beta}}{(\alpha-\beta)!}} 
= 
{\textstyle\frac{1}{\alpha!}}
\sum\limits_{\beta\leq \alpha} {\alpha \choose \beta} (p-p'')^\beta {p''}^{\alpha-\beta}
= {\textstyle\frac{p^\alpha}{\alpha!}},
\]
we see that all terms with the auxiliary function $\widetilde{w}$
drop out and we obtain for $\big[\widetilde{\textrm{ret} \, \kappa_q}\big]_{{}_{{}_{p''}}}$ the formula
\begin{equation}\label{FT(retkappaq)2}
\big[\widetilde{\textrm{ret} \, \kappa_q}\big]_{{}_{{}_{p''}}}(p) = {\textstyle\frac{1}{(2\pi)^2}} \int d^4k \widetilde{\theta}(k)
\Bigg[
\widetilde{\kappa_q}(p-k) - \sum\limits_{|\beta|=0}^{\omega}
{\textstyle\frac{(p-p'')^\beta}{\beta!}} D^\beta \widetilde{\kappa_q}(p''-k) 
\Bigg],
\end{equation}
by construction so normalized that all derivatives of $\big[\widetilde{\textrm{ret} \, \kappa_q}\big]_{{}_{{}_{p''}}}$ 
vanish at $p''$ up to order $\omega$, compare \cite{Scharf}, p. 179. 
We need to rewrite the last integral (\ref{FT(retkappaq)2}) with $\widetilde{\kappa_q}$ taken at the same point in the two terms
in (\ref{FT(retkappaq)2}), in order to combine the two terms into a single term, as these terms taken separately
are divergent. At this point the case in which we have causal $\kappa_q-(-1)^q\check{\kappa_q}$ and Lorentz convariant, instead of
$\kappa_q$, is simpler, because $\kappa_q-(-1)^q\check{\kappa_q}$ has causal support and is Lorentz convariant. 
In this causal case we can use the special $\theta(x) = \theta(x_0) = \theta(v\cdot x)$ with $v=(1,0,0,0)$, and insert
\begin{equation}\label{FTtheta}
\widetilde{\theta}(k_0,\boldsymbol{k}) = 2\pi {\textstyle\frac{i}{k_0+i\epsilon}} \,
\delta(\boldsymbol{k}),
\end{equation}
and use Lorentz frame in which $p=(p_0,0,0,0)$ and then using integration by parts we remove the differentiainion
operation in momentum variables  off $\widetilde{\kappa_q}$ in the second term in (\ref{FT(retkappaq)2}).  
Next we use invariance and analytic continuation to extend 
$\widetilde{\textrm{ret} \, \kappa_q}$ all over $p$.
Because  $\kappa_q$ is not causal, nor Lorentz invariant, we need to consider separately the case
$p=(p_0,0,0,0)$ inside the positive and negative cone and $p=(0,0,0,p_3)$ outside the cone in the momentum space
as well as $\theta(x) = \theta(v\cdot x)$ with a more general timelike unit $v$
in
\[
\widetilde{\theta}(k_0,\boldsymbol{k}) = 2\pi {\textstyle\frac{i}{k_0 +i\epsilon v_0}} \,
\delta(\boldsymbol{k}- k_0{\textstyle\frac{\boldsymbol{v}}{v_0}} ),
\]
in order to reconstruct $\widetilde{\textrm{ret} \, \kappa_q}$, now depending on $v$, as $\widetilde{\kappa_q}$
is not causal. 

Still proceeding after \cite{Scharf}, let us concentrate now on the computation of the retarded part the causal $d= \kappa_q-(-1)^q\check{\kappa_q}$,which
is simpler and moreover, on such QFT in which the normalization point $p''$ can be put equal zero. 
This is e.g. the case for QED's with massive charged fields. 
We reconstruct $\widetilde{\textrm{ret} \, d}(p)$ first for $p$ in the cone $V^+$.  
In this case computations simplify, as we can put $p''=0$, and moreover
we use Lorentz covariance of the causal $d$, and choose a Lorentz frame in which $p=(p_0,0,0,0)$.
Moreover, because $d$ is causal its retarded part is independent of the choice of the time like
unit versor $v$ in the theta function $\theta(v\cdot x)$, so that we can choose $v=(1,0,0,0)$, thus assuming that it is
always parallel  to $p\in V^+$, and thus varies with $p$. After these choices, the last formula
(\ref{FT(retkappaq)2}) for the retarded part, applied to $d$ instead of $\kappa_q$, and with substituted
Fourier transform (\ref{FTtheta}), reads
\[
\widetilde{\textrm{ret} \, d}(p_0,0,0,0) 
= {\textstyle\frac{i}{2\pi}} \int dk_0 
{\textstyle\frac{1}{k_0+i\epsilon}}
\Bigg[
\widetilde{d}(p_0-k_0,0,0,0) - \sum\limits_{a=0}^{\omega}
{\textstyle\frac{(p_0)^a}{a!}}(-1)^a D_{k_0}^{a} \widetilde{d}(q_0-k_0,0,0,0)\big|_{{}_{q_0=0}} 
\Bigg].
\]
Integrating the subtracted terms by parts and then using the intergration variable $k'_{0}= k_0-p_0$ in the first term, we get
\[
\widetilde{\textrm{ret} \, d}(p_0) = {\textstyle\frac{i}{2\pi}} \int dk'_{0} 
\,
\Bigg[
{\textstyle\frac{1}{p_0 + k'_{0}+i\epsilon}}
 - \sum\limits_{a=0}^{\omega}
{\textstyle\frac{(p_0)^a}{a!}} 
{\textstyle\frac{\partial^a}{\partial {k'}_{0}^{a}}}
{\textstyle\frac{1}{k'_{0}+i\epsilon}}
\Bigg] \,
\widetilde{d}(-k'_{0}),
\]
with
\begin{equation}
{\textstyle\frac{1}{p_0 + k'_{0}+i\epsilon}}
 - \sum\limits_{a=0}^{\omega}
{\textstyle\frac{(p_0)^a}{a!}} 
{\textstyle\frac{\partial^a}{\partial {k'}_{0}^{a}}}
{\textstyle\frac{1}{k'_{0}+i\epsilon}} 
=
\Big({\textstyle\frac{-p_0}{k'_{0}+i\epsilon}}\Big)^{\omega+1}
{\textstyle\frac{1}{p_0 +k'_{0}+i\epsilon}}.
\end{equation}
To recover the formula for arbitrary $p\in V^+$ we introduce the variable $t=k_0/p_0$ and obtain
\[
\widetilde{\textrm{ret} \, d}(p) = {\textstyle\frac{i}{2\pi}} \int\limits_{-\infty}^{+\infty} d t
\,
{\textstyle\frac{\widetilde{d}(tp),}{(t-i\epsilon)^{\omega+1}(1-t+i\epsilon)}},
\] 
or
\begin{equation}\label{FT(retCausalkappaq)}
\mathscr{F} \Big(\textrm{ret} \, \big[\kappa^{(-)}_q -(-1)^{f(q)}\kappa^{(+)}_q\big]\Big)(p) = {\textstyle\frac{i}{2\pi}} \int\limits_{-\infty}^{+\infty} d t
\,
{\textstyle\frac{\mathscr{F}\big(\kappa^{(-)}_q -(-1)^{f(q)}\kappa^{(+)}_q\big)(tp),}{(t-i\epsilon)^{\omega+1}(1-t+i\epsilon)}},
\,\,\,\, p \in V^+,
\end{equation}
and with  normalization point $p''=0$.
To compute the retarded part outside the cone $V^+$ we have several possibilities, e.g. we can use the analytic 
continuation method, or choose the normalization point $p''$ outside the cone and repeat the computation.

Below in the next Subsection we give explicit formulas for $\widetilde{\textrm{ret} \, \kappa_q}$ and
\[
\widetilde{\textrm{ret} \, d} = \mathscr{F}\big\{\textrm{ret} \, [\kappa^{(-)}_q -(-1)^{f(q)}\kappa^{(+)}_q]\big\},
\]
for all basic $\kappa_q$, $q=1,2,3$ in case of spinor QED, as an example.

Anyway, explicit formula, obtained by the method outlined above,
shows that $\big[\widetilde{\textrm{ret} \, \kappa_q}\big]_{{}_{{}_{p''}}}$ 
as well as that
\[
\widetilde{\textrm{ret} \, \kappa_q}  = \big[\widetilde{\textrm{ret} \, \kappa_q}\big]_{{}_{{}_{p''}}} - \sum\limits_{|\alpha|=0}^{\omega}
C_\alpha p^\alpha,
\]
for some constants $C_\alpha$, is a well-defined tempered distribution. 
Analogously
\[
\mathscr{F}\big(\textrm{ret} \, [\kappa^{(-)}_q -(-1)^{f(q)}\kappa^{(+)}_q]\big) 
= \Big[\mathscr{F}\big(\textrm{ret} \, [\kappa^{(-)}_q -(-1)^{f(q)}\kappa^{(+)}_q]\big)\Big]_{{}_{{}_{p''}}} - \sum\limits_{|\alpha|=0}^{\omega}
C_\alpha p^\alpha
\]
is a well-defined tempered distribution. Moreover, replacing $\theta(x)$ with $-\theta(x)$ we compute
in the same manner the Fourier transformed advanced part $\widetilde{\textrm{av} \, \kappa_q}$, which ideed
composes together with $\widetilde{\textrm{av} \, \kappa_q}$ the splitting
\[
\widetilde{\kappa_q}= \widetilde{\textrm{ret} \, \kappa_q} - \widetilde{\textrm{av} \, \kappa_q}
\]
of $\kappa_q$ into a retadred and advanced part. 
 
If there exists a regularity point $p''$ for (\ref{FT(retkappaq)1}), which we have used above as an assumption, 
then this would give a proof that  (\ref{thetakappaq(chi)}) is a well-defined
distribution, because its Fourier transform would be equal to a well-defined distribution $\widetilde{\textrm{ret} \, \kappa_q}$. 
Similarly, for the causal combination $\kappa^{(-)}_q -(-1)^{f(q)}\kappa^{(+)}_q$, if there exist a regularity point
$p''$ for (\ref{FT(retkappaq)1}) with $\kappa_q$ replaced with $\kappa^{(-)}_q -(-1)^{f(q)}\kappa^{(+)}_q$, then
its Fourier transform, being given by a formula 
\[
\mathscr{F}\big(\textrm{ret} \, [\kappa^{(-)}_q -(-1)^{f(q)}\kappa^{(+)}_q]\big) 
\]
obtained by the method outlined above and defining a function-like distribution, would be a well-defined distribution.

We know that this is not entirely arbitrary assumption, as for any distribution with cone-shaped support, as e.g. for the retarded
parts, their Fourier transform has to be a boundary value of an analytic function, regular in the tube $\mathbb{R}^4+iV$,
with $V$ being equal to the corresponding support of  $\textrm{ret} \, \kappa_q$, or, $\textrm{ret} \, [\kappa^{(-)}_q -(-1)^{f(q)}\kappa^{(+)}_q]$,
by a well-known theorem on distributions with cone-shaped supports, compare \cite{Reed_SimonII}, Theorem IX.16.
The expression, given by the formula (\ref{thetaASkappa2(chi)}) or (\ref{thetakappaq(chi)}), 
if it makes any sense at all, evidently must be
supported within the future light cone or half-space, respectively, in case of $\kappa^{(-)}_q -(-1)^{f(q)}\kappa^{(+)}_q$
or $\kappa_q$. Therefore, if  (\ref{thetakappaq(chi)}) or (\ref{thetaASkappa2(chi)}) make any sense at all,
their Fourier transform must be equal  (\ref{FT(retkappaq)2}) or (\ref{FT(retkappaq)2}) with $\kappa_q$
replaced by $\kappa^{(-)}_q -(-1)^{f(q)}\kappa^{(+)}_q$, up to a distribution (\ref{freedom}).
Below we give explicit expressions for the Fourier transform  $\widetilde{\kappa_q}$ for particular
examples of products of pairings. Becuse $\widetilde{\kappa_q}$ is analytic except the characteristic submanifold, where it has finite jump,
so in general it is expected that (\ref{FT(retkappaq)1}) indeed possess regularity points $p''$ whenever the integral  (\ref{FT(retkappaq)1})
is convergent. This is indeed the case for the final formula $\textrm{ret} \, \kappa_q$ obtained by the outlined method.  
  
Finally, we replace $\theta(x\cdot v)$, in the above computation of $\widetilde{\textrm{ret} \, \kappa_q}$, with
\begin{gather*}
\theta_{{}_{\lambda}}(x \cdot v) = {\textstyle\frac{1}{2}} +{\textstyle\frac{1}{\pi}} \textrm{arctan} \, (\lambda(x \cdot v) ) 
\\
\widetilde{\theta_{{}_{\lambda}}}(k) = 
2\pi \big[\sqrt{{\textstyle\frac{\pi}{2}}} \delta(k_0) + {\textstyle\frac{1}{\sqrt{2\pi}}}
{\textstyle\frac{ik_0}{k_{0}^{2} +v_{0}^2}} e^{-|k_0|/(\lambda v_0)} \big] \, \delta\big(\boldsymbol{k} - k_0 {\textstyle\frac{\boldsymbol{v}}{v_0}}\big)
\\
=
2\pi \Big[ {\textstyle\frac{i}{k_0+i\epsilon v_0}} - {\textstyle\frac{ik_0}{k_{0}^{2}+\epsilon^2 v_{0}^{2}}}
\big(e^{-|k_0|/(\lambda v_0)}\big) \Big] \, \delta\big(\boldsymbol{k} - k_0 {\textstyle\frac{\boldsymbol{v}}{v_0}}\big),
\end{gather*}
obtaining $\widetilde{\kappa_q}_{{}_{\lambda}}$ and show convergence 
\[
\big\langle \widetilde{\kappa_q}_{{}_{\lambda}} ,\widetilde{\phi} \big\rangle \overset{\lambda \rightarrow \infty}{\longrightarrow}
\big\langle \widetilde{\kappa_q} ,\widetilde{\phi} \big\rangle,
\]
for each test function $\widetilde{\phi}$. The integral (\ref{thetakappaq(chi)}) with 
 $\theta(x\cdot v)$ replaced by $\theta_{{}_{\lambda}}(x \cdot v)$, coincides with
$\widetilde{\varphi}(0) \, \big\langle \widetilde{\kappa_q}_{{}_{\lambda}} ,\widetilde{\phi}\big\rangle$. In this distributional
sense the integral (\ref{thetakappaq(chi)}) is convergent.

In particular, using the continuity (\ref{productkernelsmultipliers'}), we arrive at the conclusion that there exist natural $k$ and finite $c$, such that 
\begin{equation}\label{existencethetakappaq}
\Bigg|
\Big[
\theta_y\big(\kappa'_{0,1} \dot{\otimes} \kappa''_{0,1} \dot{\otimes} \kappa'''_{0,1} \dot{\otimes} \ldots \dot{\otimes} \kappa^{(q)}_{0,1} \big)
\, \otimes_{q} \,\, \big(\kappa'_{1,0} \dot{\otimes} \kappa''_{1,0}  \dot{\otimes} \kappa'''_{1,0} \dot{\otimes} 
\ldots \dot{\otimes} \kappa^{(q)}_{1,0} \big)
\Big] (\chi)
\Bigg|
\leq c \big|\phi \big|_{{}_{k}}\big|\varphi \big|_{{}_{k}}
\end{equation}
for all test functions $\chi(x,y) = \phi(x-y)\varphi(y)$, $\phi,\varphi\in \mathscr{E}$, 
with $\phi$ which respects (\ref{up-to-omegaorder-derivatives-of-phi=0}) 
and with $\omega$ in (\ref{up-to-omegaorder-derivatives-of-phi=0})
equal at least to the singularity order of $\kappa_q$, in fact equal to the singularity order of
\[
\kappa_q(x)-(-1)^q\kappa_q(-x).
\]

Therefore, the valuation integral
\[
\Big( \theta_y \big(\kappa'_{0,1} \dot{\otimes} \kappa''_{0,1} \dot{\otimes} \kappa'''_{0,1} \dot{\otimes} \ldots \dot{\otimes} \kappa^{(q)}_{0,1} \big)
\, \otimes_q \,\, \big(\kappa'_{1,0} \dot{\otimes} \kappa''_{1,0}  \dot{\otimes} \kappa'''_{1,0} \dot{\otimes} 
\ldots \dot{\otimes} \kappa^{(q)}_{1,0} \big) \big)(\chi)
\]
is convergent for
\[
\chi(x,y) = \phi(x-y)\varphi(y), \,\,\,\,\,\,\,\, \phi, \varphi \in \mathscr{E},
\]
and with $\phi$ which respect  (\ref{up-to-omegaorder-derivatives-of-phi=0})
with $\omega$ in (\ref{up-to-omegaorder-derivatives-of-phi=0})
equal at least to the singularity order of $\kappa_q$.

Still more generally, with $q'\leq q$, 
using the continuity (\ref{productkernelsmultipliers'}), 
and the fact that the retarded part of the scalar contraction (with $q'=q$)
is a well-defined distribution
we arrive at the inequality
\[
\Big|\Big\langle \Big( \theta_y \big(\kappa'_{0,1} \dot{\otimes} \kappa''_{0,1} \dot{\otimes} \kappa'''_{0,1} \dot{\otimes} \ldots \dot{\otimes} \kappa^{(q)}_{0,1} \big)
\, \otimes_{q'} \,\, \big(\kappa'_{1,0} \dot{\otimes} \kappa''_{1,0}  \dot{\otimes} \kappa'''_{1,0} \dot{\otimes} 
\ldots \dot{\otimes} \kappa^{(q)}_{1,0} \big) \Big)(\chi), \,\, \widehat{\xi} \, \Big\rangle\Big|
\]
\begin{equation}\label{InequalityForq'<qContraction}
\leq c \big|\phi \big|_{{}_{k}}\big|\varphi \big|_{{}_{k}} |\widehat{\xi}|_m, 
\end{equation}
for
\[
\chi(x,y) = \phi(x-y)\varphi(y), \,\,\, \phi,\varphi \in \mathscr{E}
\]
with $\phi$ which respects
 (\ref{up-to-omegaorder-derivatives-of-phi=0}), with $\omega$ equal at least to the singularity order
of $\kappa_{q'}$.
Here $|\cdot|_{m}$ being one of the norms defining the nuclear topology of 
\[
E_{{}_{'}} \otimes E_{{}_{''}}  \otimes  \widehat{\ldots}   \otimes  E_{{}_{(q)}}  
E_{{}_{'}} \otimes E_{{}_{''}}\otimes  \widehat{\ldots}    \otimes  E_{{}_{(q)}}. 
\]
Hat-character $\widehat{\ldots}$ means that $q'$ contracted pairs of nuclear spaces 
$E_{{}_{'}}^{*}, E_{{}_{''}}^{*}, \ldots$ are removed from the tensor product.

Recall that here we have considered the contraction distributions 
with the massless
plane wave distribution kernels $\kappa'_{0,1}, \kappa'_{1,0}, \kappa''_{0,1}, \kappa''_{1,0}, \ldots$
defining massless free fields $\mathbb{A}', \mathbb{A}'', \ldots$. Of course exactly the same analysis is valid
if some free fields, $\mathbb{A}'$, or all, are replaced with massive fields $\mathbb{A}'$. In this case the functions
$p_0(\boldsymbol{\p}') = |\boldsymbol{\p}'|$ in the exponents of the contraction integrals are of course replaced
by the functions $p_0(\boldsymbol{\p}')$ such that $(p_0(\boldsymbol{\p}'), \boldsymbol{\p}') \in \mathscr{O}' = \mathscr{O}_{m',0,0,0}$
for the orbit $\mathscr{O}' = \mathscr{O}_{m',0,0,0}$ corresponding to the massive field $\mathbb{A}'$. Similarly,
the multipliers $u'(\boldsymbol{\p}'),v'(\boldsymbol{\p}')$ are replaced with the multipliers which are present in the massive
kernels
\[
\kappa'_{0,1}(\boldsymbol{\p}';x)= u'(\boldsymbol{\p}') e^{-ip'x}, \,\,\,\,
\kappa'_{1,0}(\boldsymbol{\p}'; x) = v'(\boldsymbol{\p}') e^{ip'x},
\,\,\,\,\, p'= (p_0(\boldsymbol{\p}'), \boldsymbol{\p}') \in \mathscr{O}'.
\]

Inspired by \cite{Epstein-Glaser} we can extend the contraction kernels
\[
\theta_{y} \big(\kappa'_{0,1} \dot{\otimes} \kappa''_{0,1} \dot{\otimes} \kappa'''_{0,1} \dot{\otimes} \ldots \dot{\otimes} \kappa^{(q')}_{0,1} \big)
\, \otimes_{q'} \,\, \big(\kappa'_{1,0} \dot{\otimes} \kappa''_{1,0} \dot{\otimes} \kappa'''_{1,0} \dot{\otimes} \ldots \dot{\otimes} \kappa^{(q)}_{1,0} \big)
\]
with $q'\leq q$, over all elements $\chi$, which respect (\ref{xiInS00timesS00}). In order to do it, we use the previously defined linear continuous
operator on $\mathscr{E}$
\[
\Omega' \phi = \phi - \sum \limits_{0 \leq |\alpha| \leq \omega} D^\alpha \phi(0) \,\, \omega_{{}_{o \,\, \alpha}}.
\]
We extend this operator on functions in $\mathscr{E}^{\otimes \, 2}$ according to the general formula sated above in this Subsection.
Namely, let for any $\chi \in \mathscr{E}^{\otimes \, 2}$ the function $\chi^\natural \in \mathscr{E}^{\otimes \, 2}$ be defined by the rule
\[
\chi^\natural(x,y) = \chi(x+y,y) = (\chi \circ L) (x,y),
\]
and let
\[
\chi^\natural = \sum \limits_{j} \phi_j \otimes \varphi_j
\]
be its expansion into simple tensors $\phi_j \otimes \varphi_j$ in $\mathscr{E}^{\otimes \, 2}$, so that
\[
\chi^\natural(x,y) = \sum \limits_{j} \phi_j(x)\varphi_j(y).
\]
Then for any $\chi$ which respects (\ref{xiInS00timesS00}), $\Omega \chi$ be defined by the formula
\[
\Omega \chi(x,y) = \sum \limits_{j} \big(\Omega' \phi_j\big)(x-y)\varphi_j(y),
\]
\[
\Omega (\chi) \circ L(x,y) = \sum \limits_{j} \big(\Omega' \phi_j\big)(x)\varphi_j(y).
\]
Equivalently
\[
\Omega (\chi) \circ L (x,y) = (\chi \circ L)(x,y) - \, \sum \limits_{|\beta|=0}^{\omega} \omega_{{}_{o \,\, \beta}} (x)
\,\, D_{{}_{x}}^{\beta} \big(\chi \circ L\big) (x=0,y)
\]
\[
x,y \in \mathbb{R}^4.
\]
From this the explicit formula for $\Omega \chi (x,y)$ immediately follows
for any $\chi$ which respects (\ref{xiInS00timesS00}):
\begin{equation}\label{EpsteinGlaserOmega}
\Omega \chi (x,y) = \chi(x,y)
- \, \sum \limits_{|\beta|=0}^{\omega} \omega_{{}_{o \,\, \beta}} (x-y)
\,\,\, D_{{}_{x}}^{\beta}\chi (x=y,y).
\end{equation}

By the results obtained above for each contraction distribution 
\[
\big(\kappa'_{0,1} \dot{\otimes} \kappa''_{0,1} \dot{\otimes} \kappa'''_{0,1} \dot{\otimes} \ldots \dot{\otimes} \kappa^{(q)}_{0,1} \big)
\, \otimes_q \,\, \big(\kappa'_{1,0} \dot{\otimes} \kappa''_{1,0}  \dot{\otimes} \kappa'''_{1,0} \dot{\otimes} \ldots \dot{\otimes} \kappa^{(q)}_{1,0} \big)
\]
the contraction integral 
\[
\Big\langle \theta_y \big(\kappa'_{0,1} \dot{\otimes} \kappa''_{0,1} \dot{\otimes} \kappa'''_{0,1} \dot{\otimes} \ldots \dot{\otimes} \kappa^{(q)}_{0,1} \big)
\, \otimes||_{{}_{q}} \,\, \big(\kappa'_{1,0} \dot{\otimes} \kappa''_{1,0}  \dot{\otimes} \kappa'''_{1,0} \dot{\otimes} \ldots \dot{\otimes} \kappa^{(q)}_{1,0} \big), \,\,
\chi \Big\rangle
\]
\[
\overset{\textrm{df}}{=}
\Big\langle \theta_y \big(\kappa'_{0,1} \dot{\otimes} \kappa''_{0,1} \dot{\otimes} \kappa'''_{0,1} \dot{\otimes} \ldots \dot{\otimes} \kappa^{(q)}_{0,1} \big)
\, \otimes_q \,\, \big(\kappa'_{1,0} \dot{\otimes} \kappa''_{1,0}  \dot{\otimes} \kappa'''_{1,0} \dot{\otimes} \ldots \dot{\otimes} \kappa^{(q)}_{1,0} \big), \,\,
\Omega \chi \Big\rangle
\]
\begin{multline}\label{Int[A'(-),A'(+)]^qOmegaxi}
=
\int 
u'(\boldsymbol{\p}') v'(\boldsymbol{\p}') u''(\boldsymbol{\p}'') v''(\boldsymbol{\p}'') \ldots u^{(q)}(\boldsymbol{\p}^{(q)}) v^{(q)}(\boldsymbol{\p}^{(q)}) 
\, \times
\\
\times \,
e^{-i(|\boldsymbol{p}'| + |\boldsymbol{p}''| + \ldots +|\boldsymbol{p}^{(q)}|)(x_0-y_0) +i(\boldsymbol{p}' 
+ \boldsymbol{p}'' + \ldots + \boldsymbol{p}^{(q)})\cdot (\boldsymbol{\x} - \boldsymbol{\y})}
\,\, \times 
\\
\times \,\,
\theta(x-y) \, \Omega \chi(x,y)
\, \ud^3 \boldsymbol{p}' \, \ud^3 \boldsymbol{p}'' \ldots \ud^3 \boldsymbol{p}^{(q)} \, \ud^4 x \ud^4 y 
\end{multline}
\[
=
\int \kappa_q(x-y) \theta(x-y) \, \Omega \chi(x,y) \,\, \ud^4 x \ud^4 y
=
\int \kappa_q(x) \theta(x) \, \Omega (\chi \circ L)(x,y) \,\, \ud^4 x \ud^4 y
\]
\begin{multline*}
=
\int 
u'(\boldsymbol{\p}') v'(\boldsymbol{\p}') u''(\boldsymbol{\p}'') v''(\boldsymbol{\p}'') \ldots u^{(q)}(\boldsymbol{\p}^{(q)}) v^{(q)}(\boldsymbol{\p}^{(q)}) 
\, \times
\\
\times \,
e^{-i(|\boldsymbol{p}'| + \ldots +|\boldsymbol{p}^{(q)}|)x_0 +i(\boldsymbol{p}' 
+ \ldots + \boldsymbol{p}^{(q)})\cdot \boldsymbol{\x}}
\,\, \times 
\\
\times \,\,
\theta(x) \, \Omega (\chi \circ L) (x,y)
\, \ud^3 \boldsymbol{p}' \, \ud^3 \boldsymbol{p}'' \ldots \ud^3 \boldsymbol{p}^{(q)} \, \ud^4 x \ud^4 y 
\end{multline*}
converges absolutely and uniformly  for $\chi$ ranging over any bounded 
set in  $\mathcal{S}(\mathbb{R}^4; \mathbb{C}^{d'd''}) \otimes \mathcal{S}(\mathbb{R}^4; \mathbb{C}^{d'd''})$.

Therefore, 
\[
 \big(\theta_{y} \kappa'_{0,1} \dot{\otimes} \kappa''_{0,1} \dot{\otimes} \kappa'''_{0,1} \dot{\otimes} \ldots \dot{\otimes} \kappa^{(q)}_{0,1} \big)
\, \otimes_q \,\, \big(\kappa'_{1,0} \dot{\otimes} \kappa''_{1,0}  \dot{\otimes} \kappa'''_{1,0} \dot{\otimes} \ldots \dot{\otimes} \kappa^{(q)}_{1,0} \big)
\circ \Omega
\]
exists in 
\[
\mathscr{L}\big(\mathscr{E}^{\otimes \, 2}, \mathbb{C} \big) = \mathscr{E}^{* \, \otimes \, 2}.
\]
and analogously for $q' \leq q$
\[
\theta_{y} \big(\kappa'_{0,1} \dot{\otimes} \kappa''_{0,1} \dot{\otimes} \kappa'''_{0,1} \dot{\otimes} \ldots \dot{\otimes} \kappa^{(q')}_{0,1} \big)
\, \otimes_{q'} \,\, \big(\kappa'_{1,0} \dot{\otimes} \kappa''_{1,0}  \dot{\otimes} \kappa'''_{1,0} \dot{\otimes} \ldots \dot{\otimes} \kappa^{(q)}_{1,0} \big)
\circ \Omega
\]
in 
\[
\mathscr{L}\big(\mathscr{E}^{\otimes \, 2}, \,\, E_{{{}_{(q'+1)}}}^{*} \otimes \ldots \otimes E_{{{}_{(q)}}}^{*}\big);
\]
provided 
$\omega$ in (\ref{EpsteinGlaserOmega}) is
equal at least to the singularity degree at zero of the distribution
\[
 \big( \kappa'_{0,1} \dot{\otimes} \ldots \kappa^{(q')}_{0,1} \big)
\, \otimes_{q'}  \,\, \big( \kappa'_{1,0} \dot{\otimes} 
 \ldots \dot{\otimes}  \kappa^{(q')}_{1,0} \big)
\] 
in the sense of \cite{Epstein-Glaser}, explained above. 

However, note that the extension 
\[
\big(\theta_{y} \kappa'_{0,1} \dot{\otimes} \kappa''_{0,1} \dot{\otimes} \kappa'''_{0,1} \dot{\otimes} \ldots \dot{\otimes} \kappa^{(q)}_{0,1} \big)
\, \otimes_q \,\, \big(\kappa'_{1,0} \dot{\otimes} \kappa''_{1,0}  \dot{\otimes} \kappa'''_{1,0} \dot{\otimes} \ldots \dot{\otimes} \kappa^{(q)}_{1,0} \big)
\circ \Omega
\]
of the contraction distribution  
\[
\big(\theta_{y} \kappa'_{0,1} \dot{\otimes} \kappa''_{0,1} \dot{\otimes} \kappa'''_{0,1} \dot{\otimes} \ldots \dot{\otimes} \kappa^{(q)}_{0,1} \big)
\, \otimes_q \,\, \big(\kappa'_{1,0} \dot{\otimes} \kappa''_{1,0}  \dot{\otimes} \kappa'''_{1,0} \dot{\otimes} \ldots \dot{\otimes} \kappa^{(q)}_{1,0} \big)
\]
defined on the image of $\Omega$,
consisting of all those test functions $\chi$ which respect (\ref{Dalphaphi(x,x)=0,alpha=<omega}),
to the space of all test functions, is not unique. The singularity 
degree $\omega$ at zero of the retarded (and advanced) part of the 
product of pairings distribution ($\otimes_q$-contraction)
\[
\big(\theta_{y} \kappa'_{0,1} \dot{\otimes} \kappa''_{0,1} \dot{\otimes} \kappa'''_{0,1} \dot{\otimes} \ldots \dot{\otimes} \kappa^{(q)}_{0,1} \big)
\, \otimes_q \,\, \big(\kappa'_{1,0} \dot{\otimes} \kappa''_{1,0}  \dot{\otimes} \kappa'''_{1,0} \dot{\otimes} \ldots \dot{\otimes} \kappa^{(q)}_{1,0} \big)
\]
is the same as the singularity degree $\omega$ of that distribution at zero. It is determined up to a distribution 
supported at zero, and we can add a distribution
\[
\sum \limits_{|\alpha| =0}^{\omega} C_\alpha D^{\alpha}_{{}_{x}}\delta(x-y)
\]
to 
\begin{multline*}
\big( \theta_{y}\kappa'_{\epsilon \,\, 0,1} \dot{\otimes} \kappa''_{0,1} 
\dot{\otimes} \kappa'''_{0,1} \dot{\otimes} \ldots \big)
\, \otimes||_{{}_{q}}  \,\, \big( \kappa'_{\epsilon \,\, 1,0} \dot{\otimes} \kappa''_{1,0} 
\dot{\otimes} \kappa'''_{1,0} \dot{\otimes} \ldots \big)
\\
= \big(\theta_{y} \kappa'_{0,1} \dot{\otimes} \kappa''_{0,1} \dot{\otimes} \kappa'''_{0,1} \dot{\otimes} \ldots \dot{\otimes} \kappa^{(q)}_{0,1} \big)
\, \otimes_q \,\, \big(\kappa'_{1,0} \dot{\otimes} \kappa''_{1,0}  \dot{\otimes} \kappa'''_{1,0} \dot{\otimes} \ldots \dot{\otimes} \kappa^{(q)}_{1,0} \big)
\circ \Omega
\end{multline*}
or, equivalently, we can add the expression
\[
\sum \limits_{|\alpha| =0}^{\omega} (-1)^{|\alpha|} C_\alpha \int D^{\alpha}_{{}_{x}}\Big|_{{}_{x=0}}\chi(x+y,y) \, \ud^4 y
\]
to the contraction given by the absolutely
convergent integral (\ref{Int[A'(-),A'(+)]^qOmegaxi}), with arbitrary constants $C_\alpha$ in it. This is because
\[
\sum \limits_{|\alpha| =0}^{\omega} C_\alpha D^{\alpha}\delta(x)
\]
is the most general continuous functional on the kernel of $\Omega'$, equal zero on the image of $\Omega'$, with $\omega$ less
than or equal to the singularity order.

Note, please, that $\phi \in \mathscr{E}$ lies in the image $\textrm{Im} \, \Omega'$ 
if and only if  $\phi \in \textrm{ker} \, [\boldsymbol{1} - \Omega']$.
Therefore, the image $\textrm{Im} \, \Omega'$ of the continuous idempotent operator $\Omega' = \Omega'^{2}$, being equal
\[
\textrm{Im} \, \Omega' = \textrm{ker} \, [\boldsymbol{1} - \Omega']
\]
is closed in $\mathscr{E}$.
Thus,
\[
\mathscr{E} = \textrm{ker} \, \Omega' \oplus \textrm{Im} \, \Omega'
= \textrm{ker} \, \Omega' \oplus \textrm{ker} \, [\boldsymbol{1} - \Omega']
\]
is equal to the direct sum of closed sub-spaces $\textrm{Im} \, \Omega'$ and $\textrm{ker} \, \Omega'$. 
It is immediately seen that for each fixed $\omega$ 
the functional (distribution) 
\[
\delta^{\omega}_{{}_{1;2}}(x) = \sum \limits_{|\alpha| =0}^{\omega} C_\alpha D^{\alpha}\delta(x)
\]
is the most general functional on
$\textrm{Ker} \, \Omega' \subset \mathscr{E}$ which vanishes on $\textrm{Im} \, \Omega' \subset \mathscr{E}$.  
Because in addition $\delta^{\omega}_{{}_{1;2}}$ is equal identically zero on $\textrm{Im} \, \Omega'$ then the most general extension of the
retarded part 
\[
\kappa_q \circ \theta.\Omega'
\]
of $\kappa_q$ from $\textrm{Im} \Omega' \subset \mathscr{E}$  all over the whole $\mathscr{E}$
is equal to the sum
\begin{equation}\label{freedomInsplitting}
\kappa_q \circ \theta.\Omega' + \delta^{\omega}_{{}_{1;2}}.
\end{equation}

Equivalently the most general translationally invariant extension of the retarded part
\[
\big(\theta_{y} \kappa'_{0,1} \dot{\otimes} \kappa''_{0,1} \dot{\otimes} \kappa'''_{0,1} \dot{\otimes} \ldots \dot{\otimes} \kappa^{(q)}_{0,1} \big)
\, \otimes_q \,\, \big(\kappa'_{1,0} \dot{\otimes} \kappa''_{1,0}  \dot{\otimes} \kappa'''_{1,0} \dot{\otimes} \ldots \dot{\otimes} \kappa^{(q)}_{1,0} \big)
\circ \Omega
\]
of 
\[
\big(\kappa'_{0,1} \dot{\otimes} \kappa''_{0,1} \dot{\otimes} \kappa'''_{0,1} \dot{\otimes} \ldots \dot{\otimes} \kappa^{(q)}_{0,1} \big)
\, \otimes_q \,\, \big(\kappa'_{1,0} \dot{\otimes} \kappa''_{1,0}  \dot{\otimes} \kappa'''_{1,0} \dot{\otimes} \ldots \dot{\otimes} \kappa^{(q)}_{1,0} \big)
\] 
from the subspace $\textrm{Im} \, \Omega \subset \mathscr{E}^{\otimes \, 2}$ all over the whole 
\[
\mathscr{E}^{\otimes \, 2} = \textrm{ker} \, \Omega \oplus \textrm{Im} \, \Omega
= \textrm{ker} \, \Omega \oplus \textrm{ker} \, [\boldsymbol{1} - \Omega]
\]
is equal to the sum
\[
\big(\theta_{y} \kappa'_{0,1} \dot{\otimes} \kappa''_{0,1} \dot{\otimes} \kappa'''_{0,1} \dot{\otimes} \ldots \dot{\otimes} \kappa^{(q)}_{0,1} \big)
\, \otimes_q \,\, \big(\kappa'_{1,0} \dot{\otimes} \kappa''_{1,0}  \dot{\otimes} \kappa'''_{1,0} \dot{\otimes} \ldots \dot{\otimes} \kappa^{(q)}_{1,0} \big)
\circ \Omega + \delta^{\omega}_{{}_{1;2}},
\]
where $\delta^{\omega}_{{}_{1;2}}$ regarded as distribution, or functional, of two variables $x,y$ is equal
\[
\delta^{\omega}_{{}_{1;2}}(x,y) = \sum \limits_{|\alpha| =0}^{\omega} C_\alpha D^{\alpha}\delta(x-y).
\]

The same additional translationally invariant functional $\delta^{\omega}_{{}_{1;2}}$ will have to be added to the advanced part
\[
-\check{\theta}_{y}\kappa'_{\epsilon \,\, 0,1} \dot{\otimes} \dots  \dot{\otimes} \kappa^{(q)}_{\epsilon \,\, 0,1} \big)
\, \otimes_q  \,\, \big(\kappa'_{\epsilon \,\, 1,0} \dot{\otimes} \ldots \dot{\otimes} \kappa^{(q)}_{\epsilon \,\, 1,0} \big)  \circ \Omega
= 
-\check{\theta}_{y}\kappa'_{0,1} \dot{\otimes} \ldots  \dot{\otimes} \kappa^{(q)}_{0,1} \big)
\, \otimes||_{{}_{q}}  \,\, \big(\kappa'_{\epsilon \,\, 1,0} \dot{\otimes} \ldots \dot{\otimes} \kappa^{(q)}_{1,0} \big)
\]
where $\check{\theta}(x) = \theta(-x)$.

Recall that here $\theta_y(x) = \theta(x-y)$, $\theta_{\varepsilon \,\, y}(x) = \theta_\varepsilon(x-y)$, 
where $x$ is the space-time variable in the kernels 
\[
\kappa'_{\epsilon \,\, 0,1} \dot{\otimes} \kappa''_{\epsilon \,\, 0,1} \dot{\otimes} \kappa'''_{\epsilon \,\, 0,1} \dot{\otimes} \ldots
\,\,\,\, \textrm{and}
\,\,\,\,\,\,\,\,
\kappa'_{0,1} \dot{\otimes} \kappa''_{0,1} \dot{\otimes} \kappa'''_{0,1} \dot{\otimes} \ldots 
\]
and $y$ is the space-time variable in the kernels
\[
\kappa'_{\epsilon \,\, 1,0} \dot{\otimes} \kappa''_{\epsilon \,\, 1,0} \dot{\otimes} \kappa'''_{\epsilon \,\, 1,0} \dot{\otimes} \ldots
\,\,\,\,\, \textrm{and}
\,\,\,\,\,\,\,\,
\kappa'_{1,0} \dot{\otimes} \kappa''_{1,0} \dot{\otimes} \kappa'''_{1,0} \dot{\otimes} \ldots.
\]

Similarly, we show that for the kernels $\kappa'_{\ell',m'}, \kappa''_{\ell'',m''}$  of the Fock 
expansion of the operator $\Xi' = \mathcal{L}$, the \emph{double limit contraction} $\otimes||_{{}_{q}}$
\begin{multline*}
\theta_{{}_{'}} \,\, \kappa'_{\ell',m'} \otimes||_{{}_{q}}  \,   \kappa''_{\ell'',m''} \overset{\textrm{df}}{=}
\underset{\varepsilon, \epsilon \rightarrow 0}{\textrm{lim}} \,\,\,
\theta_{{}_{\varepsilon \,\, '}} \,\, \kappa'_{\epsilon \,\, \ell',m''} \otimes_q \,  \kappa''_{\epsilon \,\, \ell'',m''} \circ \Omega
= \theta_{{}_{ '}} \,\, \kappa'_{\ell',m''} \otimes_q \,  \kappa'';_{\ell'',m''} \circ \Omega
\end{multline*}
exists in the ordinary topology of uniform convergence on bounded sets in
\[
\mathscr{L}\big(\mathscr{E} \otimes \mathscr{E} , E_{j_1}^{*} \ldots  \otimes \ldots E_{j_{\ell'+\ell''-q}}^{*} \otimes E_{j_{\ell'+\ell''-q+1}}^{*} 
\ldots \otimes \ldots E_{j_{\ell'+\ell''+m'+m''-q}}^{*}  \big)
\]
\[
\cong \mathscr{L}\big(E_{j_1} \ldots  \otimes \ldots E_{j_{\ell'+\ell''-q}} \otimes E_{j_{\ell'+\ell''-q+1}} 
\ldots \otimes \ldots E_{j_{\ell'+\ell''+m'+m''-q}}, \, \mathscr{E}^* \otimes \mathscr{E}^*   \big),
\]
so that we can define 
\[
\theta_{{}_{'}} \,\, \kappa'_{\ell',m'} \otimes||_{{}_{q}}  \,   \kappa'_{\ell'',m''} \overset{\textrm{df}}{=}
\underset{\varepsilon, \epsilon \rightarrow 0}{\textrm{lim}} \,\,\,
\theta_{{}_{\varepsilon \,\, '}} \,\, \kappa'_{\epsilon \,\, \ell',m'} \otimes_q \,  \kappa''_{\epsilon \,\, \ell'',m''} \circ \Omega
\]
as existing in
\[
\mathscr{L}\big(\mathscr{E} \otimes \mathscr{E} , E_{j_1}^{*} \ldots  \otimes \ldots E_{j_{\ell'+\ell''-q}}^{*} \otimes E_{j_{\ell'+\ell''-q+1}}^{*} 
\ldots \otimes \ldots E_{j_{\ell'+\ell''+m'+m''-q}}^{*}  \big)
\]
\[
\cong \mathscr{L}\big(E_{j_1} \ldots  \otimes \ldots E_{j_{\ell'+\ell''-q}} \otimes E_{j_{\ell'+\ell''-q+1}} 
\ldots \otimes \ldots E_{j_{\ell'+\ell''+m'+m''-q}}, \, \mathscr{E}^* \otimes \mathscr{E}^*   \big),
\]
and determined up to the Epstein-Glaser reminder kernel of the general form given above.

Here we have put $\theta_{{}_{'}}$ for the function $\theta_{{}_{'}}(x',x'') = \theta(x'-x'')$
with the space-time variable $x'$ in the kernel $\kappa'_{\ell',m'}$ and with the space-time variable $x''$ in the kernel
$\kappa''_{\ell'',m''}$. 
 Similarly, for  $\theta_{{}_{\varepsilon \,\, '}}$ denoting the function 
$\theta_{{}_{\varepsilon \,\, '}}(x',x'') = \theta_\varepsilon(x'-x'')$, 
with the space-time variable $x'$ in the kernel $\kappa'_{\epsilon \,\, \ell',m'}$ and with the space-time variable $x''$ in the kernel
$\kappa''_{\epsilon \,\, \ell'',m''}$.

We should emphasize here that the above formula for the splitting 
\[
\kappa_q-(-1)^q\check{\kappa_q} = \textrm{ret} \, \big[ \kappa^{(-)}_q -(-1)^{f(q)}\kappa^{(+)}_q\big]
- \textrm{av} \, \big[\kappa^{(-)}_q -(-1)^{f(q)}\kappa^{(+)}_q \big]
\]
of the causal distribution (or the causal combination of the 
scalar $\otimes_q$-contractions) $\kappa^{(-)}_q -(-1)^{f(q)}\kappa^{(+)}_q$, into the part supported in the future and, respectively,
past light cones, uses in fact one more assumption: that the retarded and advanced parts of
$\kappa^{(-)}_q -(-1)^{f(q)}\kappa^{(+)}_q$
should have the same singularity order as $\kappa^{(-)}_q -(-1)^{f(q)}\kappa^{(+)}_q$ does. 
Similarly in our computation the retarded and advanced parts of $\kappa_q$ have the same singularity order as 
$\kappa_q$. Some authors call this assumtion
\emph{preservation of the Steinmann scaling degree}. In fact, in our, or rather Epstein-Glaser-Scharf computation of the retarded part,
we proceed in two steps. First we find the maximal subspace $\Omega'\mathscr{E}$ of the test space $\mathscr{E}$, 
on which the retarded (and advanced) part is defined through the natural formula, given by the ordinary multiplication 
by the step theta function and put zero on the complementary subspace $[\boldsymbol{1}-\Omega']\mathscr{E}$,
on which the natural formula does not work. This subspace $\Omega'\mathscr{E}$
turns out to be of finite codimension, meaning that its complementary 
subspace $[\boldsymbol{1}-\Omega']\mathscr{E}$, on which this formula does not work, has finite dimension. 
Then we extend this formula, by addition of the most general functional on $[\boldsymbol{1}-\Omega']\mathscr{E}$, 
which is zero on the subspace $\Omega'\mathscr{E}$.
Any such addition is supported at zero and can be added to the retarded and advanced part. Abandoning preservation of the 
singularity degree would allow addition of a functional (also supported at zero), but which would not be zero on $\Omega'\mathscr{E}$
and thus, which would modify the natural formula there, so the resulting retarded part no loner would be equal to the extension of the natural formula 
given by the multiplication by theta function on $\Omega'\mathscr{E}$. Thus our ``additional assumption'' is rather this: \emph{the retarded
part should coincide with the natural formula given by multiplication by the step theta function on a test function, whenever
the natural formula is meaningfull for this test function}. Thus, speaking intuitively: the splitiing should be ``maximally natural''.
As we have shown \cite{IF}, the remaining freedom in this  ``maximally natural'' splitting
(which preserves singularity order) can be completely eleminated in QED by the requirement that the  adiabatic limit $g\rightarrow 1$ exists for
 higher order contributions to interacting fields, and
should be equal to finite sums of well-defined integral kernel operators with vector valued kernels in the sense of \cite{obataJFA},
and this indeed holds for the so called ``natural normalization'' of the splitting. 

The formula
\[
\textrm{ret} \, \kappa_q =  \kappa_{q} \circ \theta.\Omega'
\]
replaces the naive multiplication $\theta \kappa_q$ by theta function. The point is that this formula
can also serve for the practical computation of $\textrm{ret}\kappa_q$, to which the method shown in \cite{Scharf}
works pretty well, although  $\kappa_q$ in general is not causal.
This can be done effectively for the Fourier transform $\widetilde{\textrm{ret}\kappa_q}$. Namely, denoting the 
space-time test function by $\phi$ we explicitly compute the evaluation integral
\[
\big\langle \widetilde{\textrm{ret} \, \kappa_q}, \widetilde{\phi}  \big\rangle =
\big\langle \textrm{ret} \, \kappa_q, \phi  \big\rangle = \big\langle \theta \kappa_q  \circ \Omega', \phi  \big\rangle
= \big\langle \kappa_q, \theta \Omega'\phi  \big\rangle =
\big\langle \widetilde{\kappa_q}, \widetilde{\theta} \ast \widetilde{\Omega'\phi}  \big\rangle,  
\]
which initially contains the auxiliary function(s) $w$ in
\[
\omega_{{}_{o \,\, \alpha}} = {\textstyle\frac{x^\alpha}{\alpha!}} w,
\]
used in the definition of the idempotent $\Omega'$. But using the fact that $\widetilde{\textrm{ret} \, \kappa_q}$
must have regularity regions, where it is given by ordinary function, 
with points at which it posses all derivatives, we can eliminate the auxiliary function(s) $w$  
in $\omega_{{}_{o \,\, \alpha}}$ by subtracting all terms up to order $\omega$ in the Taylor expansion
of $\widetilde{\textrm{ret} \, \kappa_q}$  around such a point (here chosen to be
zero, which is possible e.g. for spinor QED with massive charged field), 
with the very small cost that the obtained solution 
\begin{equation}\label{DispersionFormulaForRetarded}
\widetilde{\textrm{ret} \, \kappa_q}(p) = {\textstyle\frac{i}{2\pi}}  \int\limits_{-\infty}^{+\infty}
{\textstyle\frac{dt}{(t+i0)}}\big[\widetilde{\kappa_q}(p-tv) -
\sum\limits_{|\alpha|=0}^{\omega} {\textstyle\frac{p^\alpha}{\alpha!}} D^\alpha \widetilde{\kappa_q}(-tv)
\big],
\end{equation}
of the splitting problem is the one with specific normalization which is smooth around zero in
momentum space, with all derivatives (in momentum space) vanishing at zero up to order equal to the singularity
degree $\omega$ of $\kappa_q$. 
This cost is very small, as adding
\[
\sum \limits_{|\alpha| =0}^{\omega} C_\alpha p^{\alpha}
\]
to (\ref{DispersionFormulaForRetarded}) we obtain the most general solution of the splitting problem.
This choice of normalization is possible in spinor QED with the massive charged
field. For QED with massless charged field the normalization point in momenta will have to be shifted
from the zero point $p'=0$ to a point $p'\neq 0$ around which $\widetilde{\kappa_q}$ and $\widetilde{\textrm{ret} \, \kappa_q}$ 
are regular function-like distributions
and have all derivatives up to $\omega$ at $p' \neq 0$ 
and the respective formula $\widetilde{\textrm{ret} \, \kappa_q}$  
becomes slightly different
\[
\widetilde{\textrm{ret} \, \kappa_q}(p) = {\textstyle\frac{i}{2\pi}} \int\limits_{-\infty}^{+\infty}
{\textstyle\frac{dt}{(t+i0)}}\big[\widetilde{\kappa_q}(p-tv) -
\sum\limits_{|\alpha|=0}^{\omega} {\textstyle\frac{(p-p')^\alpha}{\alpha!}} D^\alpha \widetilde{\kappa_q}(p'-tv)
\big]
\]
giving solution with all derivatives up to order $\omega$ of $\widetilde{\textrm{ret} \, \kappa_q}$
equal zero at $p'\neq 0$.
Because in general the $\otimes_q$-contraction $\kappa_q$ is not causally supported, then its splitting into the retarded 
part (\ref{DispersionFormulaForRetarded}) and the advanced part is not invariant and depends on the unit time-like versor
$v$ of the coordinate system, which determines the $\theta$ function
\[
\theta(x) = \theta(v\cdot x), \,\,\, v=(1,0,0,0),
\]
with the ordinary $\theta$ function on the reals on the right hand side, so that (\ref{DispersionFormulaForRetarded})
in general depends on $v$.  

Of course dispersion integral (\ref{DispersionFormulaForRetarded}) has also its higher dimensional
analogue for translationally invariant causal distributions $d$ in more space-time variables, istead of $\kappa_q$, 
with the same derivation, given by the formulas (\ref{Def(retkappaq)})-(\ref{FT(retCausalkappaq)}), in which the step theta
function $\theta(x-y) = \theta(x_1-x_2)$ is replaced by its higher dimensional analogue
\[
\theta(x_1 -x_n) \ldots \theta(x_{n-1}-x_n) = \theta\big(v_1 \cdot (x_1-x_n)\big) \theta\big(v_2 \cdot (x_2-x_n)\big) \ldots \theta\big(v_n \cdot (x_{n-1}-x_n)\big),
\]
where on the right-hand-side we have the ordinary one dimensional step theta functions.
One can consider all the time only the scalar causal distributions $d$ in the Wick monomials
entering the Wick decomposition of the causally supported distribution $D_{(n)}(x_1, \ldots, x_n)$ in the Bogoliubov-Epstein-Glaser
method \cite{Epstein-Glaser}, as reported in \cite{Scharf} and in \cite{WN}. In this approach we are
confronted with the computation of the retarded and advanced parts of a set of multidimensional causal
distributions, which is enlarged together with the order, and cannot confine the splitting problem to a finite
set of causal distributions.

In particular for the causal distributions $d$ which are present in the Wick monomials
of the Wick decomposition of
\[
D_{(2)}(x_1,x_2) = \mathcal{L}(x_2)\mathcal{L}(x_1) - \mathcal{L}(x_1)\mathcal{L}(x_2)
\] 
these distributions $d$ are equal to the causal combinations of contractions $\kappa_q$,
because each $D_{(n)}$ is causally supported. Because these $d$ are causally supported
(within the closed future and past light cone), then the above computation of the retarded
part $\textrm{ret}d = \theta d \circ \Omega'$ simplifies and we arrive
at the formula
\begin{equation}\label{InvariantDispersionFormulaForRetarded}
\widetilde{\textrm{ret} \, d}(p) ={\textstyle\frac{i}{2\pi}} \int\limits_{-\infty}^{+\infty}
dt {\textstyle\frac{\widetilde{d}(tp)}{(t-i0)^{\omega +1}(1-t +i0)}}
\end{equation}
which is invariant, in particular independed of the time-like versor $v$ of the used Lorentz
coordinate system.  This (\ref{InvariantDispersionFormulaForRetarded}) is a solution
of the splitting problem with specific normalization point $p''=0$, which can be applid for $d$ with $\widetilde{d}$ smooth around zero in
momentum space, with all derivatives (in momentum space) of $\widetilde{d}$ vanishing at zero up to order 
equal to the singularity degree $\omega$ of $d$. Adding
\[
\sum \limits_{|\alpha| =0}^{\omega} C_\alpha p^{\alpha}
\]
to (\ref{InvariantDispersionFormulaForRetarded}) we obtain the most general solution of the splitting problem.
Again this choice of normalization is possible in spinor QED with the massive charged
field. For QED with massless charged field the normalization point in momenta will have to be shifted
from the zero point $p'=0$ to a point $p'\neq 0$ around which $\widetilde{\textrm{ret} \, d}$ is a regular function-like distribution
and has all derivatives up to $\omega$  equal zero at $p' \neq 0$.
 Dispersion integral formula  (\ref{InvariantDispersionFormulaForRetarded})
is valid for time-like momenta with $p\cdot p >0$, and in order to compute
$\widetilde{\textrm{ret} \, d}(p)$ for $p\cdot p <0$ we use the analytic continuation.

The derivation of (\ref{FT(retCausalkappaq)}) or
 (\ref{InvariantDispersionFormulaForRetarded})
for causal $d = \kappa^{(-)}_q -(-1)^{f(q)}\kappa^{(+)}_q$ with smooth $\widetilde{d}$ around zero
remains almost the same (with only minor changes) for any causal $d$ of several space-time variables, 
with smooth $\widetilde{d}$ around zero. We arrive with the the same dispersion itegral (\ref{InvariantDispersionFormulaForRetarded})
for such $d$. For some QFT, e.g. for Yang-Mills fields, the causal distributions $d$ are singular at zero
and the normalization point cannot be chosen at $p''=0$. But we have an analogous dispersion formula for such $d$
with the normalization point $p'' \neq 0$, compare \cite{DKS3}.

\subsection{Second order contribution $S_2$. Spinor QED}

Let us give the basic distributions for spinor QED with the standard realizations of the e.m. potential and Dirac free fields
(constructed with the Hida operators as the creation-annihilation operators).
These are the retarded and advanced parts of the $\otimes_q$-contractions, $q=1,2,3$, which appear in the Wick decomposition
of the product $\mathcal{L}(x_1)\mathcal{L}(x_2)$,
where $\mathcal{L}(x) = e {:} \boldsymbol{\psi}^{\sharp}\gamma^\mu \boldsymbol{\psi}A_\mu{:}(x)$ 
is the interaction Lagrangian density generalized operator for spinor QED.

Let $\kappa_{l,m}$, $\kappa^{\sharp}_{l,m}$, $\kappa'_{l,m}$, $l,m = 0,1$, 
be the plane wave kernels defining, respecively, the free Dirac field, its Dirac conjugation and the e.m. potential field.
The $\otimes_1$-contractions (pairings) are the following:

\begin{multline*}
\sum\limits_{a,b}\int \big[\boldsymbol{\psi}^{(-) \, a}(x), \boldsymbol{\psi}^{\sharp (+) \, b}(x)]_{{}_{+}} \, \phi_a(x)\varphi_b(y) \ud^4x \ud^4y
=
\sum\limits_{a,b}\int \quad\underbracket{\boldsymbol{\psi}^{a}(x) \boldsymbol{\psi}^{\sharp \, b}}(x) \, \phi_a(x)\varphi_b(y) \ud^4x \ud^4y
\\
= \sum\limits_{a,b,s}\int \kappa_{0,1}(s, \boldsymbol{\p};a,x)  \kappa_{1,0}^{\sharp}(s, \boldsymbol{\p};b,y)
\, \phi_a(x)\varphi_b(y) \, \ud^3\boldsymbol{\p} \, \ud^4x \ud^4y 
\\
= \kappa_{0,1}(\phi) \otimes_1 \kappa_{1,0}^{\sharp}(\varphi) = \big\langle \kappa_{0,1}(\phi), \kappa_{1,0}^{\sharp}(\varphi) \big\rangle
= -i S^{(-)}(\phi\otimes \varphi),
\end{multline*}
\[
\kappa_{1,0}^{\sharp}(\varphi) \otimes_1 \kappa_{0,1}(\phi) =
\kappa_{0,1}(\phi) \otimes_1 \kappa_{1,0}^{\sharp}(\varphi) = \big\langle \kappa_{0,1}(\phi), \kappa_{1,0}^{\sharp}(\varphi) \big\rangle
= -i S^{(-)}(\phi\otimes \varphi),
\]
\[
\kappa_{1,0}(\phi) \otimes_1 \kappa_{0,1}^{\sharp}(\varphi)=
\kappa_{0,1}^{\sharp}(\varphi) \otimes_1 \kappa_{1,0}(\phi) = \big\langle \kappa_{0,1}^{\sharp}(\varphi), \kappa_{1,0}(\phi) \big\rangle
= -i S^{(+)}(\phi\otimes \varphi),
\]
Similarly we have the contraction formula for the pairing
\begin{align*}
\quad\underbracket{
A_{\mu}(x)
A_{\nu}}(y) = i g_{\mu\nu} D_{0}^{(-)}(x-y) = \left[A_{\mu}^{(-)}(x),A_{\nu}^{(+)}(y) \right]_{-}
= \kappa'_{0,1}(\mu,x) \otimes_1 \kappa'_{1,0}(\nu,y)=\kappa'_{0,1}(\nu,x) \otimes_1 \kappa'_{1,0}(\mu,y),
\\
\left[A_{\mu}^{(+)}(x),A_{\nu}^{(-)}(y) \right]_{-} = i g_{\mu\nu} D_{0}^{(+)}(x-y) 
= \kappa'_{1,0}(\mu,x) \otimes_1 \kappa'_{0,1}(\nu,y)=\kappa'_{1,0}(\nu,x) \otimes_1 \kappa'_{0,1}(\mu,y),
\\\quad\underbracket{
A_{\nu}(y)
A_{\mu}}(x) 
=\kappa'_{0,1}(\nu,y) \otimes_1 \kappa'_{1,0}(\mu,x) = \kappa'_{1,0}(\mu,x) \otimes_1 \kappa'_{0,1}(\nu,y).
\end{align*}

We have the following $\otimes_2$-contractions:
\begin{multline*}
-\int \textrm{tr}\big[\gamma^\mu S^{(-)}(x-y) \gamma^\nu S^{(+)}(y-x)] \, \phi_\mu(x)\varphi_\nu(y) \ud^4x \ud^4y
\\
= 
\sum_{\substack{a,b,c,d \\ 
s,s'}} 
\int
\kappa^{\sharp}_{0,1}(s', \boldsymbol{\p}';a,x)\gamma^{\mu}_{ab}\kappa_{0,1}(s, \boldsymbol{\p}; b,x)
\kappa^{\sharp}_{1,0}(s, \boldsymbol{\p};c,y)\gamma^{\nu}_{cd}\kappa_{1,0}(s', \boldsymbol{\p}'; d,y) 
\, \times
\\
\times
\, \phi_\mu(x)\varphi_\nu(y) \ud^3\boldsymbol{\p} \ud^3\boldsymbol{\p}' \ud^4x \ud^4y
\end{multline*}
\begin{multline*}
=
\big(\kappa^{\sharp}_{0,1} \gamma^\mu \dot{\otimes} \kappa_{0,1}  \big)(\phi_\mu) \otimes_2
(\kappa^{\sharp}_{1,0} \dot{\otimes} \gamma^\nu \kappa_{1,0}\big)(\varphi_\nu)
\\
=
\sum_{\substack{a,b,c,d 
}} \int
\gamma^{\mu}_{ab} \gamma^{\nu}_{cd}
\quad\underbracket{
\boldsymbol{\psi}^{\sharp}_{a}(x)
\boldsymbol{\psi}_{d}}(y)
\quad\underbracket{
\boldsymbol{\psi}_{b}(x)
\boldsymbol{\psi}^{\sharp}_{c}}(y)
\, \phi_\mu(x)\varphi_\nu(y) \ud^4x \ud^4y
\end{multline*}
or 
\[
\sum\limits_{a,b,c,d}
\gamma^{\mu}_{ab} \gamma^{\nu}_{cd}
\quad\underbracket{
\boldsymbol{\psi}^{\sharp}_{a}(x)
\boldsymbol{\psi}_{d}}(y)
\quad\underbracket{
\boldsymbol{\psi}_{b}(x)
\boldsymbol{\psi}^{\sharp}_{c}}(y)
=
\big(\kappa^{\sharp}_{0,1} \gamma^\mu \dot{\otimes} \kappa_{0,1}\big) \otimes_2
(\kappa^{\sharp}_{1,0} \dot{\otimes} \gamma^\nu \kappa_{1,0}\big)(x, y) = \kappa^{\mu\nu}_{2}(x-y)
\]
Similarly
\begin{multline*}
\sum\limits_{a,d,\mu}\int \big[\gamma^\mu S^{(-)}\gamma_\mu\big]^{ad}(x-y)D_{0}^{(-)}(x-y) \phi_a(x)\varphi_d(y) \ud^4x \ud^4y
\\
= 
\sum_{\substack{a,b,c,d \\ \mu, \nu, s,s'}} 
\int
g_{\mu\nu}
\gamma^{\mu}_{ab}\kappa_{0,1}(s, \boldsymbol{\p}; b,x) \kappa'_{0,1}(s', \boldsymbol{\p}';\mu,x) \kappa^{\sharp}_{1,0}(s, \boldsymbol{\p};c,y)\gamma^{\nu}_{cd}
\kappa'_{1,0}(s', \boldsymbol{\p}';\nu,y)
\, \times
\\
\times
\, \phi_a(x)\varphi_d(y) \ud^3\boldsymbol{\p} \ud^3\boldsymbol{\p}' \ud^4x \ud^4y
\end{multline*}
\begin{multline*}
=
\sum\limits_{\mu, \nu}
\big(\gamma^\mu\kappa_{0,1} \dot{\otimes} \kappa'_{0,1 \,\, \mu}\big)(\phi) \otimes_2
(\kappa^{\sharp}_{1,0} \gamma^\nu \dot{\otimes} \kappa'_{1,0 \,\, \nu}\big)(\varphi)
\\
=
\sum_{\substack{a,b,c,d \\ \mu, \nu}} \int
\gamma^{\mu}_{ab} \gamma^{\nu}_{cd}
\quad\underbracket{
\boldsymbol{\psi}_{b}(x)
\boldsymbol{\psi}^{\sharp}_{c}}(y)
\quad\underbracket{
A_{\mu}(x)
A_{\nu}}(y)
\, \phi_a(x)\varphi_d(y) \ud^4x \ud^4y
\end{multline*}
or
\[
\sum_{\substack{b,c \\ \mu, \nu}}
\gamma^{\mu}_{ab} \gamma^{\nu}_{cd}
\quad\underbracket{
\boldsymbol{\psi}_{b}(x)
\boldsymbol{\psi}^{\sharp}_{c}}(y)
\quad\underbracket{
A_{\mu}(x)
A_{\nu}}(y)
=
\sum\limits_{\mu\nu}
\big(\gamma^\mu\kappa_{0,1} \dot{\otimes} \kappa'_{0,1 \,\, \mu}\big) \otimes_2
(\kappa^{\sharp}_{1,0} \gamma^\nu  \dot{\otimes}\kappa'_{1,0 \,\, \nu}\big)(a,x,d,y) = \kappa^{(-) \, ad}_{2}(x-y)
\]
\[
\sum_{\substack{b,c \\ \mu, \nu}}
\gamma^{\nu}_{ab} \gamma^{\mu}_{cd}
\quad\underbracket{
\boldsymbol{\psi}^{\sharp}_{c}(y)
\boldsymbol{\psi}_{b}}(x)
\quad\underbracket{
A_{\nu}(y)
A_{\mu}}(x)
=
\sum\limits_{\mu\nu}
\big(\gamma^\nu\kappa_{1,0} \dot{\otimes} \kappa'_{1,0 \,\, \mu}\big) \otimes_2
(\kappa^{\sharp}_{0,1} \gamma^\mu  \dot{\otimes} \kappa'_{0,1 \,\, \nu}\big)(a,x,d,y) = \kappa^{(+) \, ad}_{2}(x-y)
\]

Finally we have one $\otimes_3$-contraction:
\begin{multline*}
\sum_{\substack{a,b,c,d \\ \mu, \nu}} 
\gamma^{\mu}_{ab} \gamma^{\nu}_{cd}
\quad\underbracket{
\boldsymbol{\psi}^{\sharp}_{a}(x)
\boldsymbol{\psi}_{d}}(y)
\quad\underbracket{
\boldsymbol{\psi}_{b}(x)
\boldsymbol{\psi}^{\sharp}_{c}}(y)
\quad\underbracket{
A_{\mu}(y)
A_{\nu}}(x)
\\
= -i \textrm{tr}\big[\gamma^\mu S^{(-)}(x-y) \gamma_\mu S^{(+)}(y-x)]D_{0}^{(-)}(y-x)
\\
=
\sum\limits_{\mu\nu}
\big(\kappa^{\sharp}_{0,1} \dot{\otimes}\gamma^\mu \kappa_{0,1} \dot{\otimes} \kappa'_{0,1 \,\,\mu}\big) \otimes_3
(\kappa^{\sharp}_{1,0} \dot{\otimes} \gamma_\mu \kappa_{1,0} \dot{\otimes} \kappa'_{1,0 \,\,\nu}\big)(x, y) = \kappa_3(x-y)
\end{multline*}

Let us introduce $C^{\mu\nu}_{2}(x-y)= e^2[\kappa^{\mu\nu}_{2}(x-y)-\kappa^{\nu\mu}_{2}(y-x)] , 
K_{2}^{ad}(x-y)= e^2[\kappa^{(-) \, ad}_{2}(x-y) + \kappa^{(+) \, ad}_{2}(x-y)],
C_3(x-y) = e^2[\kappa_3(x-y) - \kappa_3(y-x)]$. It follows that $\kappa^{\mu\nu}_{2}(x-y)$ is a negative frequency $\otimes_2$-contraction,
and it is easily seen that $\kappa^{\nu\mu}_{2}(y-x)$ is the correspondinding positive frequency contraction. Similarly $\kappa_3(y-x)$ is 
the positive frequency $\otimes_3$-contraction correspnding to the negative frequency contraction $\kappa_{3}(x-y)$. Thus,
$C^{\mu\nu}_{2},K_{2}^{ad}(x-y), C_3$ are causally supported.  Recall that
\[
\kappa_{1,0} \otimes_1 \kappa_{0,1}
= \kappa_{0,1} \otimes_1 \kappa_{1,0}
= \kappa^{\sharp}_{1,0} \otimes_1 \kappa_{0,1}^{\sharp}
= \kappa^{\sharp}_{0,1} \otimes_1 \kappa_{1,0}^{\sharp}
=0,
\] 
so that in the contractions the kernels $\kappa_{0,1}, \kappa_{1,0}$ are always contracted, respectively, with the opposite frequency 
Dirac conjugated kernels $\kappa^{\sharp}_{1,0}, \kappa^{\sharp}_{0,1}$ and the kernels $\kappa'_{0,1}, \kappa'_{1,0}$, respectively, 
with the corresponding opposite frequency kernels  $\kappa'_{1,0}, \kappa'_{0,1}$.

By the application of the Wick theorem, we have 
(summation with repeated indices is understood)
\begin{multline*}
D_{(2)}(x_1,x_2) = S_1(x_2)S_1(x_1)-S_1(x_1)S_1(x_2) = -\mathcal{L}(x_2)\mathcal{L}(x_1)+\mathcal{L}(x_1)\mathcal{L}(x_2) 
\\
= e^2\Big[ 
-i \gamma^{\mu}_{ab}\gamma_{\mu \, cd} \left( D^{(-)}_{0}(x_2-x_1) - D^{(-)}_{0}(x_1-x_2) \right) \, 
{:} \boldsymbol{\psi}^{\sharp a}(x_1) \boldsymbol{\psi}^{b}(x_1)\boldsymbol{\psi}^{\sharp c}(x_2) \boldsymbol{\psi}^{d}(x_2){:}
\\
-i\big[\gamma^\mu S^{(-)}(x_1-x_2)\gamma^\nu + \gamma^\mu S^{(+)}(x_1-x_2)\gamma^\nu\big]_{ab} \, 
{:}\boldsymbol{\psi}^{\sharp a}(x_1) \boldsymbol{\psi}^{b}(x_2) 
A_\mu(x_1)A_\nu(x_2){:} 
\\
+i\big[\gamma^\mu S^{(-)}(x_2-x_1)\gamma^\nu+\gamma^\mu S^{(+)}(x_2-x_1)\gamma^\nu\big]_{ab} \, 
{:}\boldsymbol{\psi}^{\sharp a}(x_2) \boldsymbol{\psi}^{b}(x_1) 
A_\mu(x_2)A_\nu(x_1){:} \Big]
\\
+ K_{2}^{ab}(x_1-x_2) \,  {:}\boldsymbol{\psi}^{\sharp a}(x_1) \boldsymbol{\psi}^{b}(x_2){:}
- K_{2}^{ba}(x_2-x_1) \, {:}\boldsymbol{\psi}^{\sharp b}(x_2) \boldsymbol{\psi}^{a}(x_1){:}
\\
+C^{\mu\nu}_{2}(x_1-x_2) \, {:}A_\mu(x_1)A_\nu(x_2){:}
+ C_3(x_1-x_2) \, \boldsymbol{1} ,
\end{multline*}

Using the contraction formula in explicit form and the completeness relations for the plane wave kernels $\kappa_{0,1}$, $\kappa_{1,0}$ 
of the free fields we can compute the Fourier transforms 
(compare \cite{Scharf}) of these contraction distributions in explicit form:
\[
\begin{split}
\widetilde{C_{2}^{\mu\nu}}(p) =- {\textstyle\frac{e^2(2\pi)^{-3}}{3}} \big({\textstyle\frac{p^\mu p^\nu}{p^2}} - g^{\mu\nu}\big)
(p^2+2m^2)\sqrt{1-{\textstyle\frac{4m^2}{p^2}}}\theta(p^2-4m^2) \, \textrm{sgn}(p_0), \,\,\, \textrm{with $\omega = 2$}
\\
\widetilde{K_{2}}(p) = e^2 (2\pi)^{-3} \theta(p^2-m^2)  \, \textrm{sgn}(p_0) \, \big(1- {\textstyle\frac{m^2}{p^2}}\big)
\big[ m - {\textstyle\frac{\slashed{p}}{4}}\big(1+{\textstyle\frac{m^2}{p^2}}\big) \big], \,\,\, \textrm{with $\omega = 1$}
\\
\widetilde{C_{3}}(p) = - e^2 (2\pi)^{-5} \theta(p^2-4m^2) \, \textrm{sgn}(p_0) \, \Big[ 
\big({\textstyle\frac{p^4}{24}} + {\textstyle\frac{m^2}{12}}p^2 + m^4\big)\sqrt{1-{\textstyle\frac{4m^2}{p^2}}}
\\
+ {\textstyle\frac{m^4}{p^2}}(4m^2-3p^2) \textrm{ln} \big(\sqrt{{\textstyle\frac{p^2}{4m^2}}} + \sqrt{{\textstyle\frac{p^2}{4m^2}}-1}\big)
\Big], \,\,\, \textrm{with $\omega = 4$}.
\end{split}
\]

As we have seen computation of the retarded and advanced parts of the $\otimes_1$-contractions, or pairings, can be achieved by the 
ordinary multiplication by the $\theta$ function and this gives a non-invariant formula because the pairings are not causally supported,
but the retarded parts of their causal combitations -- the commutation functions (sums of the positive and negative frequency pairings) -- 
are invariant. 
In fact their singularity degree $\omega$ at zero is negative (equal $-1$ for the Dirac field or $-2$ for the e.m. potential field), 
which together with the formulas (\ref{D(+)}) and (\ref{D(-)}) justifies this fact, although they are not causal. We have given a proof 
of this fact which is not based on the singularity degree. 
In order to compute the retarded parts $\textrm{ret} \, C_{2}, \textrm{ret} \, K_{2}$ and 
$\textrm{ret} \, C_3$ of the contractions $C_{2}, K_{2}$ and $C_3$
we just insert the above formulas for the Fourier transforms $\widetilde{C_2}, \widetilde{K_2}$ and $\widetilde{C_3}$  
into the dispersion formula (\ref{InvariantDispersionFormulaForRetarded}). 

In particular we can compute the kernel of the second order contribution to the scattering operator and obtain the result
\begin{multline*}
S_2(x_1,x_2)= -e^2\Big[ {:} \boldsymbol{\psi}^{\sharp}(x_1)\gamma^\mu \boldsymbol{\psi}(x_1)A_\mu(x_1)
\boldsymbol{\psi}^{\sharp}(x_2)\gamma^\nu \boldsymbol{\psi}(x_2)A_\nu(x_2){:}
\\
-i \gamma^{\mu}_{ab}\gamma_{\mu \, cd} D^{c}_{0}(x_1-x_2) \, 
{:} \boldsymbol{\psi}^{\sharp a}(x_1) \boldsymbol{\psi}^{b}(x_1)\boldsymbol{\psi}^{\sharp c}(x_2) \boldsymbol{\psi}^{d}(x_2){:}
\\
+i\big[\gamma^\mu S^c(x_1-x_2)\gamma^\nu\big]_{ab} \, 
{:}\boldsymbol{\psi}^{\sharp a}(x_1) \boldsymbol{\psi}^{b}(x_2) 
A_\mu(x_1)A_\nu(x_2){:} 
\\
+i\big[\gamma^\mu S^c(x_2-x_1)\gamma^\nu\big]_{ab} \, 
{:}\boldsymbol{\psi}^{\sharp a}(x_2) \boldsymbol{\psi}^{b}(x_1) 
A_\mu(x_2)A_\nu(x_1){:} 
\Big]
\\
- i\Sigma_{ab}(x_2-x_1) \, {:}\boldsymbol{\psi}^{\sharp a}(x_2) \boldsymbol{\psi}^{b}(x_1){:}
- i\Sigma_{ab}(x_1-x_2) \,  {:}\boldsymbol{\psi}^{\sharp a}(x_1) \boldsymbol{\psi}^{b}(x_2){:}
\\
-i\Pi^{\mu\nu}(x_1-x_2) \, {:}A_\mu(x_1)A_\nu(x_2){:}
+ \Upsilon(x_1-x_2) \, \boldsymbol{1} ,
\end{multline*}
where 
\[
S^c(x)= S_{\textrm{ret}}(x) - S^{(+)}(x) = S_{\textrm{av}}(x) + S^{(-)}(x), 
\,\,\,\,
D^{c}_{0}(x) = D^{\textrm{ret}}_{0}(x)-D^{(+)}_{0}(x),
\]
 and the explicit expressions 
for the Fourier transforms of the distributions $\Pi, \Sigma, \Upsilon$,
are given by the formulas (\ref{Pi})-(\ref{Upsilon}). Recall, that the (always causal) Pauli-Jordan commutation functions
are equal $D(x) = D^{(-)}(x) + D^{(+)}(x)$, \emph{e.g.} $S_{ab}(x) = S^{(-)}_{ab}(x) + S^{(+)}_{ab}(x)$ and that
$D^{(-)}_{0}(x_2-x_1) - D^{(-)}_{0}(x_1-x_2) = -D_0(x_1-x_2)$,
because $D^{(-)}_{0}(x_2-x_1) = - D^{(-)}_{0}(x_1-x_2)$. 

$S_2(x_1,x_2)$ can be obtained as equal to
\[
S_2(x_1,x_2) = \textrm{ret} \, D_{(2)}(x_1,x_2) -R'_{(2)}(x_1,x_2) = \textrm{ret} \, \big[\mathcal{L}(x_1)\mathcal{L}(x_2) -\mathcal{L}(x_2)\mathcal{L}(x_1) \big] 
- \mathcal{L}(x_2)\mathcal{L}(x_1)
\]
in which case we consider the causal scalar distributions $d$ in front of the Wick monomials in the Wick decomposition of $D_{(2)}$
and apply the invariant formula (\ref{InvariantDispersionFormulaForRetarded}) for the computation of $\widetilde{\textrm{ret} \, d}$,
compare \cite{Scharf}.

For example in order to get the vacuum polarization term 
\[
-i\Pi^{\mu\nu}(x_1-x_2) \, {:}A_\mu(x_1)A_\nu(x_2){:},
\]
we apply the Wick theorem 
to the operators
\[
\mathcal{L}(x_1)\mathcal{L}(x_2) - \mathcal{L}(x_2)\mathcal{L}(x_1) 
\,\,\,\,\,
\textrm{and}
\,\,\,\,\,
\mathcal{L}(x_2)\mathcal{L}(x_1)
\]
and collect all terms proportional to
${:}A_\mu(x_1)A_\nu(x_2){:}$. Next, using the formula (\ref{InvariantDispersionFormulaForRetarded}), we compute the retarded part of the scalar factor 
$d^{\mu\nu}(x_1-x_2) = C_{2}^{\mu\nu}(x_1-x_2)$ which multiplies ${:}A_\mu(x_1)A_\nu(x_2){:}$ (which is causal) in
\[
\mathcal{L}(x_2)\mathcal{L}(x_1) - \mathcal{L}(x_1)\mathcal{L}(x_2),
\]
and subtract the scalar factor $e^2 \kappa^{\nu\mu}_{2}(x_2-x_1)$ proportional to
${:}A_\mu(x_1)A_\nu(x_2){:}$ in the Wick decomposition of $\mathcal{L}(x_2)\mathcal{L}(x_1)$, finally getting
\[
-i\Pi^{\mu\nu}(x_1-x_2) = \textrm{ret} \, d^{\mu\nu}(x_1-x_2) -e^2 \kappa^{\nu\mu}_{2}(x_2-x_1).
\]

We perform computations in the momentum space for the Fourier transform 
$\widetilde{\textrm{ret} \, d^{\mu\nu}}(p)- e^2\widetilde{c_{2}^{\nu\mu}}(-p)= \widetilde{\textrm{ret} \, C_{2}^{\mu\nu}}(p)- e^2\widetilde{\kappa_{2}^{\nu\mu}}(-p)$ of
$ret \, d^{\mu\nu}(x) - e^2 \kappa^{\nu\mu}_{2}(-x)$, 
\emph{i.e.} apply the formula (\ref{InvariantDispersionFormulaForRetarded}) with $\widetilde{d^{\mu\nu}}(p)= \widetilde{C_{2}^{\mu\nu}}(p)$
substituted for $\widetilde{d}$ with singularity order $\omega=2$.
Note here that although in general the product of tempered distributions is not a well-defined distribution,
the products $d$ of pairings in the Wick decomposition of $D_{(n)}$, $R'_{(n)}$, $A'_{(n)}$,
are always well-defined and their valuations on the test functions is given by the contraction
of the tensor products of values of the products of kernels of free fields at the test functions, 
and which are always given by absolutely converget integrals,
compare our previous paper. 

Analogously, in order to compute the self-energy term
\[
-i {:}\psi^\sharp(x_1) \Sigma(x_1-x_2) \psi(x_2){:} 
\]
in $S_2(x_1,x_2)$ we collect all terms in the Wick decomposition proportional to 
${:}\psi^{\sharp \, a}(x_1) \psi^{b}(x_2){:}$
and repeat the above stated operations. We do similarly with all terms proportional to each fixed Wick monomial
in the Wick decomposition of 
\[
\mathcal{L}(x_2)\mathcal{L}(x_1) - \mathcal{L}(x_1)\mathcal{L}(x_2)
\,\,\,\,\,
\textrm{and}
\,\,\,\,\,
\mathcal{L}(x_2)\mathcal{L}(x_1),
\]
computing in this way the full second order contribution $S_2(x_1,x_2)$.

1) For the vacuum polarization term 
\[
-i \Pi^{\mu\nu}(x_1-x_2) \, {:}A_\mu(x_1)A_\nu(x_2){:}
\]
in $S_2(x_1,x_2)$ we get with the ``natural'' splitting
\begin{equation}\label{Pi}
\widetilde{{\Pi}_{\mu \nu}}(p) = 
(2\pi)^{-4} \big({\textstyle\frac{p_\mu p_\nu}{p^2}} - g_{\mu\nu}\big) \widetilde{\Pi}(p),
\,\,\,\,\,\,\,\,\,\,\,\,\,\,\,\,\,\,\,\,\,\,
\widetilde{\Pi}(p) =
{\textstyle\frac{e^2}{3}} p^2 p^2
\int\limits_{4m^2}^{\infty} {\textstyle\frac{s+2m^2}{s^2(p^2-s+i0)}}\sqrt{1-{\textstyle\frac{4m^2}{s}}} ds.
\end{equation}

2) For the self-energy term 
\[
-i {:}\psi^\sharp(x_1) \Sigma(x_1-x_2) \psi(x_2){:} 
\]
in $S_2(x_1,x_2)$ we get with the ``natural'' splitting
\begin{equation}\label{Sigma}
\widetilde{\Sigma}(p) = 
e^2 (2\pi)^{-4}
\left\{
\left[\textrm{ln} \big|1- {\textstyle\frac{p^2}{m^2}} \big| -i\pi \, \theta(p^2 - m^2)\right]
\, 
\left[ 
m\left(1-\textstyle{\frac{m^2}{p^2}} \right)-{\textstyle\frac{\slashed{p}}{4}}\left(1-{\textstyle\frac{m^4}{p^4}}\right)\right]
+ {\textstyle\frac{m^2}{p^2}}{\textstyle\frac{\slashed{p}}{4}}
-m
+{\textstyle\frac{\slashed{p}}{8}}
\right\},
\end{equation}

3) For the vacuum term 
\[
\Upsilon(x_1-x_2) \, \boldsymbol{1}
\]
in $S_2(x_1,x_2)$ we get with the ``natural'' splitting
\begin{multline}\label{Upsilon}
\widetilde{\Upsilon}(p)= \widetilde{\Upsilon''}(p) -\widetilde{\Upsilon'}(p),
\\
\widetilde{\Upsilon''}(p) =
i e^2 (2\pi)^{-6} m^4 
\Bigg\{
{\textstyle\frac{5p^4}{48m^4}} + {\textstyle\frac{2p^2}{3m^2}} + 1 
+\big(3- {\textstyle\frac{4m^2}{p^2}}\big) \, \textrm{ln}^2\Big(\sqrt{{\textstyle\frac{-p^2}{4m^2}}} + \sqrt{1-{\textstyle\frac{p^2}{4m^2}}}\Big)
\\
+
\big({\textstyle\frac{p^4}{24m^4}} + {\textstyle\frac{p^2}{12m^2}} + 1\big) \sqrt{1-{\textstyle\frac{4m^2}{p^2}}}
\textrm{ln}{\textstyle\frac{\sqrt{1-{\textstyle\frac{4m^2}{p^2}}}-1}{\sqrt{1-{\textstyle\frac{4m^2}{p^2}}}+1}}
\Bigg\}, 
\end{multline}
\begin{multline*}
\widetilde{\Upsilon'}(p) =
e^2 (2\pi)^{-5} \, \theta(p^2 - 4m^2) \, \theta(-p_0) \, 
\Big\{
\big({\textstyle\frac{p^4}{24}} + {\textstyle\frac{m^2}{12}} p^2 + m^4\big) \sqrt{1-{\textstyle\frac{4m^2}{p^2}}}
\\
+ {\textstyle\frac{m^4}{p^2}}(4m^2-3p^2)
\textrm{ln}\Big(\sqrt{{\textstyle\frac{p^2}{4m^2}}} + \sqrt{{\textstyle\frac{p^2}{4m^2}}-1}\Big)
\Big\}. 
\end{multline*}

\section{Conclusions: Renormalizability. Adiabatic limit}\label{Ren}

The renormalizability/non-renormalizability, characterizes the quasiasymptotics of the distributional kernels
of $S_n(x_1, \ldots, x_n)$ at zero (in space-time, or at infinity in the momentum). It can be expressed in terms
of the singularity order at zero (in space-time) of these kernels, \emph{i.e.} singularity degree of the contraction scalar
factors in the kernels of the higher order contributions  $S_n(x_1, \ldots, x_n)$ to the scattering operator. By the singularity
order $\omega$ of $S_n(x_1, \ldots, x_n)$ we mean the maximum of the singularity orders of all scalar contraction factors
in the kernels of $S_n(x_1, \ldots, x_n)$. As we have seen, the singularity order $\omega$ of  
$S_n(x_1, \ldots, x_n)$ characterizes the arbitrariness in the construction of $S_n(x_1, \ldots, x_n)$, which is determined
by a finite set of arbitrary constants if $\omega\geq 0$, the number of which depends on $\omega$. 
$S_n(x_1, \ldots, x_n)$ is uniquely determined if $\omega < 0$.
A theory is renormalizable if and only if the number of arbitrary constants is bounded from above by some finite $N$, 
independent of the order $n$. In that case, theory 
is determined by a finite number of constants, and is renormalizable in ordinary sense \cite{Bogoliubov_Shirkov}.
Thus, in order to check renormalizability it is necessary to compute the singularity order $\omega$ of
$S_n(x_1, \ldots, x_n)$. Having given the general formulas (\ref{omegaOfq-contraction}) for  the singularity orders
of general contractions (\ref{q-contraction}), we then compute the singularity order $\omega$ of $S_n(x_1, \ldots, x_n)$
by induction, inspired by \cite{Epstein-Glaser}, using the fact that the singularity order of tempered distributions
(contractions, their retarded and advanced parts) behaves additively under tensor product operation, and that
singularity order of $D^\alpha \kappa$ is equal $\omega + |\alpha|$ if $\omega$ is the singularity order of $\kappa$.

In order to give an explicit form to the renormalizability condition, obtained in \cite{Epstein-Glaser},
let us introduce some auxiliary definitions after \cite{Epstein-Glaser}. Let $\alpha = (\alpha_0,\alpha_1.\alpha_2,\alpha_3)$ 
be any Schwartz quadri-index, with the standard meaning of $|\alpha|$ and $\alpha!$. Let $r$ be a superquadri-index, \emph{i.e.},
a function $r: \alpha \rightarrow r(\alpha) \in \mathbb{N} \cup \{0\}$ with values in positive integers, 
which is zero for all $|\alpha|>M$, for some finite $M$. Finally, let $X$ be a finite set (in our case it will be
a set of space-time variables). Let $(r_j)_{j\in X}$ be multi-superquadri-index, \emph{i.e.} family of superquadri-indices indexed
by $X$, with
\[
|r|= \sum_{j\in X} |r_j|, \,\,\, |r_j| = \sum_{\alpha} r_j(\alpha), \,\,\,
r! = \underset{j\in X}{\prod} r_j!, \,\,\,\, r_j! = \underset{\alpha}{\prod} r_j(\alpha)!.
\]
Let $\mathbb{A}(x)$ be a scalar field. We consider renormalizability of the theory with the scattering operator
$S(g\mathcal{L})$, with the multi-component switching-off test function $g$, and the generalized interaction
Lagrangian $g\mathcal{L}(x) = \Sigma_{j} g_{j}(x)\mathcal{L}_{j}(x)$, with the terms $\mathcal{L}_{j}(x)$
equal to the Wick monomials in derivatives of the field $\mathbb{A}(x)$, including the field itself as 
possible factors:
\[
\mathcal{L}_{j}(x) \overset{\textrm{df}}{=} 
{:} \underset{\alpha}{\prod} (D^\alpha\mathbb{A}(x)^{\nu_j(\alpha)}{:},
\]
where for each $j$, $\nu_j$ is a superquadri-index, and with $\nu$ understood as multi-superquadri-index.
Consider the $n$-th order contribution
$S_{j_1 \ldots j_n r_1 \ldots r_n}(x_1, \ldots, x_n)$ to the scattering operator
with $n$ vertices of the type $j_1, \ldots, j_n$ to each of which, respectively, are attached $|r_1|, \ldots, |r_n|$
external lines. Among the $|r_k|$ external lines attached to the vertex $j_k$, $r_k(\alpha)$ have $D^\alpha\mathbb{A}$ derivative.  
The singularity degree $\omega$ of the $n$-th order contribution
$S_{j_1 \ldots j_n r_1 \ldots r_n}(x_1, \ldots, x_n)$ is equal \cite{Epstein-Glaser}:
\[
\omega = 4 + \sum_{k=1}^{n} \left[\sum_{\alpha}(1+|\alpha|)(\nu_{j_{{}_{k}}}(\alpha) - r_k(\alpha))-4\right].
\]
Thus, the said theory is renormalizable iff
\[
\sum_{\alpha}(1+|\alpha|)\nu_{j}(\alpha)  \leq 4, \,\,\,
\textrm{for all} \,\, j,
\]
as only in this case, $\omega$ does not grow unlimitedly with the growing order $n$, and non-renormalizable
if 
\[
\sum_{\alpha}(1+|\alpha|)\nu_{j}(\alpha)  > 4, \,\,\,
\textrm{for some} \,\, j.
\]
Proceeding inductively with spinor QED with the generalized interaction Lagrangian $g\mathcal{L}(x) = \Sigma_{j} g_{j}(x)\mathcal{L}_{j}(x)$ 
 (including, after Schwinger and \cite{Bogoliubov_Shirkov}, auxiliary terms allowing to compute interacting fields) 
with multi-component switching-off test function $g$ and with
\[
\mathcal{L}_0(x) = {:} \boldsymbol{\psi}^{+}(x)\gamma^0\gamma^\mu \boldsymbol{\psi}(x)A_{\mu}(x){:},
\,\,\,
\mathcal{L}_{a}(x) = \boldsymbol{\psi}^{a}(x), \,\,\, \mathcal{L}_b(x) = [\boldsymbol{\psi}^{+}(x)\gamma^0]^b, \,\,\,
\mathcal{L}_\mu(x) = A_\mu(x),
\]
we prove that the singularity degree $\omega$ of the $n$-th order contribution $S_{j_1 \ldots j_n r_1 \ldots r_n}(x_1, \ldots, x_n)$ 
is always equal $4$ minus a positive
integer $3m+k$ depending on the numbers $|r_0|, |r_1|, |r_2|, \ldots$, of external photon, electron and positron lines, \emph{i.e.}
with $m$ equal to the total number of fermion external lines divided by two and $k$ equal to the total number of external photon lines,  
so that e.g. the spinor QED is renormalizable. We should emphasize that the number of arbitrary constants involved in the splitting is not in general equal to 
the minimal number $N$ majorizing $\omega$, as it also depends on the geometric type (scalar, spinor, fourvector, e.t.c.) of the involved fields, 
and is further reduced by invariance properties of the considered QFT.  

As to the renormalizability, characterizing the ultraviolet (UV) asymptotics of the theory, usage of Hida operators 
adds nothing new. However, usage of Hida operators and the generalized integral kernel operators is crucial for the infrared (IR) asymptotics, and 
allows solving IR problems, which are intractable within the standard 
approach, using the generalized operators in the Wightman sense. This is essential for QFT with the interaction Lagrangian $\mathcal{L}(x)$,
regarded as Wick polynomial in free fields, containing massless fields. In this case, in general, there does not exist
the adiabatic limit $g \rightarrow 1$ (or $g_0 \rightarrow 1$ in the notation of the last paragraph) for interacting fields
in the standard approach. When using Hida operators situation is different, because in this case the adiabatic limit does exist,
and moreover, its existence puts further restrictions on the arbitrary constants involved in the splitting. We have not yet investigated
these phenomena in full generality for arbitrary  Wick polynomials $\mathcal{L}(x)$ involving massless fields, but have investigated
various QED's, and have shown that for QED's with minimal $U(1)$-coupling and massive charged field the adiabatic limit does exist 
and eliminates altogether the whole arbitrariness in the splitting, \emph{i.e.} the adiabatic limit exists for such QED's only if the normalization
in the splitting is ``natural''. Compare \cite{IF}, where explicit computations are provided for the first and second order, and the general
theorem follows by induction.
In \cite{IF} we have also proved that the adiabatic limit does not exist for  QED's with $U(1)$-coupling and massless charged fields, 
which can be interpreted as a theoretical proof that charged fields should be massive.

A problem, put by prof. Kazakov, arises: if the usage of Hida operators and the condition of existence of the adiabatic limit, 
restricting further the arbitrariness in the splitting, can throw some new light on some non-renormalizable theories. 
Of course, this can be the case only for theories 
with $\mathcal{L}(x)$ involving massless fields. We have not undertaken this problem in full generality yet, but the partial results  
\cite{IF} suggest that in general for QFT with $\mathcal{L}(x)$ (renormalizable or not) containing only massless fields as the Wick factors, 
adiabatic limit does not exist (order by order) with any choice in the splitting and, thus, this would eliminate them as 
mathematically inconsistent. The general ``mixed case'' with $\mathcal{L}(x)$ containing both, massless and massive Wick factors, 
is more involved, and it is not clear yet if the number of arbitrary constants involved in the splitting (growing with the order 
for non-renormalizable case) can be eliminated (or restricted to a finite number, independent of the order) 
by the requirement of the existence of the adiabatic limit. This problem, understood in this generality, is still open.

\section*{Acknowledgements}

The author would like to express his deep gratitude to Professor D. Kazakov  and Professor I. Volovich 
for the very helpful discussions.
He also would like to thank for the excellent conditions for work at JINR, Dubna.
He would like to thank Professor M. Je\.zabek 
for the excellent conditions for work at INP PAS in Krak\'ow, Poland
and would like to thank Professor A. Staruszkiewicz and 
Professor M. Je\.zabek for the warm encouragement. The author would like to acknowledge the Referee for his
suggestions.


\begin{thebibliography}{}
\bibitem{Berezin}  Berezin, F. A.: The method of second quantization. Acad. Press, New York, London, 1966.
\bibitem{Bogoliubov_Shirkov} Bogoliubov, N. N., Shirkov, D. V.: Introduction to the Theory of Quantized Fields. New York (1959), second ed. John Wiley \& Sons, Inc., New York, Chichester, Brisbane, Toronto, 1980.
\bibitem{DKS3} D\"utsch, M., Krahe, F., Scharf, G.: Nuovo Cimento A 107, 375 (1994).
\bibitem{Epstein-Glaser} Epstein, H., Glaser, V.: The role of locality in perturbation theory. Ann. Inst. H. Poincar\'e A19, 211-295 (1973).
\bibitem{GelfandI}  Gelfand, I. M., Shilov, G. E.: Generalized Functions. Vol I. Academic Press, New York, San Francisco, London, 1964.
\bibitem{GelfandIV} 
I. M. Gelfand and N. Ya. Vilenkin: Applications of Harmonic Analysis: Generalized functions. Vol. 4. 
Acad. Press, New York, 1964.
\bibitem{obata} 
N. Obata: An analytic characterization of symbols of
operators on white noise functionals. J. Math. Soc. Japan {\bf 45}, 421 (1993).
\bibitem{obataJFA} Obata, N.: Operator calculus on vector-valued white noise functionals.  J. of Funct. Anal. 121, 185-232 (1994).
\bibitem{Rudin} Rudin, W.: Functional Analysis. McGraw-Hill, Inc. 1991.
\bibitem{Reed_SimonII} Reed, M., Simon, B.: Methods of Modern Mathematical  Physics II:
Fourier Analysis. Self-Adjointness. Acad. Press, New York 1975.
\bibitem{Schaefer} 
H. H. Schaefer: Topological vector spaces. 2nd ed.  Springer, New York 1999. 
\bibitem{Scharf} Scharf, G: Finite Quantum electrodynamics, Dover Publications, Mineola, New York, 2014.
\bibitem{WN} Wawrzycki, J: Causal Perturbative QED and white noise. Submitted to 
Infinite Dimensional Analysis, Quantum Probability and Related Topics (Aug 13, 2021). ArXiv: math-ph $\slash$ 220305884.
\bibitem{IF} Wawrzycki, J: Causal perturbative QED and white noise. Interacting fields. To appear in: Theoretical and Mathematical Physics.
ArXiv: math-ph $\slash$ 220305854.
\bibitem{QFTSG} Wawrzycki, J: Causal perturbative QFT and space-time geometry. ArXiv: math-ph $\slash$180206719.



\end{thebibliography}
\end{document}